\documentclass[useAMS,usenatbib]{mn2e}
\usepackage{longtable,lscape,array}
\usepackage{amsmath}

\usepackage{graphicx}
\usepackage{aas_macros}
\usepackage{amssymb}

\newcommand{\about}{$\sim\!\!$~}
\newcommand{\kms}{km~s$^{-1}$}
\newcommand{\etal}{et~al.\ }

\newcommand{\bvri}{\protect\hbox{$BV\!RI$} }

\newcommand{\bvmax}{\protect\hbox{$\left(B-V\right)_{\rm max}$}}
\newcommand{\lcdm}{\protect\hbox{$\Lambda {\rm CDM}~$}}
\newcommand{\scriptr}{\protect\hbox{$\mathcal{R}$}}
\newcommand{\scriptrc}{\protect\hbox{$\mathcal{R}^c$}}
\mathchardef\mhyphen="2D

\newcommand{\be}{\begin{displaymath}}
\newcommand{\ee}{\end{displaymath}}

\def\lsim{\hbox{\rlap{\raise 0.425ex\hbox{$<$}}\lower 0.65ex\hbox{$\sim$}}}
\def\gsim{\hbox{\rlap{\raise 0.425ex\hbox{$>$}}\lower 0.65ex\hbox{$\sim$}}}

\newcommand{\ion}[2]{#1$\;${\small{#2}}\relax}

\graphicspath{{/Users/jsilv/BSNIP/jkong/Dropbox/BSNIPIII/plots/}{/Users/jsilv/BSNIP/jkong/Dropbox/BSNIPIII/ratios_SNF/}{/Users/jsilv/BSNIP/jkong/Dropbox/BSNIPIII/misc_indicators/}{/Users/mganesh/Desktop/Dropbox/BSNIPIII/plots/}{/Users/mganesh/Desktop/Dropbox/BSNIPIII/ratios_SNF/}{/Users/mganesh/Desktop/Dropbox/BSNIPIII/misc_indicators/}}

\title[BSNIP III: Spectral and Photometric Analysis]{Berkeley Supernova 
Ia Program III: Spectra Near Maximum Brightness Improve the Accuracy
of Derived Distances to Type Ia Supernovae}

\author[Silverman, et~al.]{Jeffrey~M.~Silverman,$^{1,2}$\thanks{E-mail:
    JSilverman@astro.berkeley.edu} Mohan~Ganeshalingam,$^{1}$ 
  Weidong~Li,$^{1}$\dag
\newauthor
Alexei~V.~Filippenko$^{1}$ \\
$^{1}$Department of Astronomy, University of California, Berkeley, CA 94720-3411, USA \\
$^{2}$Marc J. Staley Fellow \\
\dag Deceased 12 December 2011
}

\begin{document}
\date{Accepted  . Received   ; in original form  }
\pagerange{\pageref{firstpage}--\pageref{lastpage}} \pubyear{2012}
\maketitle
\label{firstpage}


\begin{abstract}
In this third paper in a series we compare spectral feature
measurements to photometric properties of 108 low-redshift ($z < 0.1$,
$<z> \approx 0.023$) Type~Ia supernovae 
(SNe~Ia) for which we have optical spectra within 5~d of maximum
brightness. The spectral data were obtained from 1989 through 
the end of 2008 as part of the Berkeley SN~Ia Program (BSNIP) and are
presented in BSNIP~I (Silverman et~al. 2012), and the photometric
data come mainly from the Lick Observatory Supernova Search (LOSS)
and are published by Ganeshalingam \etal (2010). The spectral
measurements are presented and discussed in BSNIP~II (Silverman, Kong
\& Filippenko 2012), and the light-curve fits and photometric parameters can be
found in Ganeshalingam \etal (in preparation). A variety of previously
proposed correlations between spectral and photometric parameters are
investigated using the large and self-consistent BSNIP dataset. 
We find {\it the
pseudo-equivalent width (pEW) of the \ion{Si}{II} $\lambda$4000 line
to be a good indicator of light-curve width, and the pEWs of
the \ion{Mg}{II} and \ion{Fe}{II} complexes are relatively good
proxies for SN colour.}
We also employ a combination of light-curve parameters (specifically the
SALT2 stretch and colour parameters $x_1$ and $c$, respectively) and spectral
measurements to calculate distance moduli. The residuals from these
models are then compared to the standard model which uses only
light-curve stretch and colour. Our investigations show that {\it a
  distance model that uses $x_1$, $c$, and the velocity of the
  \ion{Si}{II} $\lambda$6355 feature does not lead to a 
  decrease in the Hubble residuals.} We also find that {\it distance
  models with flux ratios alone or in conjunction with light-curve
  information rarely perform better than the standard
  $\left(x_1,c\right)$ model.} However, {\it when adopting
a distance model which combines the ratio of fluxes near
\about3750~\AA\ and \about4550~\AA\ with both $x_1$ and $c$, the
Hubble residuals are decreased by \about10~per~cent, which is found to
be significant at about the 2$\sigma$ level. The weighted
root-mean-square of the residuals using this model is $0.130 \pm
0.017$~mag (as compared with $0.144 \pm 0.019$~mag when using the same
sample with the standard model).} This Hubble diagram fit has one of
the smallest scatters ever published and at the highest significance
ever seen in such a study. Finally, these results are discussed with
regard to how they can improve the cosmological accuracy of future,
large-scale SN~Ia surveys.
\end{abstract}



\section{Introduction}\label{s:intro}

Type~Ia supernovae (SNe~Ia) have been used in the recent past to
discover the accelerating expansion of the Universe
\citep{Riess98:lambda,Perlmutter99}, as well as to measure
cosmological parameters with increasing accuracy and precision
\citep[e.g.,][]{Astier06, Riess07, Wood-Vasey07, Hicken09:cosmo,
Kessler09, Amanullah10,Sullivan11,Suzuki12}. In the most general terms,
thermonuclear explosions of carbon/oxygen (C/O) white dwarfs (WDs) are
thought to give rise to SNe~Ia (e.g., \citealt{Hoyle60, Colgate69,
  Nomoto84}; see \citealt{Hillebrandt00} for a review). However, after
decades of observations and theoretical work, a detailed understanding
of both the SN progenitors and explosion mechanisms is still
missing. In addition, there is very little known about how differences
in the initial conditions in SNe~Ia give rise to the measured range of
observables. A large, self-consistent dataset is needed in order to
solve these problems.

The ability to do precision cosmology using SNe~Ia requires that one
is able to calibrate or standardise their
luminosity. \citet{Phillips93} showed a tight correlation between
light-curve decline rate and luminosity at peak brightness for the
majority of SNe~Ia, the so-called ``Phillips relation.'' However, the
addition of spectral observations to the light-curve data complicates
the picture far beyond the simple assumption underlying the Phillips
relation. Many comparisons of spectral and photometric data of
low-redshift SNe~Ia have been performed in the past
\citep[e.g.,][]{Nugent95,Benetti05,Bongard06,Hachinger06,Arsenijevic08,Walker11,Nordin11a,Blondin11,Chotard11}.
In addition, there has been similar work with SNe~Ia at higher
redshifts
\citep[e.g.,][]{Blondin06,Altavilla09,Nordin11a,Walker11}. These
studies were often aimed at finding a ``second parameter'' in SN~Ia
spectral or photometric data which would increase the accuracy of
their distance measurements.

Most of these previous studies utilised relatively small and
heterogeneous datasets. The data studied here, in contrast, were
self-consistently observed and reduced, and constitute one of the
largest datasets to be analysed in this manner. Low-redshift ($z \le
0.1$, $<z> \approx 0.023$) optical SN~Ia spectra from the
Berkeley Supernova Ia Program \citep[BSNIP;][]{Silverman12:BSNIPI} are
used along with complementary 
photometric data, largely from \citet{Ganeshalingam10:phot_paper}. The
spectral features have been accurately and robustly measured
\citep*[BSNIP~II;][]{Silverman12:BSNIPII}, and the light curves have
been fit using a variety of methods (Ganeshalingam et~al., in
preparation).

We summarise both the spectral and photometric
datasets used for this analysis in Section~\ref{s:data}, and we
describe our procedure for measuring spectral features, fitting light
curves, and producing Hubble diagrams in
Section~\ref{s:procedure}. How these measured values correlate with 
each other and with previously determined classifications is presented
in Section~\ref{s:analysis}, along with our Hubble diagram results
using various models for the distances to SNe~Ia. We present our
conclusions in Section~\ref{s:conclusions}, where the main results of
our analysis are summarised and the most accurate and useful spectral
indicators are discussed. Other, future BSNIP papers will
expand on the analysis performed here with the addition of host-galaxy
properties and late-time SN spectra.


\section{Dataset}\label{s:data}

\subsection{Spectral Data}

The same SN~Ia spectral data are analysed in the current study as were
used in BSNIP~II.  The spectra are all originally published in
BSNIP~I. Most of the spectra were obtained using the Shane 3~m
telescope at Lick Observatory with the Kast double spectrograph
\citep{Miller93}, and the typical wavelength coverage is
3300--10,400~\AA\ with resolutions of \about11~\AA\ and \about6~\AA\ on the
red and blue sides (crossover wavelength \about5500~\AA), 
respectively. See BSNIP~I For more information regarding the 
observations and data reduction.

In BSNIP~II, we required that a spectrum be within 20~d
(rest frame) of maximum brightness and we {\it a priori} ignored the
extremely peculiar SN~2000cx \citep[e.g.,][]{Li01:00cx}, SN~2002cx
\citep[e.g.,][]{Li03:02cx,Jha06:02cx}, SN~2005hk
\citep[e.g.,][]{Chornock06,Phillips07}, and SN~2008ha
\citep[e.g.,][]{Foley09:08ha,Valenti09}. BSNIP~II contains 432 spectra
of 261 SNe~Ia with a ``good'' fit for at least one spectral
feature. However, only a subset of these data are used in the current
study since not all of them have reliable photometric
observations and we are currently only considering spectra within 5~d
of maximum brightness. It was shown in BSNIP~II that the spectral
measurements do not evolve significantly during these epochs. For the
11 SNe~Ia that had more than one spectrum within 5~d of 
maximum brightness and photometric information, we only use the
spectrum closest to the date of maximum in the current analysis.

For our investigation using arbitrary ratios of fluxes (see
Section~\ref{ss:ratios}) we also use spectra within 5~d of maximum
brightness, even though previous studies only investigated spectra
within 2.5~d of maximum \citep{Bailey09,Blondin11}. When we employ spectra 
only in this narrower age range, our sample size for the flux-ratio analysis
decreases from 62 to 38 objects. While our overall results are mostly
unchanged when considering only spectra within 2.5~d of maximum brightness,
their significance is weakened due to the smaller number of objects
used. Adopting a larger age range than within 5~d of maximum increases
the sample size only moderately and introduces much larger scatter in
spectral measurements (see BSNIP~II).

\subsection{Photometric Data}

A majority of the SNe in our spectral sample were discovered as part
of the Lick Observatory Supernova Search (LOSS). LOSS is a transient
survey utilising the 0.76-m Katzman Automatic Imaging Telescope (KAIT)
at Lick Observatory (\citealt{li00:kait,Filippenko01}; see also
Filippenko, Li, \& Treffers, in preparation). KAIT is a robotic
telescope that monitors a sample of \about15,000 galaxies in the
nearby Universe (redshift $z < 0.05$) with the goal of finding transients
within days of explosion. Fields are imaged every 3--10~d and compared
to archived template images, after which potential new transients are 
flagged. These images are examined by human image checkers and the
best candidates are re-observed the following night. Candidates that
are present on two consecutive nights are reported to the community
using the International Astronomical Union Circulars (IAUCs) and the
Central Bureau of Electronic Telegrams (CBETs). The statistical power
of the LOSS sample is well demonstrated by the series of papers
deriving nearby SN rates \citep{Leaman11:ratesI, Li11a, Li11b}.

In addition to the SN search, KAIT monitors active SNe of all types in
broad-band \bvri filters. The first data release of \bvri light curves
for 165 SNe~Ia along with details about the reduction procedure have
been published by \cite{Ganeshalingam10:phot_paper}. In summary,
point-spread function (PSF) fitting photometry is performed on images
from which the host galaxy has been subtracted using templates
obtained $> 1$~yr after explosion. Photometry is transformed to the
Landolt system \citep{Landolt83,Landolt92} using averaged colour terms
determined over many photometric nights. Calibrations for each SN
field are obtained on photometric nights with an average of 5
calibrations per field. 

We also include SN~Ia light curves obtained from the literature to
maximise the overlap between our photometric and spectroscopic
samples. We include 29 objects from the Cal\'an-Tololo sample
\citep{Hamuy96}, 22 objects from the CfA1 sample \citep{Riess99:lc},
44 objects from CfA2 \citep{Jha06}, and 185 objects from CfA3
\citep{Hicken09}. In instances where we have data for the same SN from 
multiple samples, we use the light curve that is most densely sampled
and best captures the light-curve evolution. We also include light
curves for SNe 1999aw \citep{Strolger02}, 1999ee, 2000bh, 2000ca,
2001ba \citep{Krisciunas04a}, 2001bt, 2001cn, and 2001cz
\citep{Krisciunas04b}. Our final photometry sample consists of 335 
multi-colour light curves, though we note that not all the objects in
this sample have corresponding spectroscopy. 

Of the data within 5~d of maximum brightness investigated in BSNIP~II,
115 SNe have light-curve width or colour information and are included
in the present study. The redshift range spanned by this sample is $0
< z < 0.1$. A complete list of these SNe~Ia, their ages, spectral
classifications, and spectral feature measurements can be found in
BSNIP~II. The photometric parameters of these objects are presented by
Ganeshalingam \etal (in preparation).

\section{Measurement Procedures}\label{s:procedure}

\subsection{Spectral Measurements}

The algorithm used to measure each of nine spectral features and the
features themselves are described in detail in BSNIP~II. Here we give
a brief summary of the procedure.

Each spectrum has its host-galaxy recession velocity removed and is 
corrected for Galactic reddening (according to the values presented in 
Table~1 of BSNIP~I), and it is smoothed using a Savitzky-Golay smoothing
filter \citep{Savitzky64}. If the signal-to-noise ratio (S/N) is larger 
than 6.5~pixel$^{-1}$ over the entire spectral range, we attempt to define a
pseudo-continuum for each spectral feature. This is done by
determining where the local slope changes sign on either side of the
feature's minimum. Quadratic functions are fit to each of these
endpoints and the peaks of the parabolas (assuming that they are both
concave downward) are used as the endpoints of the feature; they
are then connected with a line to define the pseudo-continuum.
We record the flux at the blue and red endpoints of the feature
($F_b$ and $F_r$, respectively) as well as the pseudo-equivalent width 
\citep[pEW; e.g.,][]{Garavini07}.

Once a pseudo-continuum is calculated, a cubic spline is fit to the
smoothed data between the endpoints of the spectral
feature.\footnote{No attempt is made to fit any of the Mg~II or Fe~II
  features in this manner; these complexes
consist of so many blended spectral lines that it is unclear which
reference wavelength to use when attempting to define an expansion
velocity.}  From the wavelength at which
the spline fit reaches its minimum ($\lambda_\textrm{min}$) the
expansion velocity ($v$) is calculated.  The flux is then normalised
to the pseudo-continuum, and the relative depth of the feature ($a$)
and its full-width at half-maximum (FWHM) are computed. Finally,
every spectral feature in each spectrum is visually inspected by more
than one person and removed from the study if the spline fit and/or
pseudo-continuum do not accurately reflect the spectral feature.

\subsection{Light-Curve Fitting}

A variety of methods have been developed to measure the photometric
properties of SN~Ia light curves. Here, we describe three different
light-curve fitting methods adopted in this paper to characterise the
SN~Ia light curves. While the light-curve parameters derived from
each of these methods are degenerate to some extent, it is useful to
perform all three fitting techniques for the purpose of comparing our
results to previous results in the literature. There are also cases,
for certain spectroscopic subtypes, where one method is superior to
the other two.

\subsubsection{Template and Polynomial Fitting}

Our most direct measurement of light-curve properties makes use of the
template-fitting routine introduced by \cite{Prieto06}. For a given
photometric bandpass, a set of template light curves is used to
construct models which match the light-curve data. The
model light curves are linear combinations of the template light
curves using the weighting scheme described by 
\cite{Ganeshalingam10:phot_paper}. A $\chi^2$-minimisation fitting
routine is used to determine the combination of templates that best
fits the data. For band $X$,  we measure the date of maximum
brightness, the apparent peak magnitude ($m_{X}$), and the light-curve
width parametrised as the difference in magnitudes between maximum and 
fifteen days past maximum, $\Delta m_{15}(X)$. We independently
fit the $B$- and $V$-band light curves for the SNe in our sample. 

In instances where we have a well-sampled light curve, but cannot
achieve an acceptable fit with our template-fitting routine, we fit
the data with a fourth-order polynomial.

For both the template- and polynomial-fitting routines, we use a Monte 
Carlo routine to measure the uncertainty in our derived parameters. We
simulate realisations of our dataset by randomly perturbing each data
point by its $1\sigma$ photometric error assuming a Gaussian
distribution centred at 0~mag. We fit the dataset realisation with our
fitting routine and measure light-curve properties for that
simulation. We estimate the uncertainty in our derived parameters to
be the standard deviation of 50 dataset realisations.

\subsubsection{MLCS2k2}

The Multi-colour Light Curve Shape (MLCS) distance-fitting software
was first introduced by \cite{Riess96} to simultaneously fit all
light-curve data for a given SN~Ia to produce a distance
estimate. This method relies on the observation that more luminous SNe~Ia
have broader light curves \citep{Phillips93} and also have bluer colours
during the photospheric phase. MLCS parametrises light-curve width
using the parameter $\Delta$, which measures the difference in absolute
magnitude of the SN with reference to a fiducial SN~Ia. MLCS attempts
to disentangle intrinsic colour variations from host-galaxy effects to
also produce an estimate for host-galaxy extinction, $A_{V}$.
MLCS2k2.v006 \citep[referred to as simply MLCS2k2 for the rest of this
work;][]{Jha07} is the the most current publicly available
implementation of this fitting routine. In comparison to the original
version, it has an expanded set of training templates and improvements 
in the treatment of host reddening and $K$-corrections.

For our analysis with MLCS2k2, we use the galactic line-of-sight prior
which models the distribution of host-galaxy extinction values as a
decaying exponential with a peak value of 0~mag \citep{Hatano98}. We
also set the host-galaxy $R_{V} = 1.7$ based on the cosmological
analysis of \cite{Hicken09:cosmo}. Their analysis found that a lower
host-galaxy $R_{V}$ reduced the scatter in the Hubble diagram compared
to a more typical Galactic value of $R_{V} = 3.1$. A fit using
MLCS2k2 is considered reliable (and thus its parameters are used in
the current analysis) only when the reduced $\chi^2 \le 1.6$.

\subsubsection{SALT2}

Spectral Adaptive Light-curve Template (SALT) was first developed by
the SuperNova Legacy Survey (SNLS) \citep{Guy05}.  Calculating
distances with SALT is a two-step process. SALT first measures
light-curve parameters that are expected to correlate with the
intrinsic brightness of individual SNe (i.e., light-curve width and
colour). Then a model for the corrected apparent magnitude of a SN,
$m_{B{\rm ,corr}}$, is adopted which applies linear corrections for the
light-curve width and colour to the measured apparent magnitude,
$m_B$. Thus, the corrected apparent magnitude has the form 
\begin{equation}
m_{B{\rm ,corr}} = m_B  + \alpha \times({\rm light \mhyphen curve~width})  - \beta \times({\rm colour}).
\end{equation}
The constants $\alpha$ and $\beta$ are found by minimising $\chi^2$
using distance estimates from a large sample of SNe~Ia compared to a 
cosmological model (see Section~\ref{ss:hubble} for more details).

SALT2 is an updated version of SALT with an expanded training set of
light-curve templates and is the version implemented here. SALT2 is
trained on light curves and spectra from low-$z$ SNe compiled from the
literature and high-$z$ SNe from the first two years of the SNLS 
\citep{Guy07}.  SALT2 measures a parametrisation of the light-curve
width ($x_1$), the SN colour ($c$), and the apparent $B$-band magnitude
at maximum light ($m_{B}$).

In fitting our light curves, we exclude $I$-band data, which are not included 
in the SALT2 template set. We also exclude subluminous SNe~Ia, often of the
spectral subclass of SN~1991bg-like objects 
\citep[e.g.,][]{Filippenko92:91bg, Leibundgut93}, since SALT2 was not developed
to fit this particular subtype. This is achieved by using only SNe~Ia
with $-3 \le x_1 \le 2$ \citep[as in][]{Blondin11}. Finally, the
results of SALT2 fits are utilised here only when the reduced
$\chi^2 < 2$. 

\subsection{Hubble Diagrams}\label{ss:hubble}

In this section we present the methodology used to standardise SNe~Ia
for cosmological application.  We use a model that applies linear
corrections for light-curve width and colour. The width of a light 
curve correlates with the intrinsic luminosity in the sense that SNe
with broader light curves are also more luminous. This correlation has
been well established \citep{Phillips93}. The colour parameter
combines the effects of intrinsic colour variations and host-galaxy
reddening. We use the SALT2 parameters $x_1$ and $c$ as the
parametrisations of light-curve width and SN colour, respectively. We
will also generalise this approach to allow for linear corrections
using spectroscopic parameters.

The distance modulus for each SN can be estimated from its redshift by
$\mu(z) = 25 + 5\log_{10}\left[D_L\left(z\right)\right]$, where
$D_L$ is the luminosity distance expressed in units of Mpc. The
distance modulus including 
linear corrections for light-curve width and colour can be expressed
as 
\begin{equation}\label{eq:mu}
\mu_{\rm SN} = m_B - M + \alpha x_1  - \beta c.
\end{equation}
The variables $\alpha$, $\beta$, and $M$ (the fiducial absolute
magnitude of a SN~Ia) are determined by using a custom version of
{\tt cosfitter} (A.~Conley, 2011, private communication) based on the {\tt
  Minuit} minimisation package \citep{Roos75}. The software minimises
the function
\begin{equation}
\chi^2 =\sum_{s=1}^N \frac{(\mu (z_s) -\mu_{{\rm SN},s})^2}{\sigma_{{\rm m},s}^2 + \sigma_{{\rm pec},s}^2 + \sigma_{\rm int}^2},
\end{equation}
where $\mu(z) $ is the distance modulus of the galaxy in the cosmic microwave 
background (CMB) rest-frame redshift $z$, $\sigma_{\rm m}$ is the measurement 
error in light-curve properties accounting for covariances between measured
parameters, $\sigma_{\rm pec}$ is the uncertainty due to deviations
from Hubble's law induced by gravitational interactions from
neighbouring galaxies, and $\sigma_{\rm int}$ is a constant intrinsic
scatter added to each SN to achieve a reduced $\chi^2 \approx 1$.
We adopt 300~\kms\ as the peculiar velocity for each SN. The intrinsic
scatter, $\sigma_{\rm int}$, can be considered as the uncertainty
associated with a model that attempts to standardise SNe using the
parameters $x_1$ and $c$.  

Only objects with $z_{\rm helio} > 0.01$ are used in the Hubble
diagrams in order to avoid including SNe with motions dominated by
peculiar velocities. We also adopt the same Hubble diagram colour cut
as \citet{Blondin11}, who exclude objects with $c > 0.50$. Finally, as
mentioned above, all analysis using SALT2 fits is restricted to
objects with a reduced $\chi^2 < 2$.

In this work we consider nearby SNe (median $z_{\rm CMB} \approx
0.021$) and are not attempting to find a best-fitting cosmology. A goal
of this study is to combine photometric and spectroscopic properties
such that SNe~Ia become more accurate standardisable candles. We also
aim to quantify the amount of improvement when using a variety of
observed measurements. To that end, we adopt the standard \lcdm
cosmology with $\Omega_{\rm m} = 0.27$, $\Omega_{\Lambda} = 0.73$, and
$w = -1$ when calculating $\mu(z)$.

\subsubsection{Models for Predicting SN~Ia Distances}

Here we describe the generalisation of our calculation of the distance
modulus for each SN to allow for linear corrections using measured
spectral parameters (such as velocity and pEW; see
Sections~\ref{ss:vel} and \ref{ss:ew}) or the ratios of pEWs and
fluxes (such as $\Re$(\ion{Si}{II}) and \scriptr; see
Sections~\ref{ss:si_ratio} and \ref{ss:ratios}). We consider five models
for predicting distances to SNe~Ia using combinations of the SALT2
measured light-curve parameters ($x_1$ and $c$) and spectral
measurements:
\begin{align}
\mu_{\rm SN} = ~&  m_B - M + \gamma \mathcal{S},\label{eq:m1} \\
\mu_{\rm SN} = ~&  m_B - M + \alpha x_1 + \gamma \mathcal{S}, \label{eq:m2} \\
\mu_{\rm SN} = ~& m_B - M - \beta c + \gamma \mathcal{S}, \label{eq:m3} \\
\mu_{\rm SN} = ~&  m_B - M + \alpha x_1 - \beta c  + \gamma \mathcal{S}, \label{eq:m4} \\
\mu_{\rm SN} = ~&  m_B - M + \alpha x_1 - \beta c. \label{eq:m5}
\end{align}
Here, $\mathcal{S}$ represents any spectral measurement: $v$, pEW,
$\Re$(\ion{Si}{II}), \scriptr, etc. The last model included in our
study was already mentioned above (Equation~\ref{eq:mu}) and is the
model usually adopted in cosmological studies of SNe~Ia using only
light-curve parameters 
\citep[e.g.,][]{Astier06,Kowalski08,Hicken09:cosmo,Amanullah10}. In  
the following analysis the so-called $\left(x_1,c\right)$ model will
be the one to which we compare the other cosmological models that 
include spectral information.

\subsubsection{Cross-Validation}

Ideally, with a sufficiently large sample, the predictive abilities of a
model could be inferred by inspecting the dispersion of the
residuals. However, for samples of limited size, the dispersion of the
residuals is prone to statistical fluctuations and may not accurately
reflect the true predictive nature of the model. Furthermore, for a
fixed sample, one can always reduce the dispersion of the residuals by
adding more variables to the model. However, it is not clear whether
the added variables are actually improving the model itself or simply
fitting to the noise inherent in the observables. For analysing the
predictive nature of a model, it is useful to perform some form of
cross-validation (CV) in which a subset of the entire sample is used
to train the model and another subset is used to validate the
predictive ability of that model. 

\cite{Bailey09} use a sample of 58 SNe, 28 of which are adopted as a
training set to train a model and the other 30 are used as a
validation set to assess the predictivity of the model. Using a
smaller sample of 26 SNe, \cite{Blondin11} use a $K$-fold CV method
that allows all of the SNe to be used in the training and validation
procedure. For this study, and following \citet{Blondin11}, we adopt
the $K$-fold CV with $K = 10$. We also tested CV with $K = 2$, 5, and
62 and find that our final results are mostly unchanged (however, see
Section~\ref{sss:model1} for more on our $K = 2$ run). The basics of
the procedure are best illuminated by an example, as follows.

Let us begin with a sample of 60 SNe. 10-fold CV starts by randomly
dividing the sample into 10 subgroups of 6 SNe each. The first
subgroup is set aside; this will be our first validation set. We
combine the remaining 9 subgroups and train our model on the 54 other
SNe to determine the best-fitting parameters (i.e.,
$\alpha$, $\beta$, $\gamma$, and $M$). Using the best-fitting
parameters found with the training set, we apply our model to the
validation set and calculate the Hubble residual for each of the 6
SNe in our validation set. We repeat this process using the second
subgroup as the validation set and the union of the other 9 subgroups
as the training set. This process is repeated a total of 10 times
(once for each subgroup as a validation set) until we have calculated
a residual for every SN in the sample.

\subsubsection{Comparing the Models}

As in \citet{Blondin11}, the dispersion in each model is estimated
using the weighted root-mean-square (WRMS), 
\begin{equation}
{\rm WRMS}^2 =\frac{ \sum_{s =1}^Nw_s [\mu(z_s) - \mu_{{\rm SN},s}]^2}{\sum_{s =1}^Nw_s},
\end{equation}
where the weights, $w_s$, are given by
\begin{equation}
w_s = \sigma_{{\rm m},s}^2 + \sigma_{{\rm pec},s}^2 + \sigma_{\rm int}^2.
\end{equation}
The variance in WRMS is estimated as
\begin{equation}
{\rm Var[WRMS]} = \Big[\sum_{s = 1}^Nw_s\Big]^{-1}.
\end{equation}
The $1\sigma$ uncertainty is found by taking the square root of the
variance. This is a more appropriate estimator of the dispersion in
the model then simply taking the standard deviation in the residuals
since we are not guaranteed that the mean residual will be zero
\citep{Blondin11}. 

Following \citet{Blondin11}, for each model which uses a spectral
measurement (Equations~\ref{eq:m1}--\ref{eq:m4}) we also calculate the
intrinsic prediction error ($\sigma_{\rm pred}$), the intrinsic
correlation ($\rho_{x_1,c}$) of the residuals with residuals using the 
$\left(x_1,c\right)$ model (Equation~\ref{eq:m5}), and the difference
($\Delta_{x_1,c}$) in intrinsic prediction error with respect to the
$\left(x_1,c\right)$ model. An uncertainty can be computed for 
$\Delta_{x_1,c}$ \citep[Appendix B of][]{Blondin11}, and thus the
significance of the difference between a given model and the standard
$\left(x_1,c\right)$ model can also be computed. This 
parameter is the most direct comparison of how much better (or worse)
a model which utilises a spectral 
measurement is compared to the $\left(x_1,c\right)$ model, and how
significant the change is. Note that  
$\Delta_{x_1,c} < 0$ represents an improvement over the
$\left(x_1,c\right)$ model.

\section{Analysis}\label{s:analysis}

\subsection{Velocity Gradients}\label{ss:vdot}

\citet{Benetti05} defined the velocity gradient, $\dot{v} = -\Delta v
/ \Delta t$,  as the ``average daily rate of decrease of the expansion
velocity'' of the \ion{Si}{II} $\lambda$6355 feature and used this
parameter to place each of their 26 SNe~Ia into one of three
categories. The high velocity gradient (HVG) group had the largest
velocity gradients ($\dot{v} \ga 70$~\kms~d$^{-1}$) and the low
velocity gradient (LVG) group had the smallest velocity gradients. The
third subclass (FAINT) had the lowest expansion velocities, yet
moderately large velocity gradients, and consisted of subluminous
SNe~Ia with the narrowest light curves ($\Delta m_{15}(B) \ga
1.6$~mag).

\begin{figure}
\centering$
\begin{array}{c}
\includegraphics[width=3.5in]{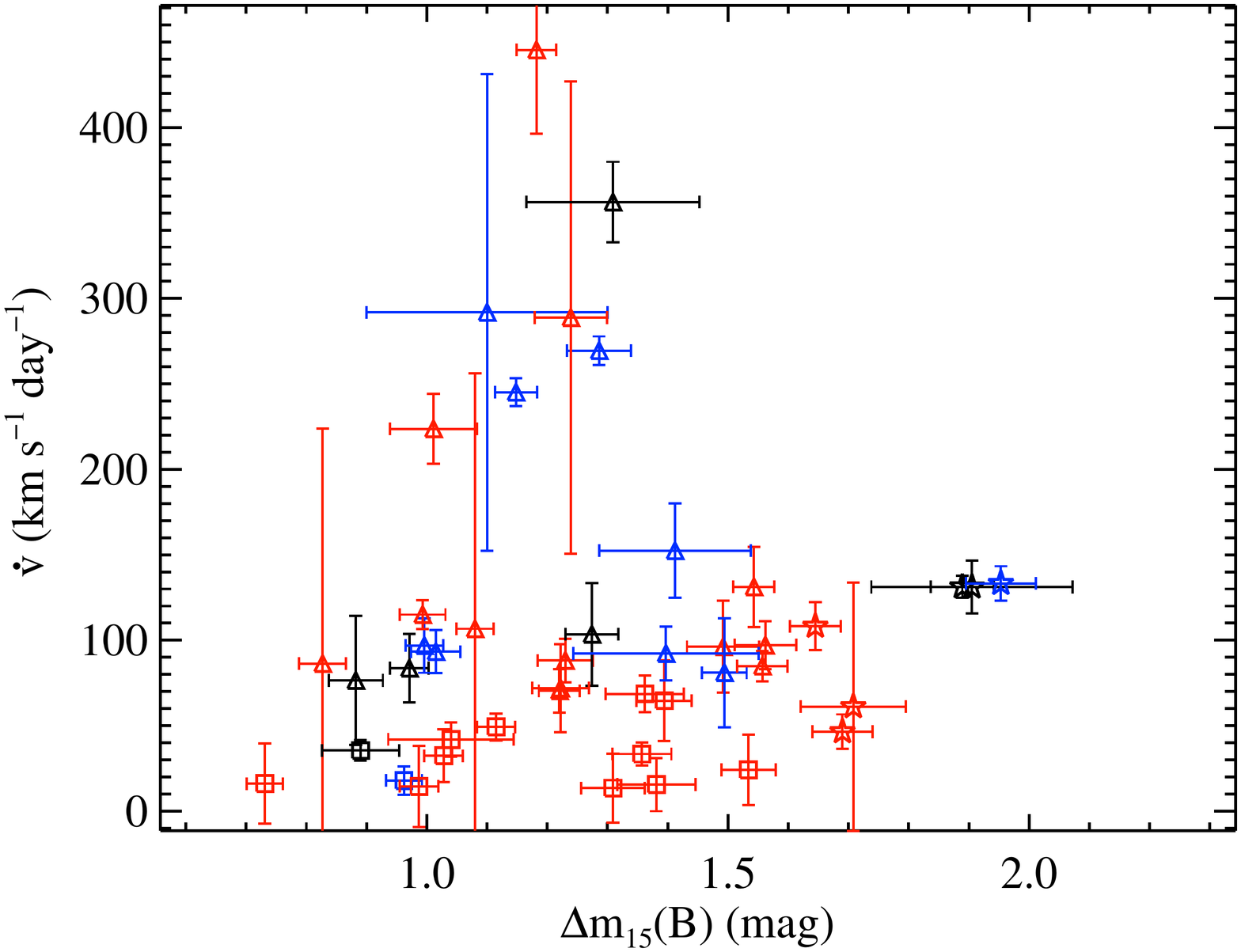} \\
\includegraphics[width=3.5in]{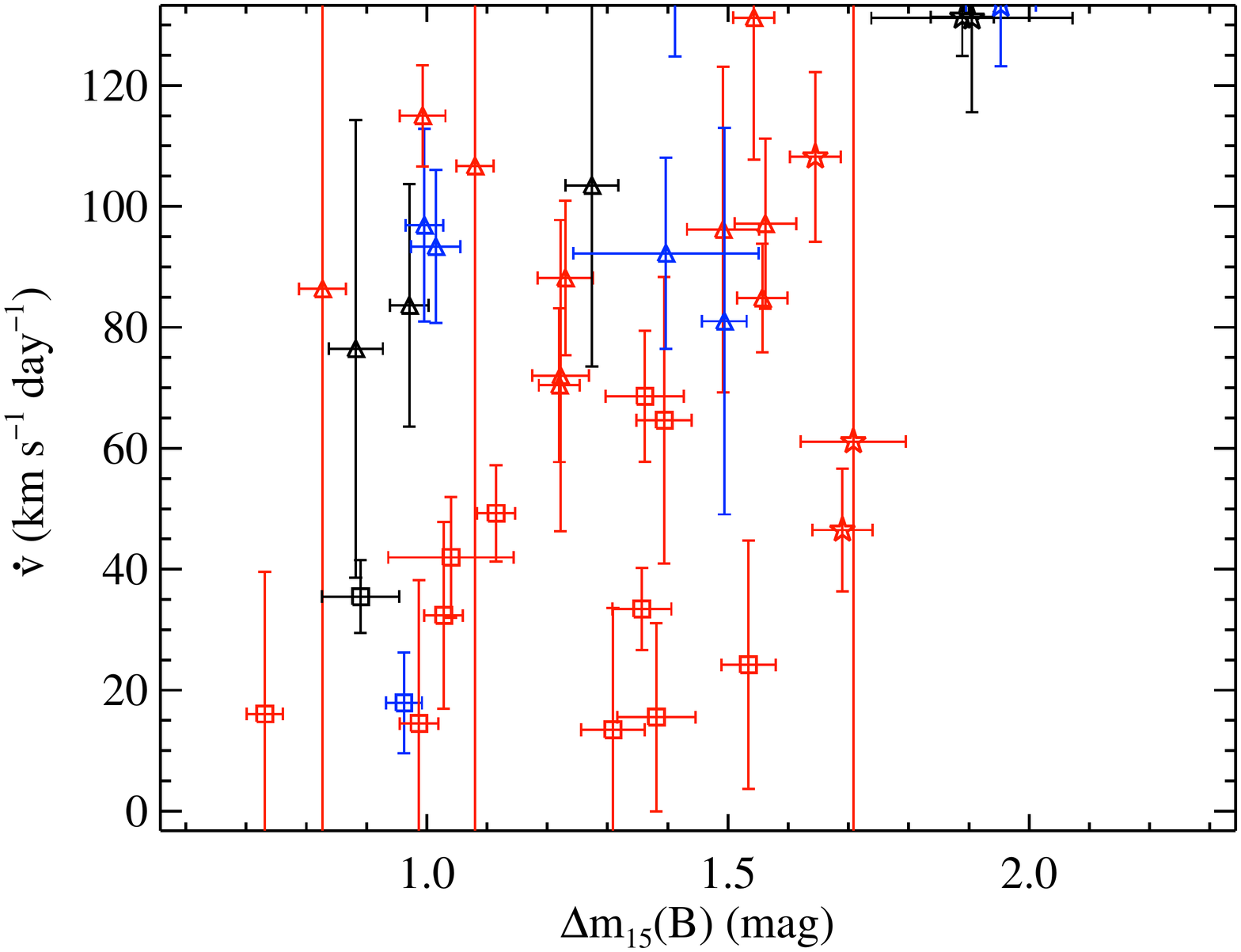}
\end{array}$
\caption[$\dot{v}$ versus $\Delta m_{15}(B)$]{The velocity gradient
  versus $\Delta m_{15}(B)$ of 44 SNe ({\it top}) and a close-up view 
  of the low-$\dot{v}$ objects ({\it bottom}). Blue
  points are high-velocity (HV) objects, red points are 
  normal-velocity objects, and black points are objects for which we
  could not determine whether the SN was normal or high velocity (see
  BSNIP~II for further details regarding how HV SNe are
  defined). Stars are FAINT objects, squares are low velocity gradient
  (LVG) objects, and triangles are high velocity gradient (HVG)
  objects (see BSNIP~II for further details
  regarding how these subclasses are defined).}\label{f:v_dot_dm15} 
\end{figure}

Even though the BSNIP sample is not well suited to velocity gradient
measurements (the average number of spectra per object is
\about2, shown in BSNIP~I), we are still able to calculate a
$\dot{v}$ value for many of our SNe~Ia. Figure~\ref{f:v_dot_dm15}
shows 44 objects and their $\dot{v}$ measurements plotted against
their $\Delta 
m_{15}(B)$ values. The points are colour-coded by their near-maximum
\ion{Si}{II} $\lambda$6355 velocity, with red points representing
normal-velocity objects and blue points representing high-velocity
(HV) objects. These so-called ``Wang types'' were first presented by
\citet{Wang09}, and in BSNIP~II we discuss our definition of these
subclasses in more detail. The data are also shape-coded by the
aforementioned ``Benetti types'' (FAINT are stars, LVG are squares,
and HVG are triangles). The top panel of Figure~\ref{f:v_dot_dm15}
shows all objects for which a $\dot{v}$ and $\Delta m_{15}(B)$ are
measured, while the bottom panel shows a close-up view of the same data such
that the axis ranges match those of Figure~3b of \citet{Benetti05}.

As a result of the definitions of the different subclasses, the FAINT
SNe are 
found to the right in Figure~\ref{f:v_dot_dm15} (i.e., large $\Delta
m_{15}(B)$ values) and the HVG SNe are found in the upper part of the
Figure (i.e., large $\dot{v}$ values), though there is no obvious
break between the various classes. As pointed out in BSNIP~II and
confirmed in Figure~\ref{f:v_dot_dm15}, the HVG and LVG objects have
similar average $\Delta m_{15}(B)$ values and ranges of values.

It was stated by \citet{Benetti05} that $\dot{v}$ is weakly correlated
with $\Delta m_{15}(B)$, though there is no evidence of such a
correlation in the BSNIP data. They also claim that there are three
distinct families of SNe~Ia (LVG, HVG, and FAINT) based partially on
their plot of $\dot{v}$ versus $\Delta m_{15}(B)$. With almost
70~per~cent more objects, the BSNIP data fill in this parameter space
and cast serious doubt on the existence of truly distinct families of
SNe~Ia based on velocity-gradient measurements (see Table~4 of
BSNIP~II for the median values of $\dot{v}$ and $\Delta m_{15}(B)$ for
each of these three subclasses).

\subsection{Expansion Velocities}\label{ss:vel}

\subsubsection{Velocities at Maximum Brightness}\label{sss:v0}

\begin{figure}
\centering
\includegraphics[width=3.5in]{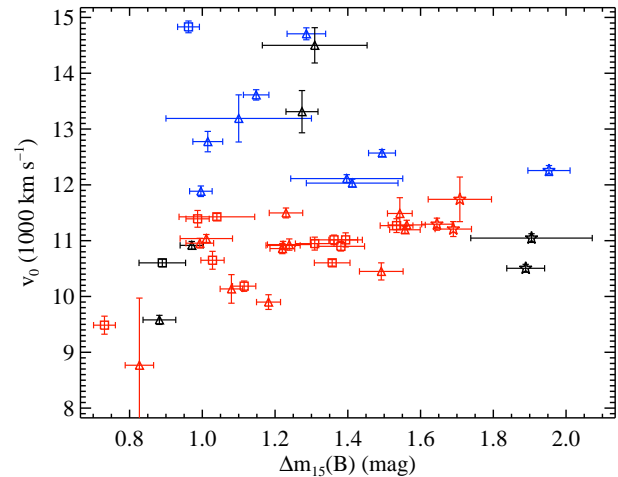}
\caption[$v_0$ of Si~II $\lambda$6355 versus $\Delta
m_{15}(B)$]{The velocity at maximum brightness of the \ion{Si}{II} 
  $\lambda$6355 feature versus $\Delta m_{15}(B)$ of 44 SNe. Colours
  and shapes of data points are the same as in
  Figure~\ref{f:v_dot_dm15}.}\label{f:v0_dm15}
\end{figure}

Expansion velocities of SNe~Ia are calculated from the minima of
various absorption features (see BSNIP~II for more information on how
this measurement is performed on the data presented here). These
velocities 
have been compared to light-curve width measurements and photometric
colours in a variety of ways
\citep[e.g.,][]{Hachinger06,Blondin11,Nordin11a}. As discussed in
BSNIP~II, \citet{Hachinger06} interpolate/extrapolate their expansion
velocities to the time of maximum brightness (i.e., $t=0$~d), and $v_0$
was defined as the expansion velocity of \ion{Si}{II} $\lambda$6355 at
maximum brightness. They then compare these velocities to the
light-curve shape parameter $\Delta
m_{15}(B)$. Figure~\ref{f:v0_dm15} presents the 44 SNe in the BSNIP
data for which both $v_0$ and $\Delta m_{15}(B)$ are calculated. As
above, the points are colour-coded by ``Wang type'' and shape-coded by
Benetti type.

As in Figure~\ref{f:v_dot_dm15}, the FAINT objects are (by definition)
found at the right-hand edge of Figure~\ref{f:v0_dm15}. Similarly, the HV
objects are all in the upper half of the figure. All of the objects,
except for a few of the highest velocity SNe and the object with
the lowest velocity in the figure, are within \about1500~\kms\ of 
11,000~\kms. This is remarkably similar to what was found by
\citet{Hachinger06}, except the BSNIP data have a slightly larger
scatter around the typical velocity of
\about11,000~\kms. \citet{Hachinger06} also note that the majority of
the scatter in velocity comes from the HVG objects, which is also true
in Figure~\ref{f:v0_dm15}. However, there is a major difference between 
the two results. The BNSIP data in Figure~\ref{f:v0_dm15} show
that HVG SNe have a huge range of $v_0$, spanning well above average
to significantly below average values, whereas the data presented by
\citet{Hachinger06} exhibit evidence of the (oft-quoted) one-to-one
relationship between HVG and HV SNe~Ia. As mentioned in BSNIP~II,
while most HVG objects are found to have expansion velocities above
the LVG objects, this is not an exclusive feature of HVG SNe~Ia.

\begin{figure}
\centering
\includegraphics[width=3.5in]{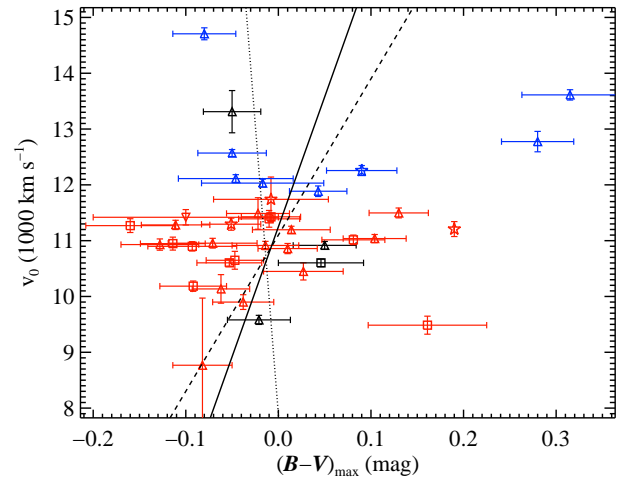}
\caption[$v_0$ of Si~II $\lambda$6355 versus \bvmax]{The velocity
  at maximum brightness of the \ion{Si}{II} $\lambda$6355 feature
  versus observed \bvmax\ (i.e., the difference between observed
  $B$-band magnitude and observed 
  $V$-band magnitude at the time of $B$-band maximum brightness) of
  39 SNe. Colours and shapes of data points are the same as in
  Figure~\ref{f:v_dot_dm15}. The solid line is the fit to all of the
  data while the dotted line is the fit only to objects with
  $(B-V)_{\rm max} < 0.25$~mag. The dashed line is the
  relationship between $v_0$ and $B_{\rm max} - V_{\rm max}$ from the
  model spectra of \citet{Kasen07}, as shown by 
  \citet{Foley11:vel}. }\label{f:v0_bv}
\end{figure}

A relationship between the calculated intrinsic ``pseudo-colour''
$B_{\rm max} - V_{\rm max}$ (i.e., the $B$-band magnitude at $B$-band
maximum minus the $V$-band magnitude at $V$-band maximum corrected for
host-galaxy reddening) and $v_0$ has been seen
previously \citep{Foley11:velb}. Figure~\ref{f:v0_bv} shows 39 BSNIP
SNe for which we measure $v_0$ and the difference between observed
$B$-band magnitude and observed $V$-band magnitude at the time of
$B$-band maximum brightness (what we refer to as \bvmax\ in this
work). We opt to use (observed) \bvmax\ since it is an actual,
physical colour of the SN at a discrete period of time, though keep in
mind that in this study we do not attempt to correct for host-galaxy
reddening. Following \citet{Foley11:velb}, we are only presenting
SNe~Ia with $(B-V)_{\rm max} < 0.319$~mag in Figure~\ref{f:v0_bv}.

The linear least-squares fit to all of the points is shown by the
solid line. We see effectively no evidence for any overall correlation
(Pearson correlation coefficient of 0.17\footnote{The Pearson
  correlation coefficient is the correlation coefficient used
  throughout this work unless otherwise noted.}), and the linear fit
seems 
to be driven solely by the two outliers with $(B-V)_{\rm max} >
0.25$~mag. If those two objects are removed, the correlation
coefficient drops to 0.05 (the best-fitting line to the remaining
data is shown by the dotted line). Finally, the dashed line is the
relationship between $v_0$ and $B_{\rm max} - V_{\rm max}$ from the
model spectra of \citet{Kasen07}, as shown by
\citet{Foley11:vel}. When including all of the data in
Figure~\ref{f:v0_bv}, the correlation between $v_0$ and \bvmax\ is
weaker that what was seen by \citet{Foley11:velb}, where
they derive correlation coefficients of 0.28 and 0.39 for two
different datasets.\footnote{The first dataset, yielding a correlation
  coefficient of 0.28, comes from \citet{Wang09} and includes some of
  the data used here. However, the two samples {\it are}
  distinct and the methods of velocity determination are significantly
  different.} However, no correlation is present whatsoever if the two
significant outliers are removed. The difference between the results
shown here and those of \citet{Foley11:velb} is likely due to the fact
that we do not correct the BSNIP colours for any possible host-galaxy
reddening while \citet{Foley11:velb} attempt to convert the observed
colours into intrinsic colours. We will delve deeper into this
colour conversion in future BSNIP studies (Ganeshalingam et~al., in
preparation).

\subsubsection{Velocities Near Maximum Brightness}\label{sss:vels}

If instead of $v_0$ we plot the actual measured velocity of the
\ion{Si}{II} $\lambda$6355 feature for each object having a spectrum
within 5~d of maximum brightness versus $\Delta m_{15}(B)$, the same
basic trends are seen (but with nearly twice as many data points). A
comparison of the velocity of the \ion{Si}{II} $\lambda$5972 feature
(within 5~d of maximum brightness) with $\Delta m_{15}(B)$ yields
nearly identical results. The biggest difference is that the
velocities are clustered around 10,300~\kms, lower than that of the
\ion{Si}{II} $\lambda$6355 feature. This difference between these 
features has been seen in previous studies as well
\citep{Hachinger06}. The same analysis using the velocity of the
\ion{S}{II} ``W'' once again shows the same behaviour, but with an even
lower typical velocity (\about9000~\kms). This has also been pointed
out in earlier work \citep{Hachinger06}. The velocity of the
\ion{S}{II} ``W'' feature is further discussed below.

\citet{Blondin11} present a Hubble diagram that is corrected by SALT2
light-curve width parameter $x_1$ and colour parameter $c$ {\it in
  addition to} the velocity of the \ion{Si}{II} $\lambda$6355
feature. This yielded approximately a 10~per~cent decrease in the
scatter of their Hubble diagram. It has also been shown that the
\ion{Si}{II} $\lambda$6355 velocity is uncorrelated with both  $x_1$
and $c$, and thus it gives information beyond light-curve width and
colour. However, the anticorrelation of this velocity with Hubble
residuals (corrected for light-curve width and colour) is relatively
small \citep{Blondin11}.

\begin{figure}
\centering$
\begin{array}{c}
\includegraphics[width=3.4in,angle=180]{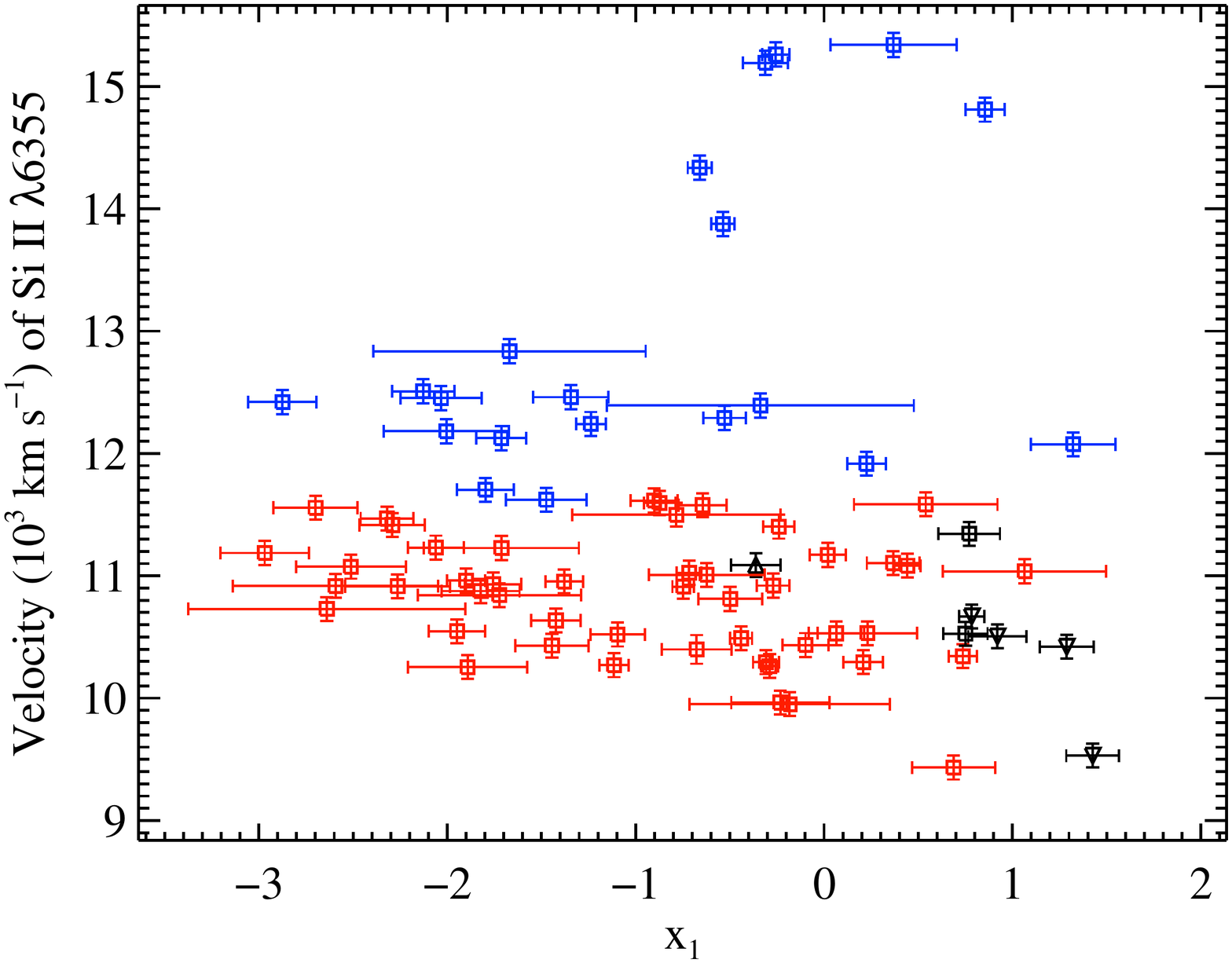} \\
\includegraphics[width=3.4in,angle=180]{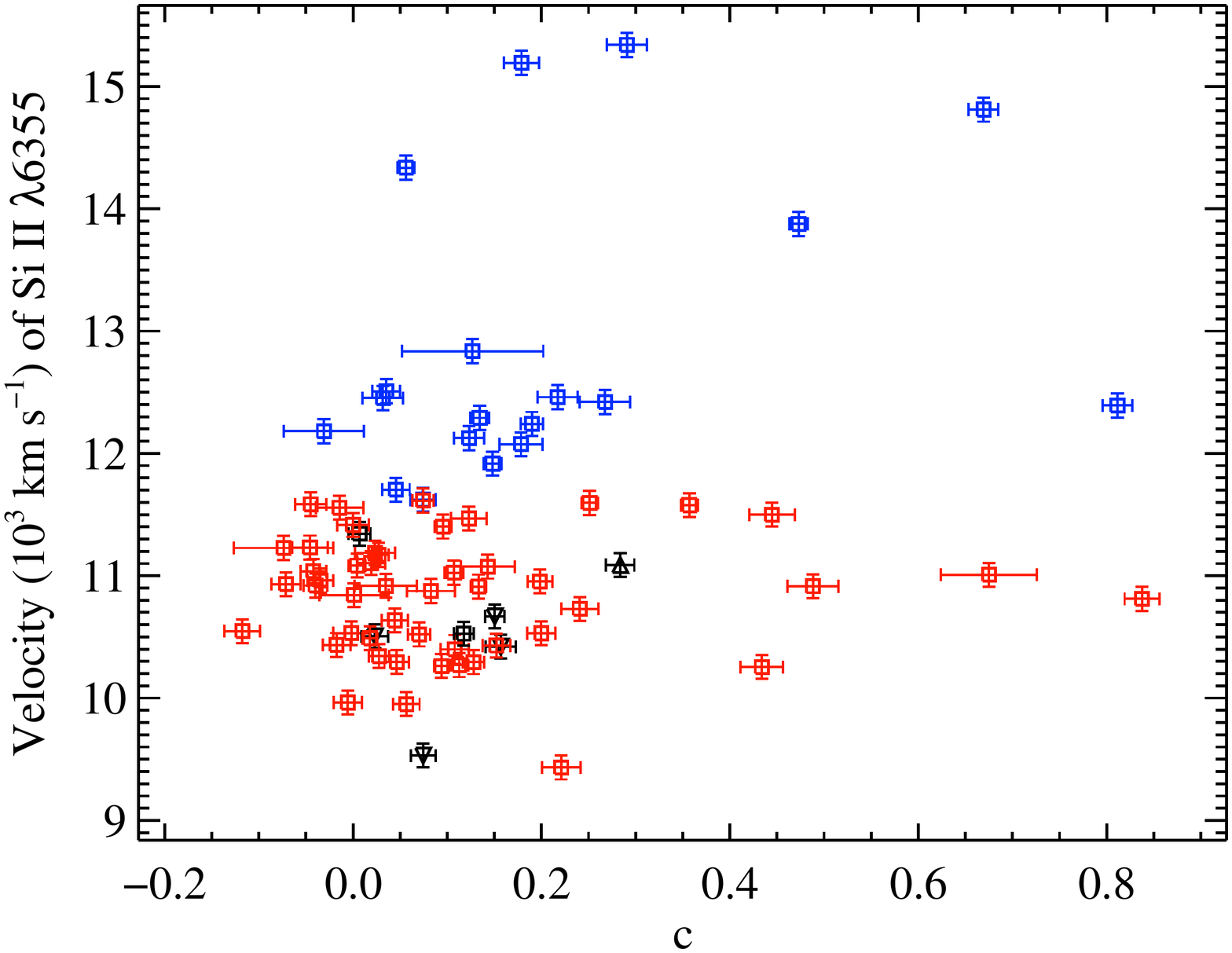} \\
\includegraphics[width=3.4in]{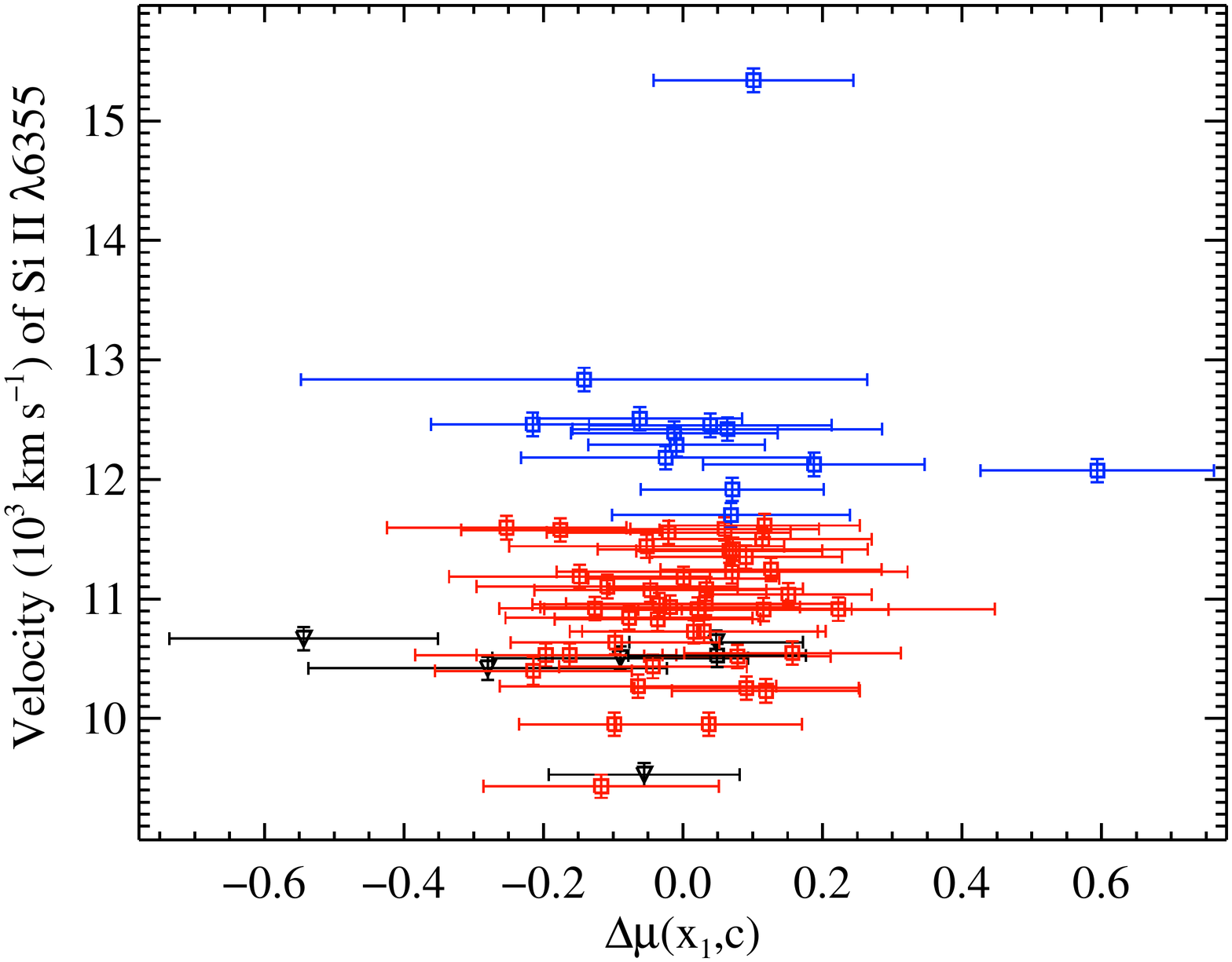} 
\end{array}$
\caption[Velocity of Si~II $\lambda$6355 versus $x_1$, $c$, and
Hubble residual]{The velocity of the \ion{Si}{II} $\lambda$6355
  feature versus SALT2 light-curve width parameter $x_1$ ({\it top}), SALT2
  colour parameter $c$ ({\it middle}), and Hubble residuals corrected for
  light-curve width and colour ({\it bottom}). Blue points are
  HV objects, red points are normal-velocity objects, and black points
  are objects for which we could not determine whether the SN was
  normal or high velocity. Squares are Ia-norm,
  upward-pointing triangles are Ia-91T,
  downward-pointing triangles are Ia-99aa, and circles are
  objects which do not have a SNID subtype (see BSNIP~I for further
  details regarding how these subclasses are
  defined).}\label{f:v_SALT_si6355}
\end{figure}

Plotted in Figure~\ref{f:v_SALT_si6355} are the 66 SNe~Ia in the BSNIP
sample which have SALT2 fits and measured \ion{Si}{II}
$\lambda$6355 velocities within 5~d of maximum brightness. The
velocities are plotted against $x_1$, $c$, and Hubble residuals
corrected for light-curve width and colour (only for objects which are
used to make the Hubble diagram). As above, the points are
colour-coded by ``Wang type.'' In Figure~\ref{f:v_SALT_si6355} the
data are shape-coded by what is referred to in BSNIP~II as the ``SNID
type.'' The SuperNova IDentification code \citep[SNID;][]{Blondin07},
as implemented in BSNIP~I, was used to determine the spectroscopic
subtype of each SN used in BSNIP~II (as well as this
study). SNID compares an input spectrum to a library of spectral
templates in order to determine the most likely spectroscopic
subtype. Spectroscopically normal objects are objects classified as
``Ia-norm'' by SNID.

The spectroscopically peculiar SNID subtypes used
here include the often underluminous SN~1991bg-like objects
\citep[``Ia-91bg,'' e.g.,][]{Filippenko92:91bg,Leibundgut93}, and the often
overluminous SN~1991T-like objects \citep[``Ia-91T,''
e.g.,][]{Filippenko92:91T,Phillips92} and SN~1999aa-like objects
\citep[``Ia-99aa,''][]{Li01:pec,Strolger02,Garavini04}. See BSNIP~I for more
information regarding our implementation of SNID and the various
spectroscopic subtype classifications. In 
Figure~\ref{f:v_SALT_si6355}, Ia-norm objects are plotted as squares,
Ia-91T objects are shown as upward-pointing triangles, Ia-99aa objects
are displayed as downward-pointing triangles, and objects with no
determined SNID subtype are circles.

The data in Figure~\ref{f:v_SALT_si6355} are extremely similar
to those of \citet{Blondin11} and show the same correlations (or lack
thereof). The velocity of the \ion{Si}{II} $\lambda$6355 feature as
measured from the BSNIP data is uncorrelated with $x_1$ (correlation
coefficient $-0.21$). This velocity is also 
uncorrelated with $c$, as seen in \citet{Blondin11}. Removing SNe with
$c > 0.5$ from the BSNIP data (which is done when Hubble diagrams are
produced using this parameter) yields a correlation
coefficient of 0.13 between \ion{Si}{II} $\lambda$6355 velocity and
$c$.

The bottom panel of Figure~\ref{f:v_SALT_si6355} shows the velocity of
the \ion{Si}{II} $\lambda$6355 feature versus the
$\left(x_1,c\right)$-corrected Hubble residuals. If the correction
term in a given model 
(here, the velocity of \ion{Si}{II} $\lambda$6355) is well correlated
with the SALT2-corrected Hubble residuals, then the extra term is
likely providing new information that is actually in the data. Thus, the
model is improving the fit not by fitting to noise, but to physical
information contained in the data. However, the correction term here 
is uncorrelated with the residuals (correlation coefficient
0.19). 

No improvement is found (i.e., the WRMS increases) when adding the
\ion{Si}{II} $\lambda$6355 velocity to the standard
$\left(x_1,c\right)$ model ($\Delta_{x_1,c} = 0.0083 \pm
0.0084$). \citet{Blondin11} found that there was a $\la 10$~per~cent decrease
in the WRMS when using the $x_1$, $c$, and \ion{Si}{II} $\lambda$6355
velocity, but their $\Delta_{x_1,c}$ is consistent with 0 (as is
ours). They also find only a ``modest'' correlation between
\ion{Si}{II} $\lambda$6355 velocity and $\left(x_1,c\right)$-corrected
residuals \citep[correlation coefficient
0.4;][]{Blondin11}. Therefore, it seems that adding the \ion{Si}{II}
$\lambda$6355 velocity to the standard $\left(x_1,c\right)$ model does
{\it not} significantly improve the precision of SN~Ia distance calculations.

\begin{figure}
\centering$
\begin{array}{c}
\includegraphics[angle=180,width=3.5in]{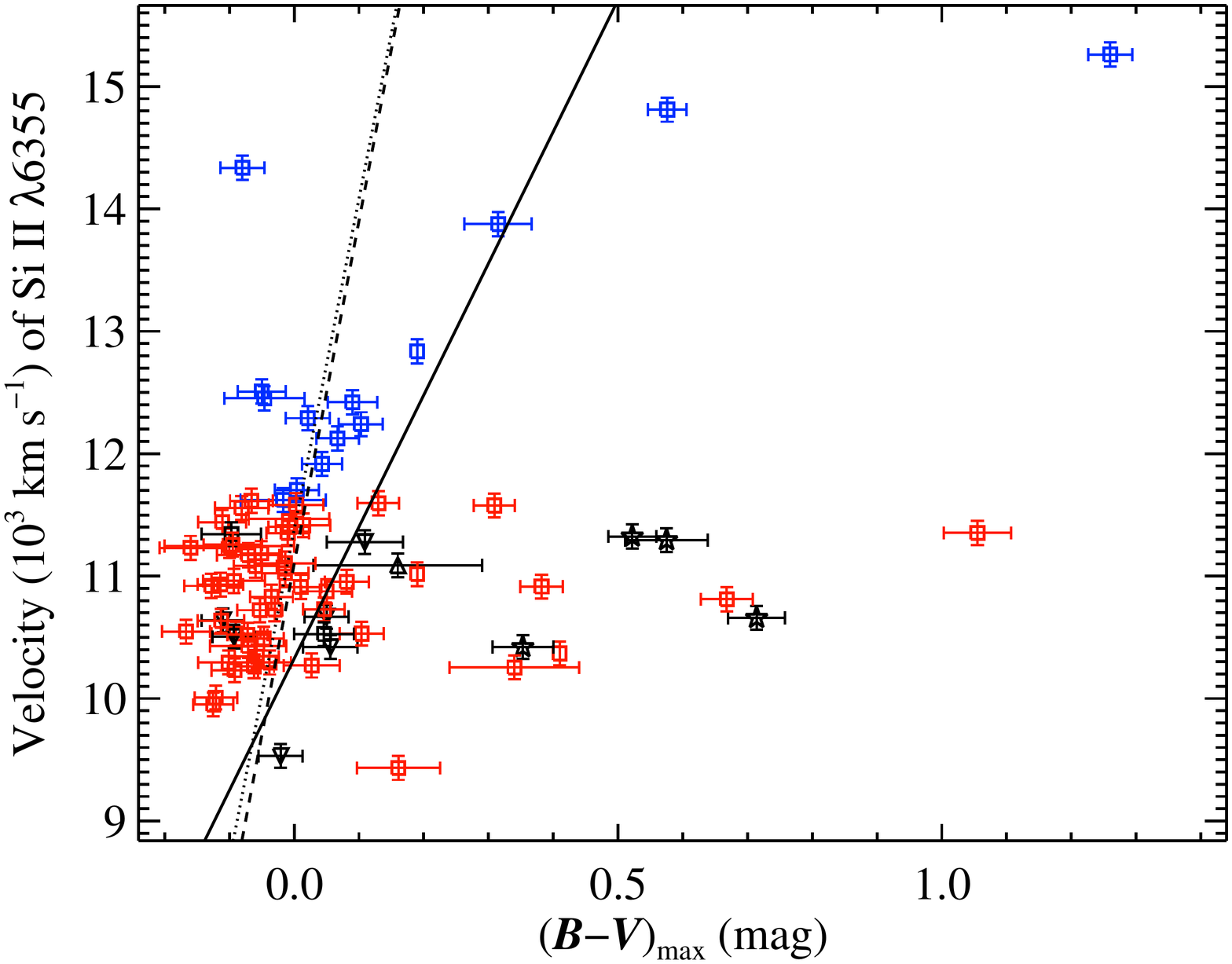} \\
\includegraphics[width=3.5in]{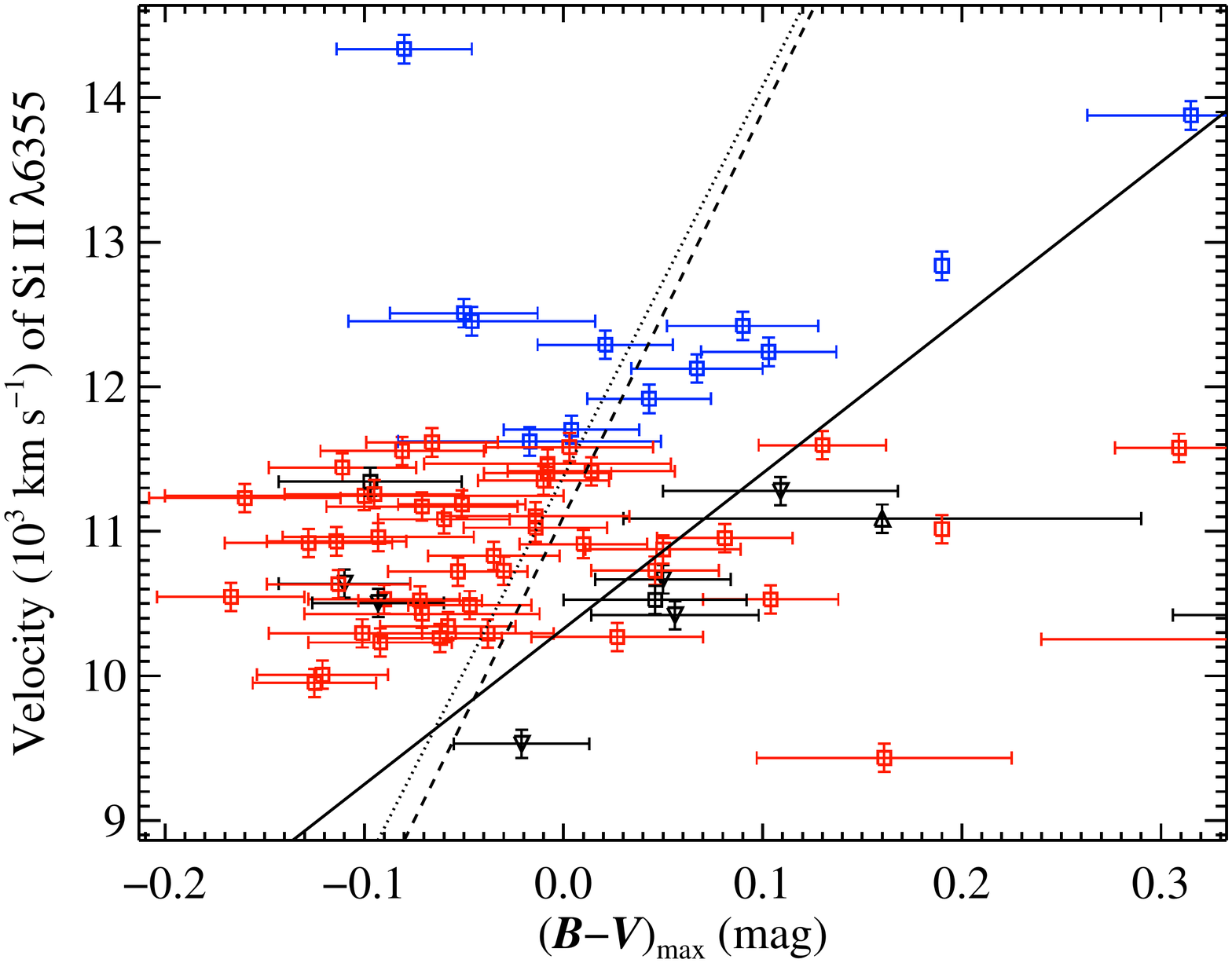} 
\end{array}$
\caption[Velocity of Si~II $\lambda$6355 versus
$\left(B-V\right)_\textrm{max}$]{The velocity of the \ion{Si}{II}
  $\lambda$6355 feature versus observed \bvmax\ ({\it top}) and a
  close-up view of objects with $\left(B-V\right)_\textrm{max} < 0.319$~mag
  ({\it bottom}). Blue points are
  HV objects, red points are normal-velocity objects, and black points
  are objects for which we could not determine whether the SN was
  normal or high velocity. Squares are Ia-norm, stars are 
  Ia-91bg, upward-pointing triangles are Ia-91T,
  downward-pointing triangles are Ia-99aa, and circles are
  objects which do not have a SNID subtype (see BSNIP~I for further
  details regarding how these subclasses are defined). The solid line
  is the fit to all of the data while the dotted line is the fit only
  to objects with $(B-V)_{\rm max} < 0.319$~mag. The dashed line is
  the relationship between $v_0$ and $B_{\rm max} - V_{\rm max}$ from
  the model spectra of \citet{Kasen07}, as shown by 
  \citet{Foley11:vel}.}\label{f:v_bv_si6355}
\end{figure}

In Figure~\ref{f:v_bv_si6355} we plot all measured velocities of the
\ion{Si}{II} $\lambda$6355 feature (within 5~d of maximum brightness)
against $\left(B-V\right)_\textrm{max}$. This differs from
Figure~\ref{f:v0_bv} since the velocities in that plot were
interpolated/extrapolated to $t = 0$~d (i.e., $v_0$) and this plot
shows the actual velocities measured from the spectra.
The top panel displays all 77 SNe from
the BSNIP data for which both of these values have been measured and
the bottom panel provides a close-up view of objects with
$\left(B-V\right)_\textrm{max} < 0.319$~mag \citep[in order to match
the sample fit by][]{Foley11:velb}.

The linear least-squares fit to
all of the points is shown by the solid line and the fit to SNe with
$\left(B-V\right)_\textrm{max} < 0.319$~mag is shown by the dotted
line; the correlation coefficients are 0.23 and 0.30,
respectively, and slightly lower than the value found by
\citet{Foley11:velb}, 0.39 (though the linear fit shown here matches
well to what was found in their study). As with Figure~\ref{f:v0_bv},
there is only marginal evidence for a correlation and there is a large
amount of scatter around the linear fit. Again, this is unsurprising
since we are measuring {\it observed} \bvmax\ and \citet{Foley11:velb}
are plotting pseudo-colours that have been corrected for host-galaxy
reddening. 

The typical $\left(B-V\right)_\textrm{max}$ for HV objects is larger
than for Ia-norm and Ia-91T/99aa, but smaller than that of
Ia-91bg (which appear in the Figure as highly reddened, significant
outliers). However, the range of $\left(B-V\right)_\textrm{max}$
spanned by Ia-norm and HV objects is similar, with a significant amount of
overlap. The scatter in \bvmax\ of the HV objects (0.098~mag) is
effectively equal to that of the other objects in the bottom panel of
Figure~\ref{f:v_bv_si6355} (0.101~mag). While this matches the scatter
in the HV objects found previously \citep[0.095~mag;][]{Foley11:velb},
it differs from the SNe with velocities $< 11,800$~\kms, which they
found to have a smaller intrinsic colour scatter (0.072~mag). The
correction for host-galaxy reddening applied by \citet{Foley11:velb}
is likely responsible for the decrease in colour scatter of the
normal-velocity objects.

The dashed line in Figure~\ref{f:v_bv_si6355} is the
relationship between $v_0$ and $B_{\rm max} - V_{\rm max}$ from the
model spectra of \citet{Kasen07}, as shown in Figure~8 of
\citet{Foley11:vel}. Interestingly, even though \citet{Foley11:velb}
plot {\it intrinsic} colours and we plot {\it observed} colours, both
studies match
these predictions very well. One difference between
Figure~\ref{f:v_bv_si6355} and Figure~8 of \citet{Foley11:vel} is that
the BSNIP data contain a handful of objects that are extremely
reddened (i.e., they have relatively large values of
$\left(B-V\right)_\textrm{max}$). However, this can easily be
explained. All of the Ia-norm and HV SNe in 
Figure~\ref{f:v_bv_si6355} with $\left(B-V\right)_\textrm{max} >
0.31$~mag have been observed to have significant reddening from their
host galaxies \citep[which is not taken into account in the models
of][]{Kasen07}. The other objects with $\left(B-V\right)_\textrm{max}
> 0.31$~mag are Ia-91bg, which were also not discussed in the models of 
\citet{Kasen07}.


Figure~8 of \citet{Foley11:vel} also presents the theoretical
relationship between velocity of the \ion{Ca}{II}~H\&K feature and
intrinsic 
$\left(B-V\right)_\textrm{max}$. A comparison of this velocity and
observed \bvmax\ 
measured from the BSNIP data 
shows effectively no correlation with a correlation coefficient of only 
0.14. Again, the scatter in
$\left(B-V\right)_\textrm{max}$ is similar for the HV and
normal-velocity objects. However, this is unsurprising since, as
pointed out 
in BSNIP~II, the \ion{Ca}{II}~H\&K velocities of HV and
normal-velocity objects (determined using the \ion{Si}{II}
$\lambda$6355 
velocity) are highly overlapping. 


A distance model involving $x_1$, $c$, and the velocity of the
\ion{Ca}{II}~H\&K feature was calculated, and while the WRMS
technically decreased with the addition of this velocity, it was not
found to be significant ($\Delta_{x_1,c} = -0.0085 \pm
0.0141$). Other models involving this velocity were all found to
degrade the accuracy of distance measurements when compared to the
standard $\left(x_1,c\right)$ model.

Furthermore, we used Equations~\ref{eq:m1}--\ref{eq:m4} along with
velocities of all seven spectral features for which velocities were
measured and compared the results to the $\left(x_1,c\right)$
model. The vast majority of these models predicted a larger scatter
than the standard model corrected for light-curve width and colour. 
However, both the \ion{O}{I} $\lambda$7773 triplet and the \ion{Ca}{II} 
near-IR triplet, when combined with $x_1$ and $c$, were found to perform
equally as well as when using just $x_1$ and $c$. Thus, adding either
of these velocities did not degrade the distances calculated, but they
did not significantly improve them either.

\begin{figure}
\centering$
\begin{array}{c}
\includegraphics[width=3.4in,angle=180]{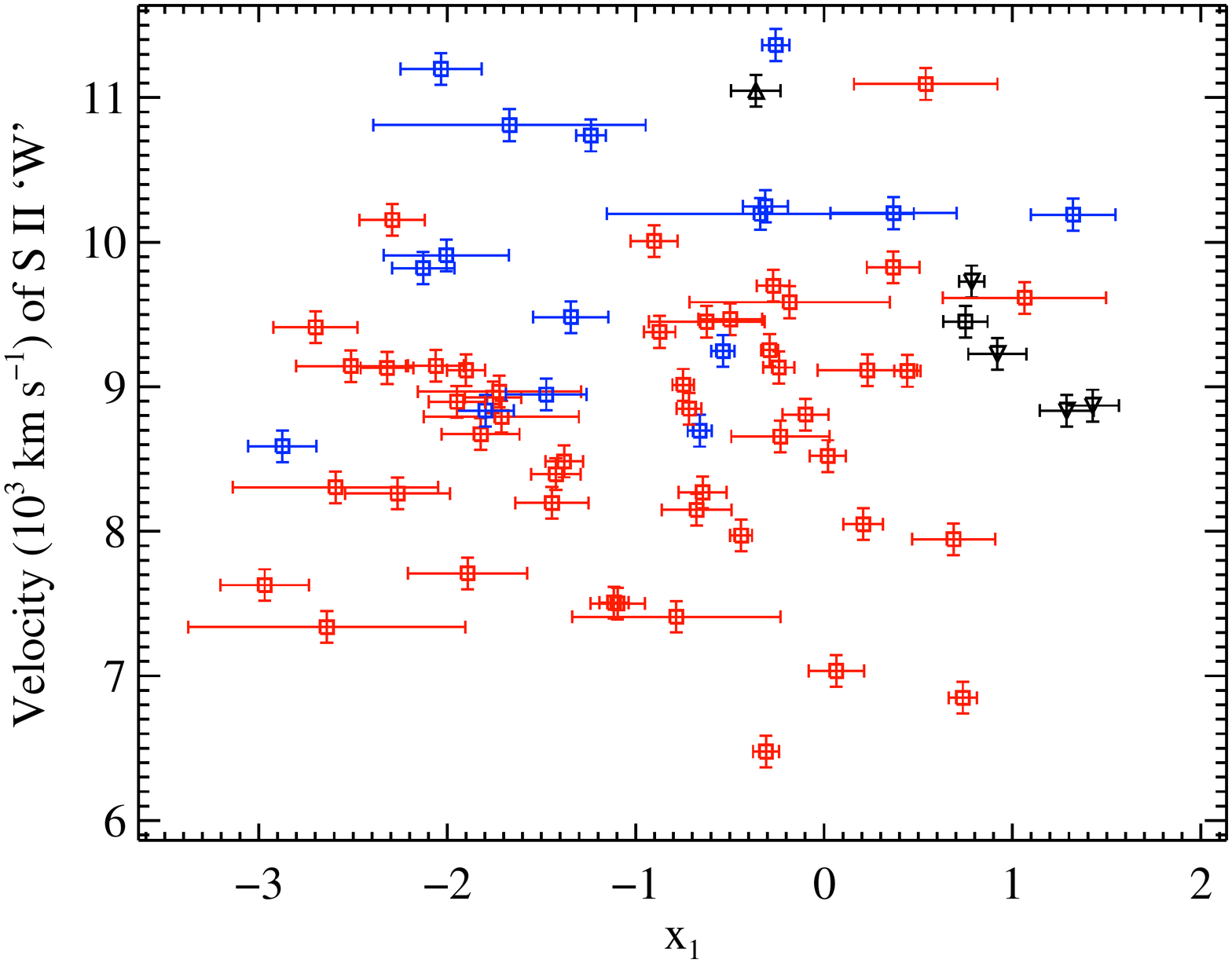} \\
\includegraphics[width=3.4in,angle=180]{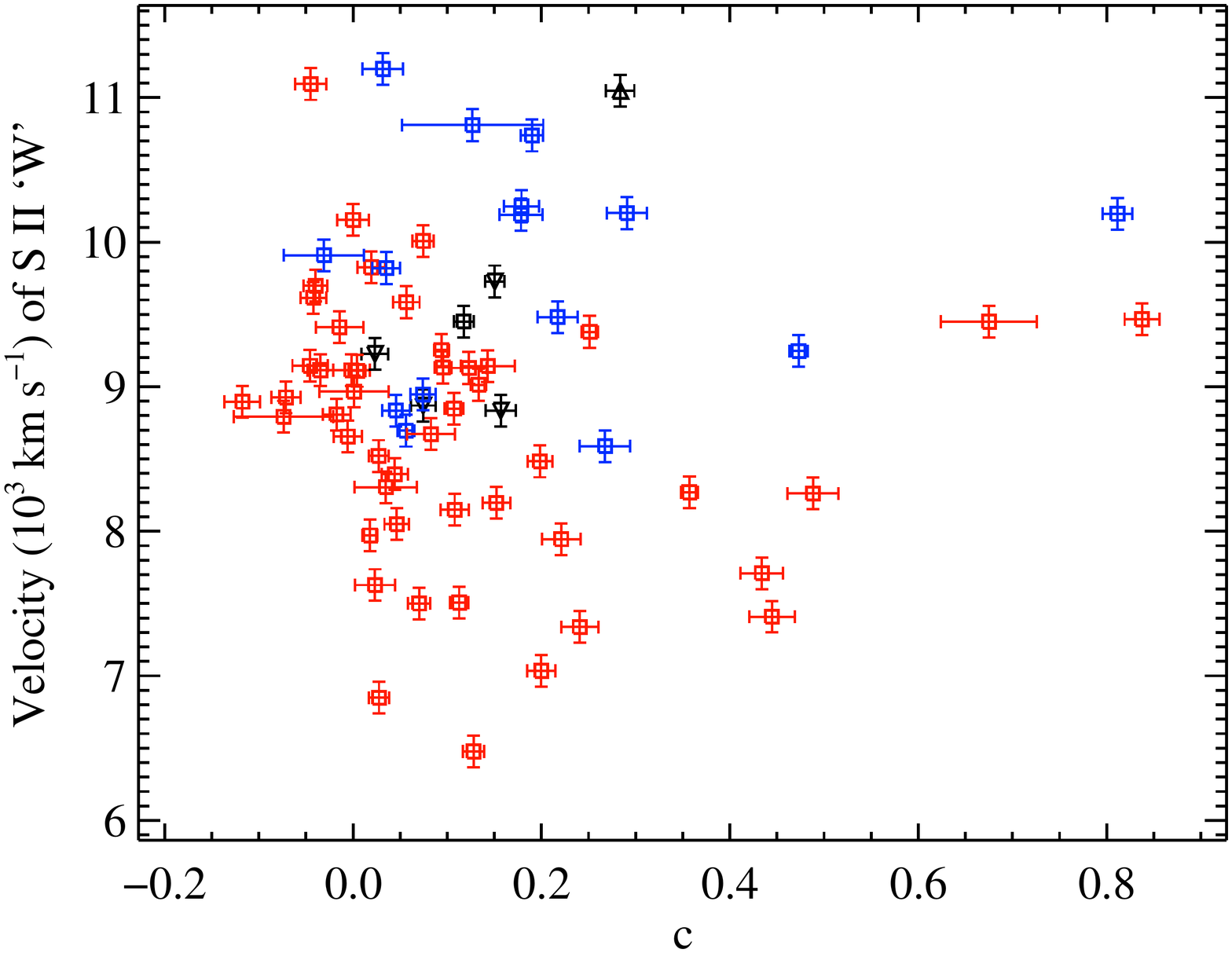} \\
\includegraphics[width=3.4in]{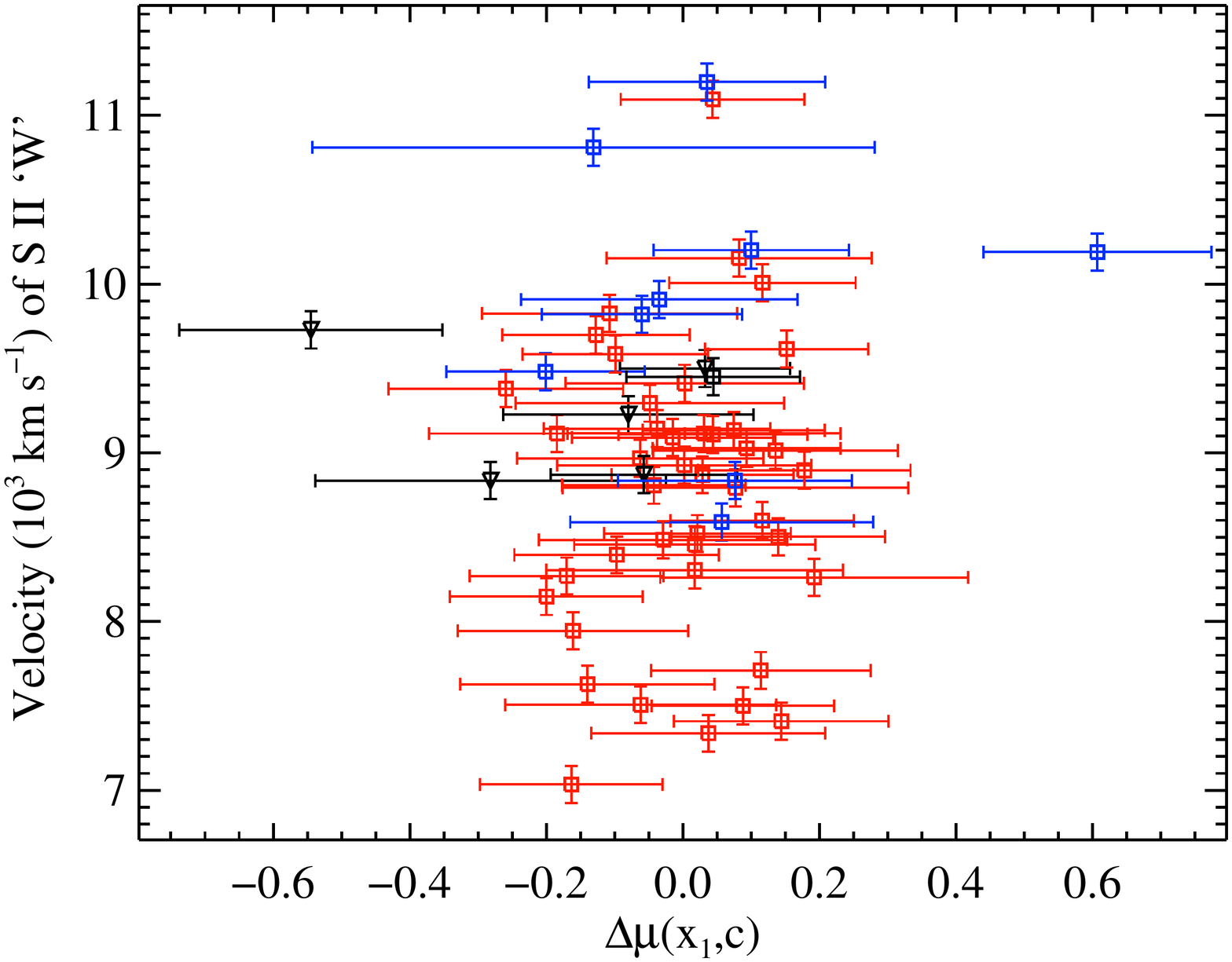} 
\end{array}$
\caption[Velocity of S~II ``W'' versus $x_1$, $c$, and
Hubble residual]{The velocity of the \ion{S}{II} ``W''
  feature versus SALT2 light-curve width parameter $x_1$, SALT2
  colour parameter $c$, and Hubble residuals corrected for
  light-curve width and colour. Colours and shapes of data points are
  the same as in 
  Figure~\ref{f:v_bv_si6355}.}\label{f:v_SALT_sw}
\end{figure}

On the other hand, the velocity of the \ion{S}{II} ``W,'' when used in
conjunction with $x_1$ and $c$, decreased the WRMS by \about3~per~cent
and the $\sigma_{\rm pred}$ by \about14~per~cent, at the 1.8$\sigma$
level ($\Delta_{x_1,c} = -0.0119 \pm
0.0066$). Figure~\ref{f:v_SALT_sw} shows the 64 SNe~Ia in the BSNIP
sample which have SALT2 fits and measured \ion{S}{II} ``W''
velocities within 5~d of maximum brightness. The velocities are again
plotted against $x_1$, $c$, and Hubble residuals corrected for
light-curve width and colour (only for SNe that are part of the Hubble
diagram).

\begin{figure}
\centering$
\begin{array}{c}
\includegraphics[width=3.45in]{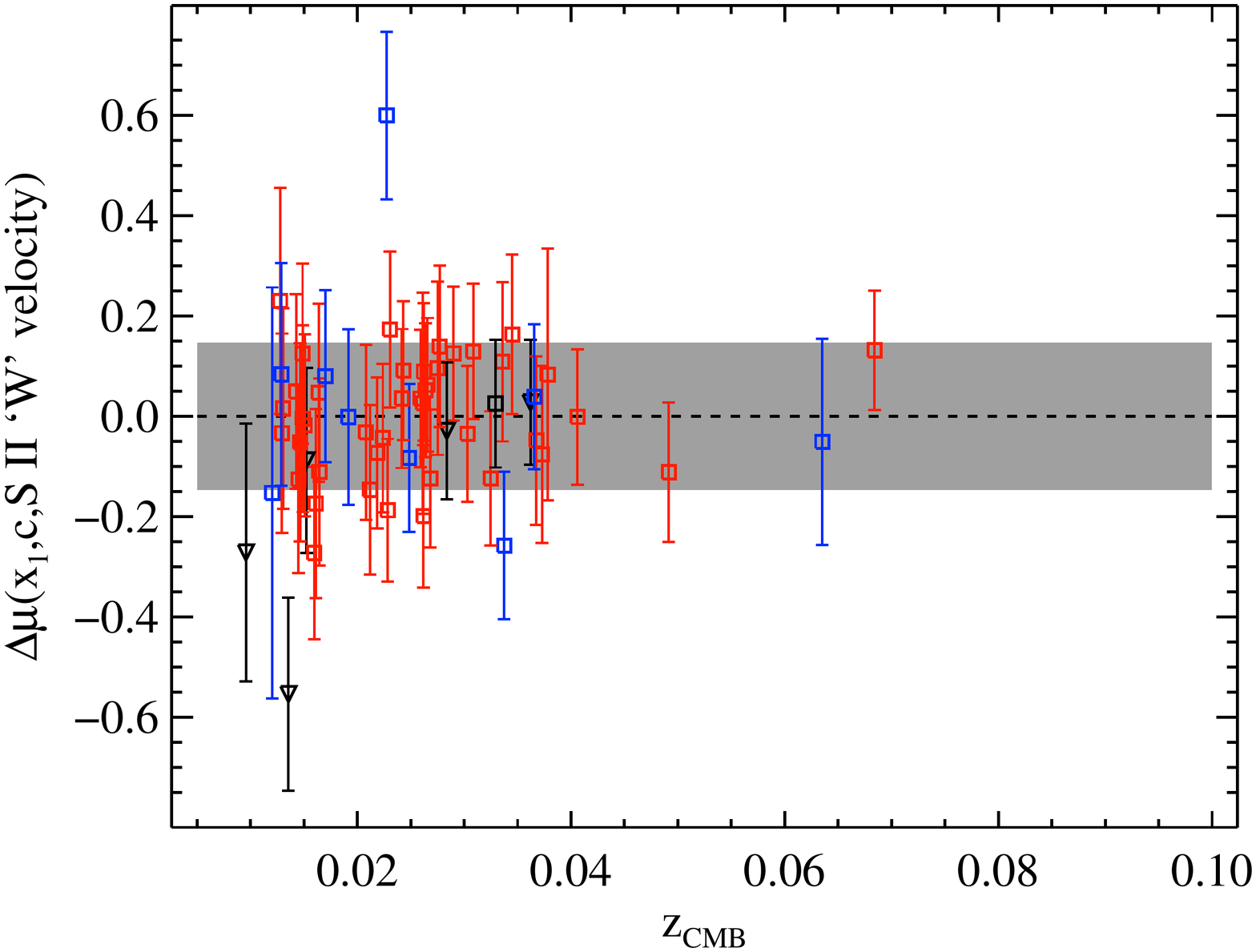} \\
\includegraphics[width=3.45in]{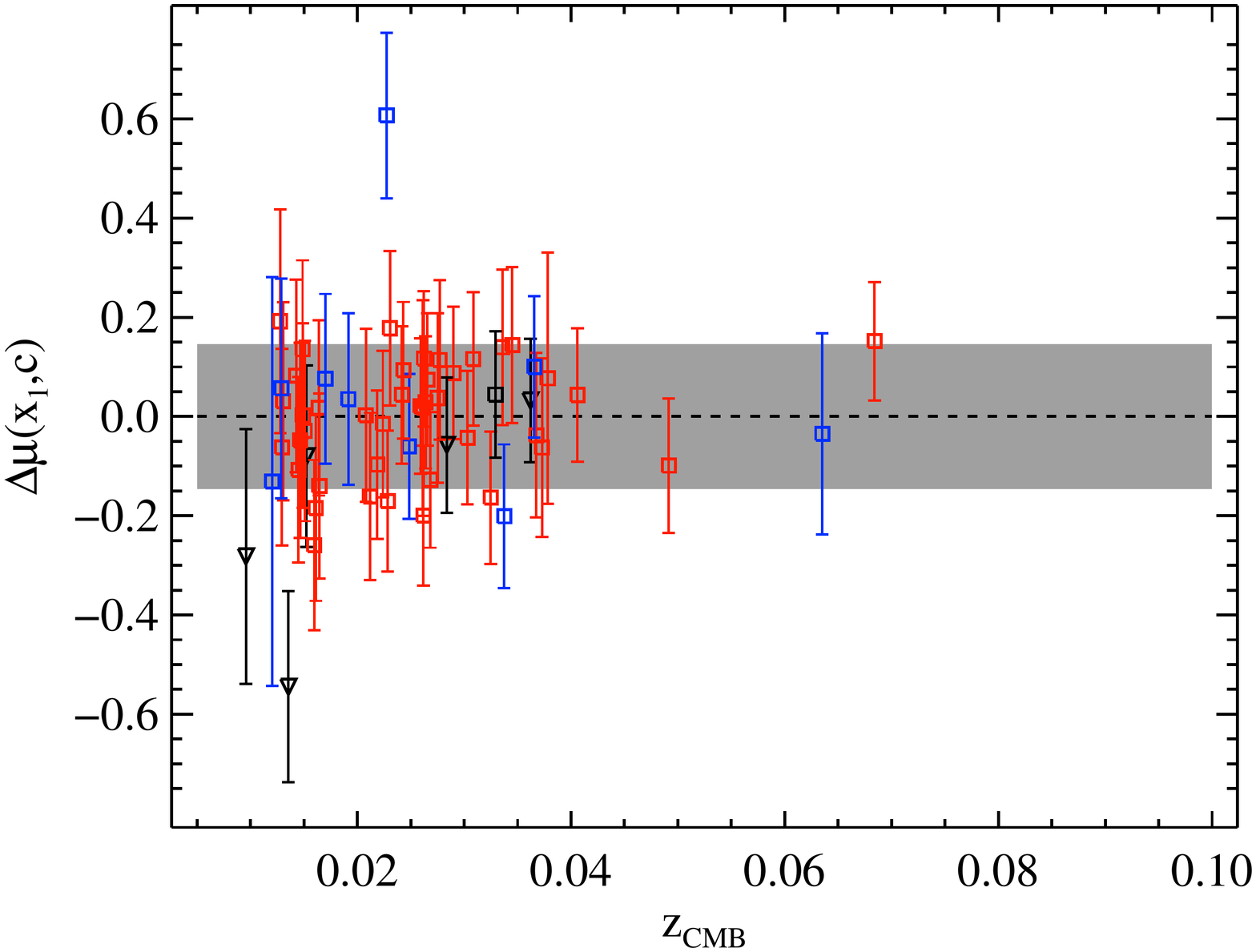} 
\end{array}$
\caption[Residuals versus $z_{\rm CMB}$ for the
$\left(x_1,c, {\rm S~II ``W'' velocity}\right)$ and $\left(x_1,c\right)$
models]{Hubble diagram residuals versus $z_{\rm CMB}$ for the
  $(x_1,c, $\ion{S}{II} ``W'' velocity$)$ model ({\it top}) and the standard
  $\left(x_1,c\right)$ model (Equation~\ref{eq:m5}, {\it bottom}). The
  grey band is 
  the WRMS for each model. Colours and shapes of data points are the
  same as in Figure~\ref{f:v_bv_si6355}.}\label{f:v_sw_z}
\end{figure}

Neither $x_1$ nor $c$ show any correlation with the velocity of the
\ion{S}{II} ``W'' (correlation coefficients of 0.13 and 0.17,
respectively). Even when removing SNe with $c > 0.5$ from the BSNIP
data (as done for the Hubble diagrams), the correlation
coefficient becomes only $-0.15$. The bottom panel of
Figure~\ref{f:v_SALT_sw} shows the velocity of the \ion{S}{II} ``W''
versus the $\left(x_1,c\right)$-corrected Hubble residuals. The
correction term is uncorrelated with the residuals 
(correlation coefficient 0.003). Figure~\ref{f:v_sw_z} shows the
actual Hubble diagram residuals for this model (top panel) as well as
the standard  $\left(x_1,c\right)$ model (using the same set of
objects; bottom panel) versus redshift. Also shown, as the grey band,
is the WRMS for both models.

While the relative depth of this feature has been seen to improve
Hubble diagrams \citep[][ and Section~\ref{ss:depth} of this
work]{Blondin11}, the velocity of this feature has not previously been
shown to do this. When adding the velocity of the \ion{S}{II} ``W''
feature to the standard $\left(x_1,c\right)$ model, the overall
decrease in WRMS is relatively small, but the effect appears to be
fairly significant. This distance model should be explored further
using future, larger datasets.

\subsection{Relative Depths}\label{ss:depth}

\begin{figure}
\centering$
\begin{array}{c}
\includegraphics[width=3.4in]{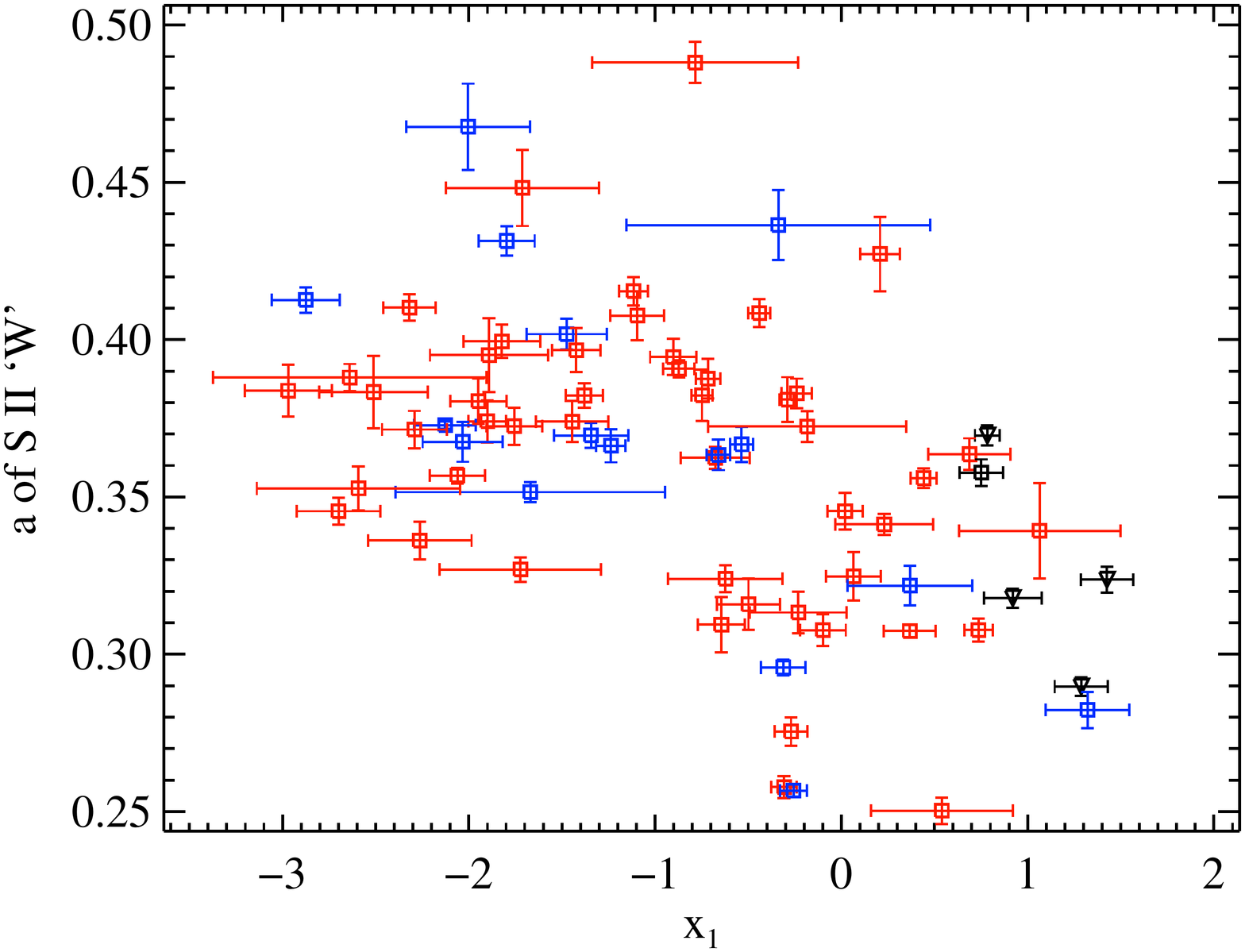} \\
\includegraphics[width=3.4in]{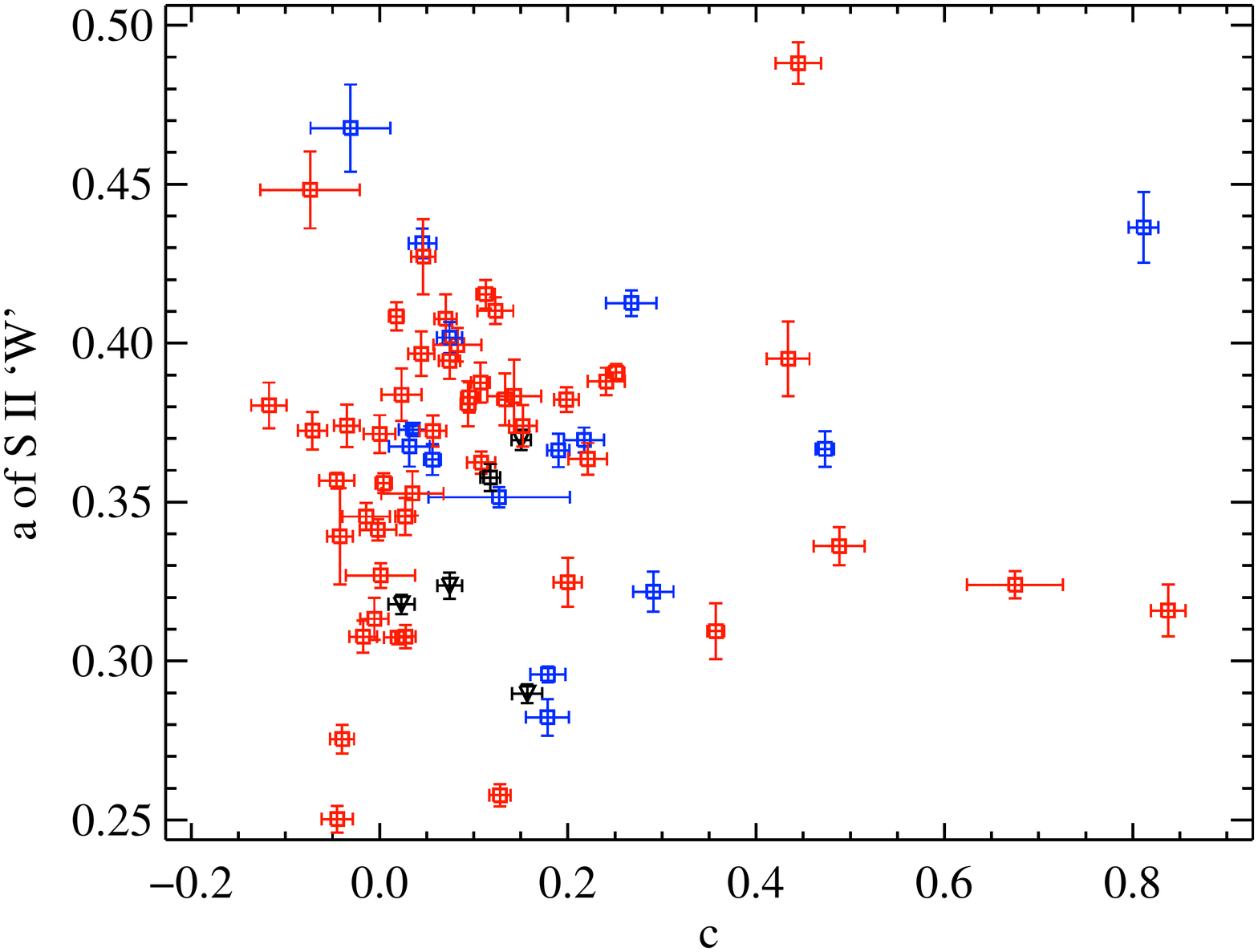} \\
\includegraphics[width=3.4in]{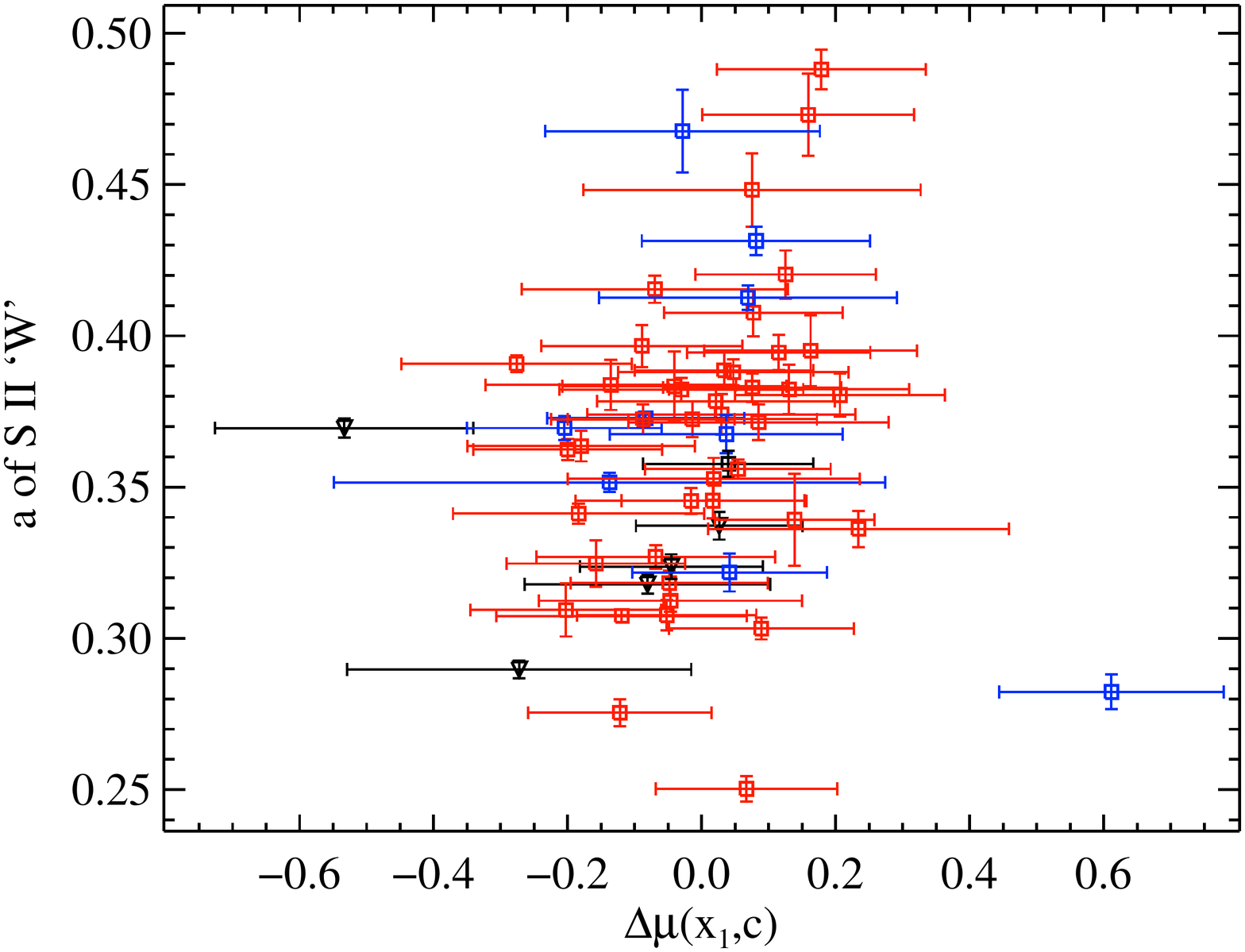} 
\end{array}$
\caption[Depth of S~II ``W'' versus $x_1$, $c$, and Hubble
residual]{The relative depth of the \ion{S}{II} ``W'' feature versus
  SALT2 light-curve width parameter $x_1$ ({\it top}), SALT2 colour
  parameter $c$ ({\it middle}), and Hubble residuals corrected for
  light-curve width 
  and colour ({\it bottom}). Colours and shapes of data points are the
  same as in 
  Figure~\ref{f:v_bv_si6355}.}\label{f:a_SALT_sw}
\end{figure}

The depth of the {\it bluer absorption} of the \ion{S}{II} ``W'' feature 
relative to the pseudo-continuum has been shown to decrease the
scatter of Hubble residuals by about 10~per~cent
\citep{Blondin11}. The relative depth ($a$) of this feature was found
to be uncorrelated with both $x_1$ and $c$ by \citet{Blondin11}, and
its correlation with Hubble residuals (corrected for light-curve
width and colour) is relatively small. Figure~\ref{f:a_SALT_sw}
presents the 64 BSNIP SNe~Ia which have SALT2 fits and measured
relative depths of the {\it redder absorption} of the \ion{S}{II} ``W''
feature within 5~d of maximum brightness. The depths are plotted
against $x_1$, $c$, and Hubble residuals corrected for light-curve
width and colour (for objects that are in the Hubble diagram). It
should be noted 
that whereas \citet{Blondin11} measure the bluer absorption of this
feature ($\lambda$5454), we measure only the redder absorption
($\lambda$5624) in BSNIP~II. While these absorptions are 
separated by $<200$~\AA, there may be differences between the relative
depths of the two. Nevertheless, we will compare the results presented
here to those of \citet{Blondin11}, with the caveat that this may be
analogous to comparing ``Red Delicious apples'' to ``Granny Smith apples.''

The BSNIP data show a fairly significant correlation between the
relative depth of \ion{S}{II} ``W'' and $x_1$ (a correlation
coefficient of $-0.47$, which is significant at the
$>3\sigma$ level\footnote{This significance comes from using the Pearson
  correlation coefficient for each correlation tested in this study to
  calculate a $p$-value. This value tests the null hypothesis that there
  is no correlation versus the alternative that a correlation exists. Then,
  based on the $p$-values of all of the tested correlations, we
  calculate the 3$\sigma$ cutoff by employing the False-Discovery Rate
  Method \citep{Miller01}.}). This 
is in stark contrast to \citet{Blondin11}, who found no evidence of
such a correlation (correlation coefficient
$-0.04$). As opposed to $x_1$, both studies agree that $c$ is
uncorrelated with the relative depth 
$a$ of the \ion{S}{II} ``W'' feature (the BSNIP data having a 
correlation coefficient of 0.09). The Hubble residuals corrected
for $x_1$ and $c$ show weak evidence of a correlation with the relative
depth of the \ion{S}{II} ``W'' (correlation coefficient
0.29). \citet{Blondin11} found
that a model which includes $x_1$, $c$, and $a$ of the \ion{S}{II} ``W''
will decrease the WRMS by $\la 10$~per~cent, while we find the WRMS to be
effectively unchanged whether or not one adds in the relative depth of
\ion{S}{II} ``W'' ($\Delta_{x_1,c} = 0.0012 \pm 0.0047$).

All other spectral features' relative depths were used, along with
Equations~\ref{eq:m1}--\ref{eq:m4}, to create Hubble diagrams. No
model significantly decreased the residuals over the standard
$\left(x_1,c\right)$ model. Models involving the relative depth of
\ion{Ca}{II}~H\&K, \ion{Si}{II} $\lambda$6355, and the \ion{O}{I}
triplet, each in combination with $x_1$ and $c$, were found to be as
accurate as the $\left(x_1,c\right)$ model.

\subsection{Pseudo-Equivalent Widths}\label{ss:ew}

\citet{Nordin11a} fit the temporal evolution of their pEW measurements
to attempt to ``remove'' the age dependence of the pEW values. To do
this, an epoch-independent quantity called the ``pEW difference''
($\Delta$pEW) was defined; it is simply the measured pEW minus the
expected pEW at the same epoch using the linear or quadratic fit. In
BSNIP~II we calculated $\Delta$pEW for the BSNIP sample. However, the
relationships seen in BSNIP~II involving $\Delta$pEW values were also
seen when simply using the pEW values (within 5~d of maximum
brightness). This is due to the fact that pEWs do not evolve much
within 
a few days of maximum. Furthermore, the $\Delta$pEW values rely on
defining a fit to the measurements which adds another assumption to
the analysis. Thus, the current study will focus solely on pEW values
within 5~d of maximum brightness and will not further investigate
$\Delta$pEW values. Note that comparisons to \citet{Nordin11a} will be 
made, despite the fact that their study uses $\Delta$pEW values almost 
exclusively.

\subsubsection{\ion{Si}{II} $\lambda$4000}\label{sss:ew_si4000}

The pEW of the \ion{Si}{II} $\lambda$4000 feature has recently been
found to be an indicator of light-curve width due its relatively tight
anticorrelation with the SALT2 $x_1$ parameter \citep{Arsenijevic08,
  Walker11,Blondin11,Nordin11a,Chotard11}. Curiously, in BSNIP~II,
only a weak 
correlation was found between this pEW and another often-used SN~Ia 
luminosity indicator, the so-called ``\ion{Si}{II} ratio'',
$\Re$(\ion{Si}{II}) (originally defined by Nugent \etal 1995; see also BSNIP~II
and Section~\ref{ss:si_ratio} for more information on this spectral
parameter). Here the pEW of \ion{Si}{II} $\lambda$4000 is
compared directly to photometric parameters.

\begin{figure}
\centering$
\begin{array}{c}
\includegraphics[width=3.4in]{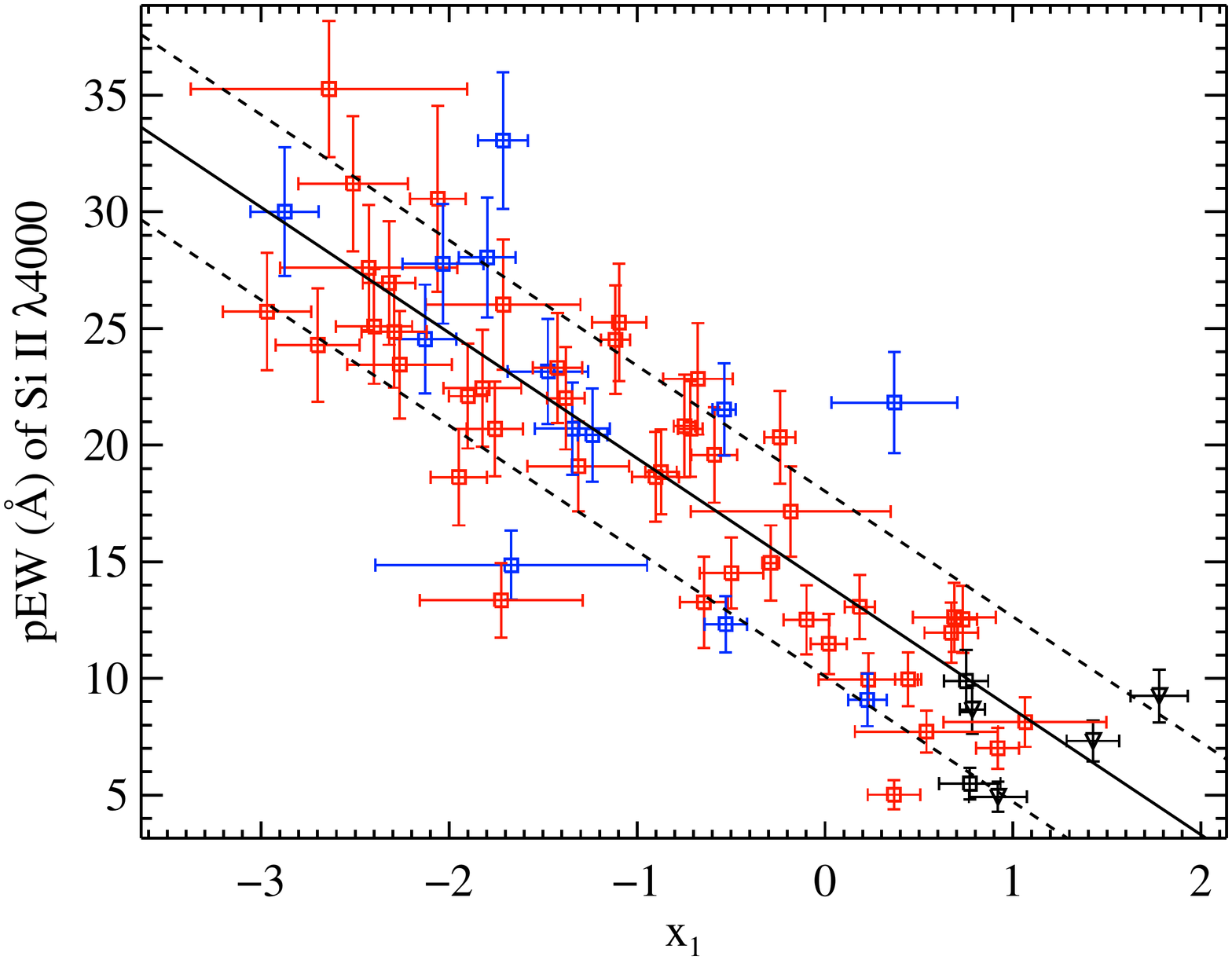} \\
\includegraphics[width=3.4in]{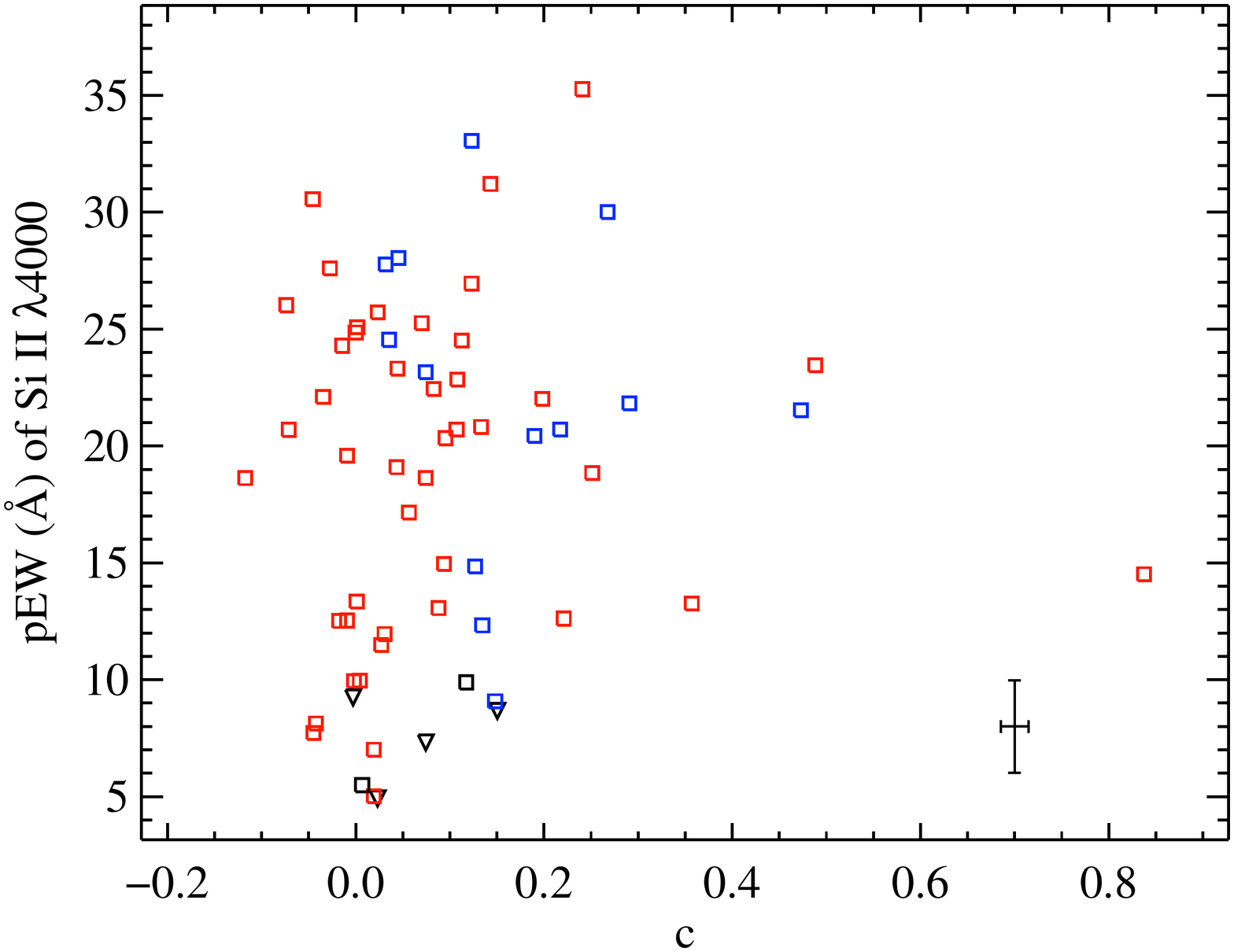} \\
\includegraphics[width=3.4in]{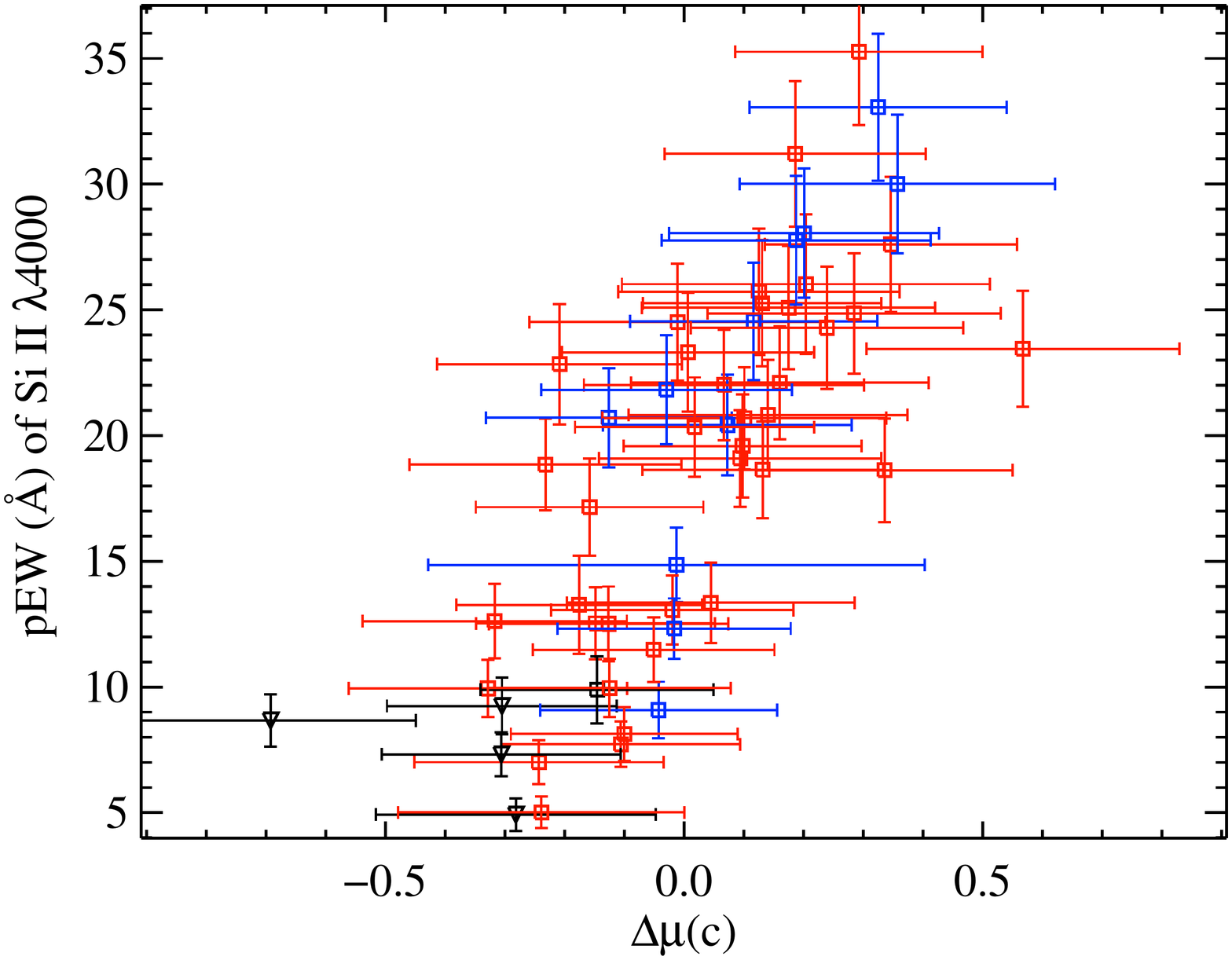} 
\end{array}$
\caption[pEW of Si~II $\lambda$4000 versus $x_1$, $c$, and
Hubble residual]{The pEW of the \ion{Si}{II} $\lambda$4000 feature versus
  SALT2 light-curve width parameter $x_1$ ({\it top}), SALT2 colour
  parameter $c$ ({\it middle}), and Hubble residuals corrected for colour
  only ({\it bottom}). Colours and shapes of data points are the same as in
  Figure~\ref{f:v_bv_si6355}. In the top plot, the solid line
  is the linear least-squares fit and the dashed lines are the
  standard error of the fit. In the middle plot, the median
  uncertainty in both 
  directions is shown in the lower-right
  corner.}\label{f:ew_SALT_si4000}
\end{figure}

In Figure~\ref{f:ew_SALT_si4000} we present the 57 BSNIP SNe which
have SALT2 fits and measured pEW values for the \ion{Si}{II}
$\lambda$4000 feature within 5~d of maximum brightness. The pEWs are
plotted against $x_1$, $c$, and Hubble residuals corrected only for
colour (for SNe~Ia that are used when constructing the Hubble
diagram).

The pEW of \ion{Si}{II} $\lambda$4000 is highly correlated with $x_1$
(Figure~\ref{f:ew_SALT_si4000}, top plot) with a 
correlation coefficient of $-0.86$ (which is significant at
$>3\sigma$). This is in agreement with many
previous studies and is actually a stronger correlation than has been
seen before
\citep[e.g.,][]{Arsenijevic08,Walker11,Blondin11,Nordin11a,Chotard11}. The 
least-squares linear fit to the data is shown as the solid line in the
top plot of Figure~\ref{f:ew_SALT_si4000} (the dashed lines are
the standard error of the fit). Note that the Ia-99aa objects in the
figure also lie along the relationship.

\citet{Arsenijevic08} plotted $x_1$ versus the pEW of \ion{Si}{II}
$\lambda$4000 and coded each low-redshift point based on its Benetti type;
they found that the FAINT objects fell below the linear
relationship (i.e., they had smaller than expected pEW values). In
BSNIP~II it was shown that FAINT (and similarly, 
Ia-91bg) objects have, if anything, {\it larger} than average pEW
values (especially for the \ion{Si}{II} $\lambda$4000
feature). Figure~\ref{f:ew_SALT_si4000} shows no Ia-91bg objects since
SALT/2 is unable to fit that spectral subtype. However, when the BSNIP
values of $x_1$ are plotted against the pEW of \ion{Si}{II} $\lambda$4000
and coded by Benetti type, the two FAINT objects fall at the
upper-left end of the linear correlation. The cause of
this discrepancy between the two studies is unclear.

The middle plot of Figure~\ref{f:ew_SALT_si4000} shows no real
evidence that the pEW of the \ion{Si}{II} $\lambda$4000 feature is
correlated with $c$. The correlation coefficient we find for
all of the objects is 0.095, and when removing objects with $c > 0.5$,
the coefficient only increases to 0.20. This is slightly smaller than what
was found by \citet{Blondin11}, and significantly smaller than
what was found by \citet{Nordin11b}. While the former claim no
observed correlation, the latter do claim that the pEW of \ion{Si}{II}
$\lambda$4000 is correlated with $c$. 

\begin{figure}
\centering$
\begin{array}{c}
\includegraphics[width=3.45in]{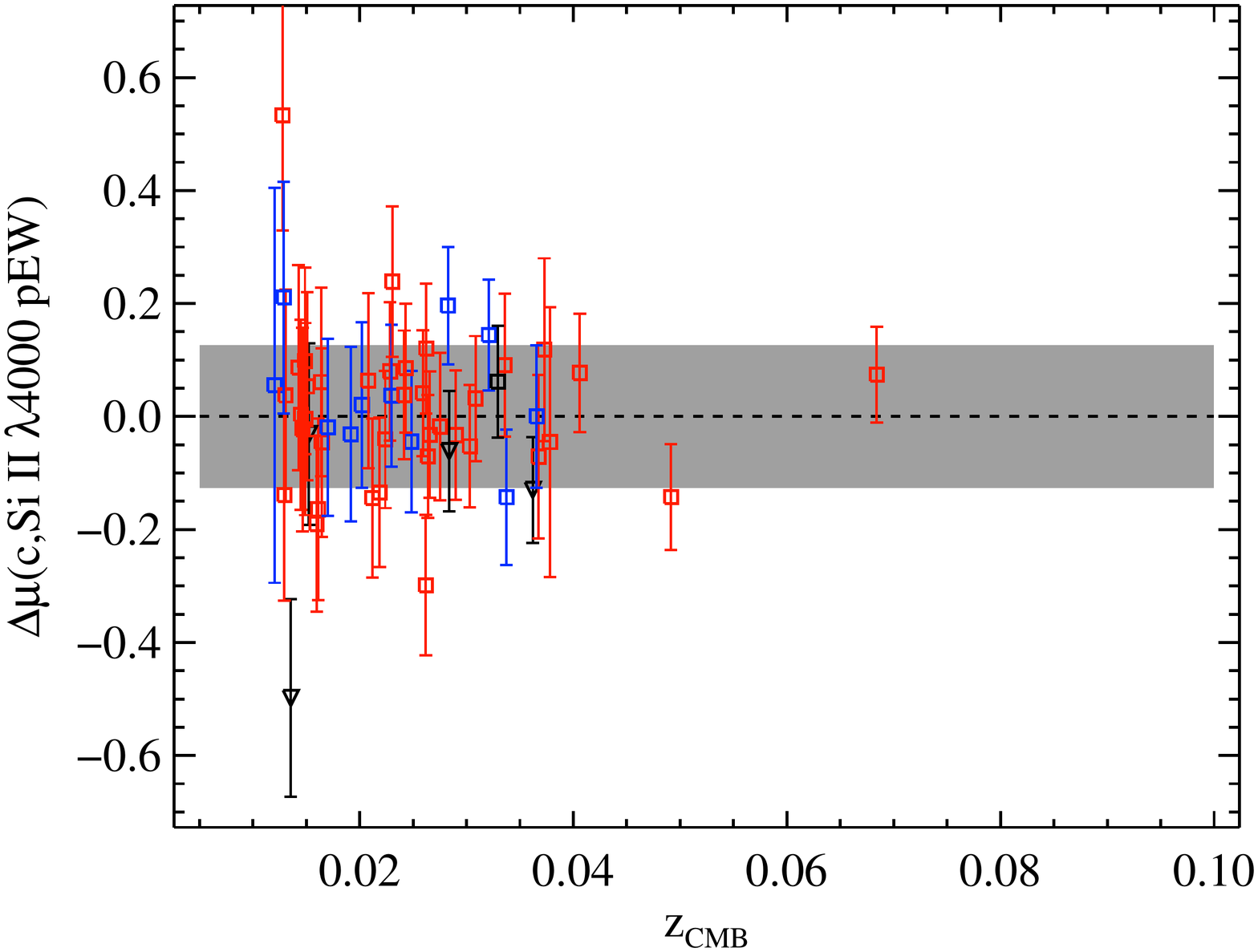} \\
\includegraphics[width=3.45in]{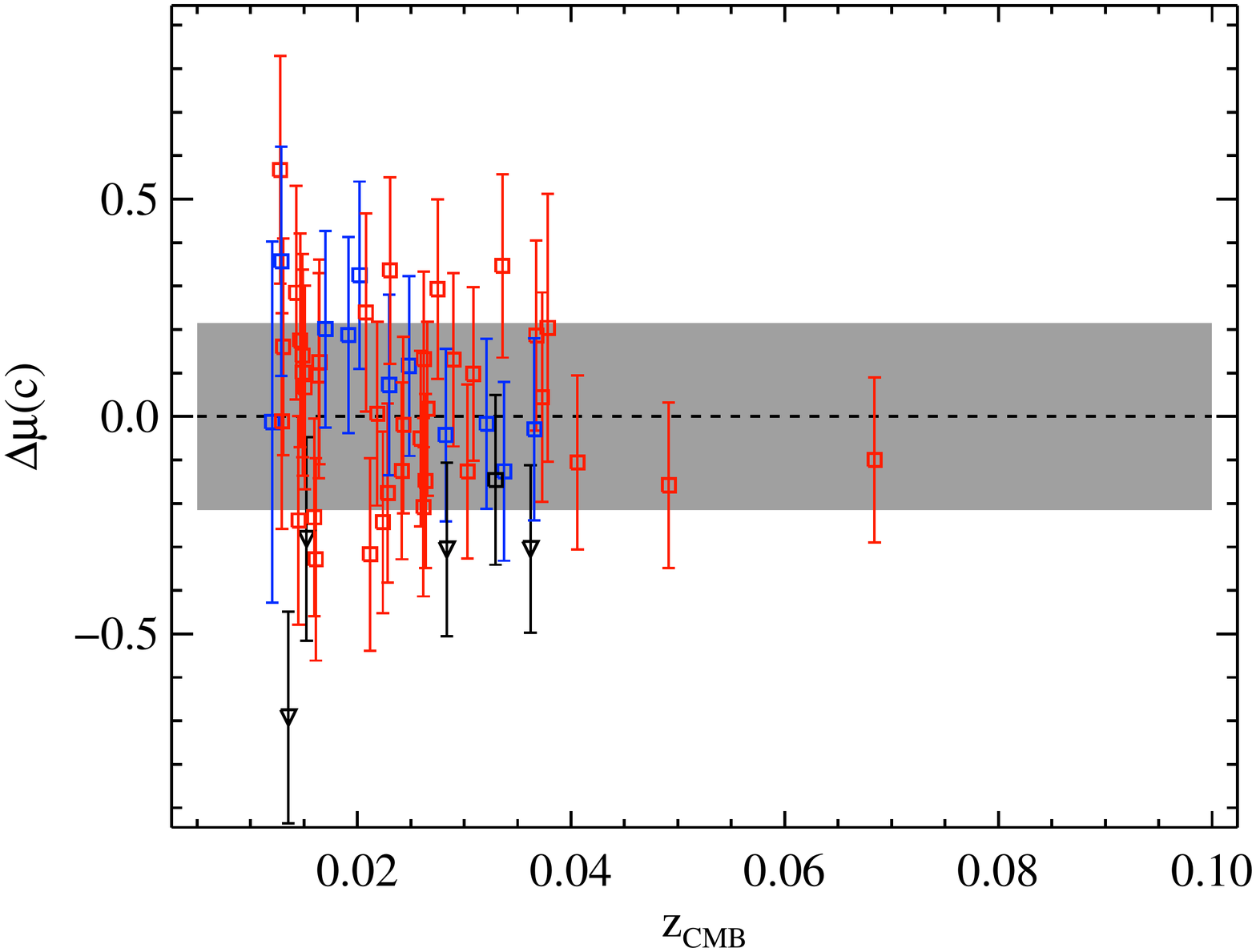} 
\end{array}$
\caption[Residuals versus $z_{\rm CMB}$ for the
$(c,$ Si~II $\lambda$4000 pEW$)$ and $(c)$
models]{Hubble diagram residuals versus $z_{\rm CMB}$ for the
  $(c,$ \ion{Si}{II} $\lambda$4000 pEW$)$ model ({\it top}) and the
  colour-corrected-only model ({\it bottom}). The grey band is the WRMS for each 
  model. Colours and shapes of data points are the
  same as in Figure~\ref{f:v_bv_si6355}.}\label{f:ew_si4000_z_c} 
\end{figure}

The bottom panel of Figure~\ref{f:ew_SALT_si4000} shows the Hubble
residuals when corrected only for SALT2 colour versus the pEW of
\ion{Si}{II} $\lambda$4000. \citet{Blondin11} saw a relatively weak
correlation between these parameters and found that a distance model
involving $c$ and the pEW of the \ion{Si}{II} $\lambda$4000 feature led to
a ``marginal improvement'' over the standard $\left(x_1,c\right)$
model. We find a strong correlation (coefficient of 0.81, significant
at $>3\sigma$), 
and the $(c,$ \ion{Si}{II} $\lambda$4000 pEW$)$ model
performs nearly as well as the $\left(x_1,c\right)$ model
($\Delta_{x_1,c} = 0.012 \pm 0.036$). Thus, the BSNIP data are in
agreement with the finding of \citet{Blondin11} that the pEW of the
\ion{Si}{II} $\lambda$4000 feature is essentially a replacement for
the $x_1$ parameter and is an accurate measurement of light-curve
width. Figure~\ref{f:ew_si4000_z_c} shows the Hubble diagram residuals
for the $(c,$ \ion{Si}{II} $\lambda$4000 pEW$)$ model and the
colour-corrected-only model verus redshift, with the WRMS for each
model as the grey band.

\begin{figure}
\centering$
\begin{array}{c}
\includegraphics[width=3.45in]{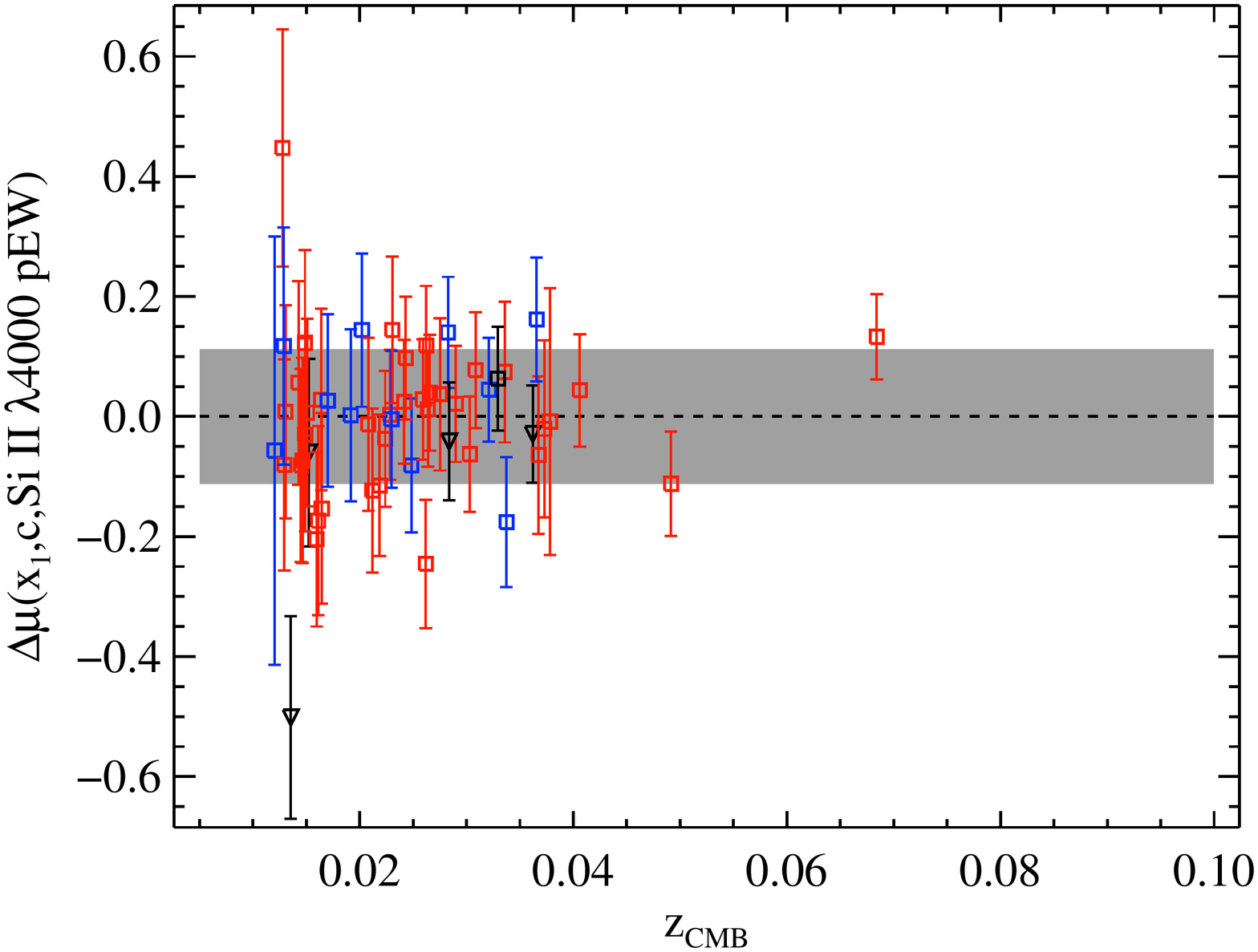} \\
\includegraphics[width=3.45in]{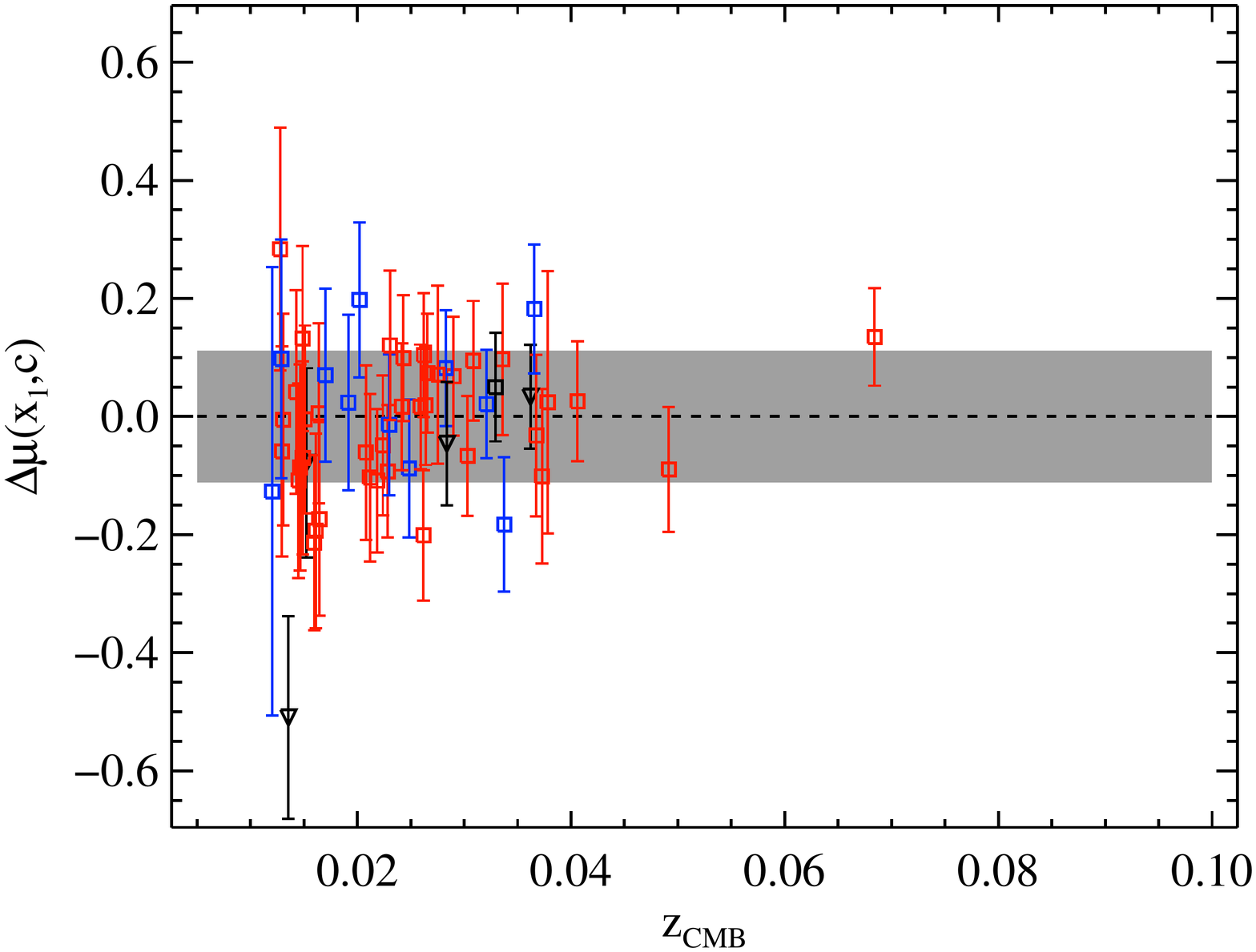} 
\end{array}$
\caption[Residuals versus $z_{\rm CMB}$ for the
$(x_1,c,$ Si~II $\lambda$4000 pEW$)$ and $(x_1,c)$
models]{Hubble diagram residuals versus $z_{\rm CMB}$ for the
  $(x_1,c,$ \ion{Si}{II} $\lambda$4000 pEW$)$ model ({\it top}) and the 
  standard $\left(x_1,c\right)$ model ({\it bottom}). The grey band is the WRMS for
  each model. Colours and shapes of data points are the
  same as in Figure~\ref{f:v_bv_si6355}.}\label{f:ew_si4000_z} 
\end{figure}

Interestingly, we found that a model involving $c$, $x_1$, {\it and} the
pEW of the \ion{Si}{II} $\lambda$4000 feature actually leads to a
\about10~per~cent decrease in WRMS and a \about28~per~cent decrease in $\sigma_{\rm
  pred}$. For this model, $\Delta_{x_1,c} = -0.026 \pm 0.15$, which
implies that the improvement has a significance of about 1.8$\sigma$. The
correlation between pEW of \ion{Si}{II} $\lambda$4000 and Hubble
residuals corrected for colour and light-curve width is only
0.20, and thus perhaps the combination of $c$, $x_1$, and pEW of the
\ion{Si}{II} $\lambda$4000 feature is not actually adding much new
information. However, the utility of a model including all three
of these parameters should be investigated with other SN
samples. Figure~\ref{f:ew_si4000_z} contains Hubble residuals for the 
$(x_1,c,$ \ion{Si}{II} $\lambda$4000 pEW$)$ model as well as
the standard $\left(x_1,c\right)$ model (using the same set of
objects) versus redshift. Also shown, as the grey band, is the WRMS
for each model.

\begin{figure}
\centering$
\begin{array}{c}
\includegraphics[width=3.45in]{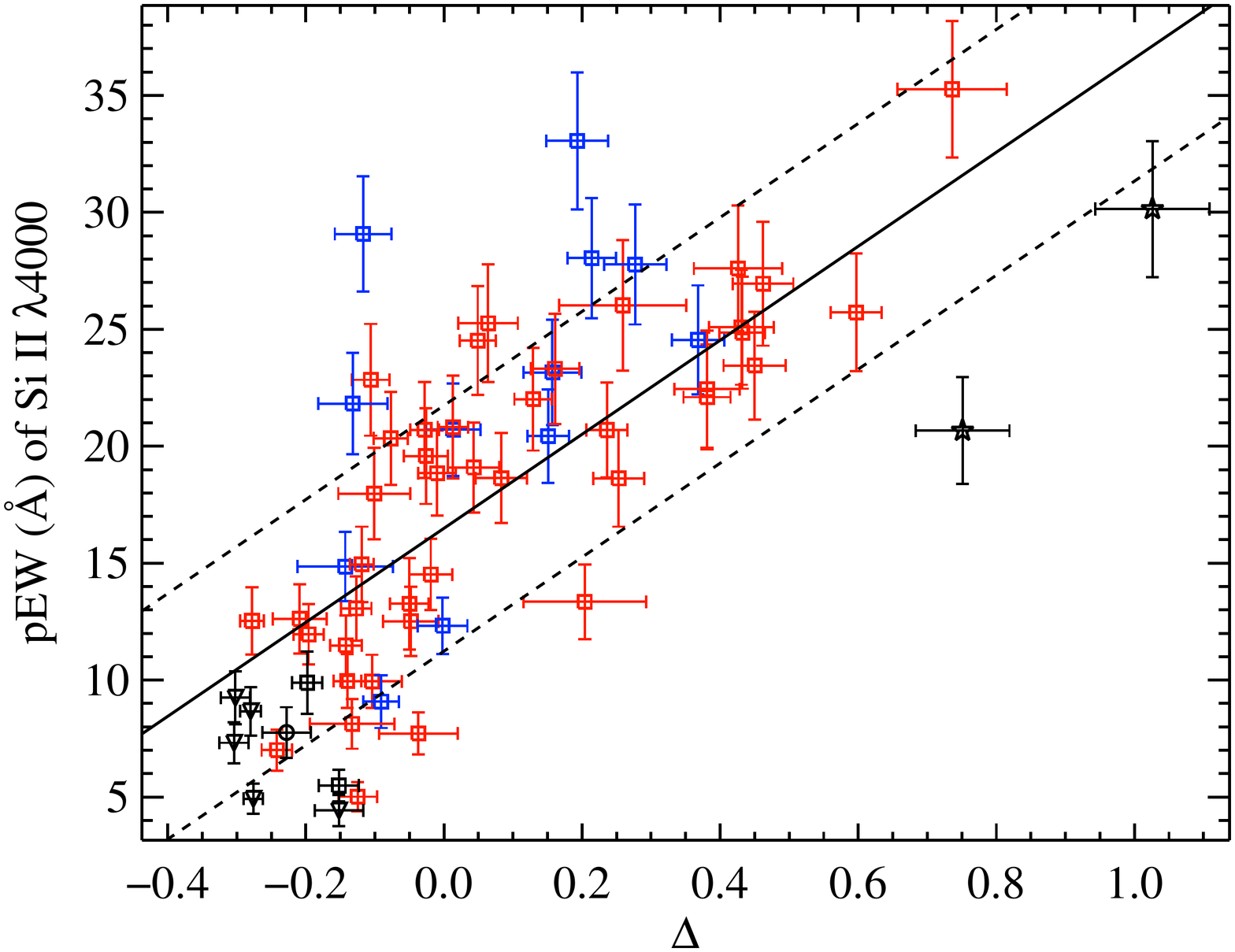} \\
\includegraphics[width=3.45in]{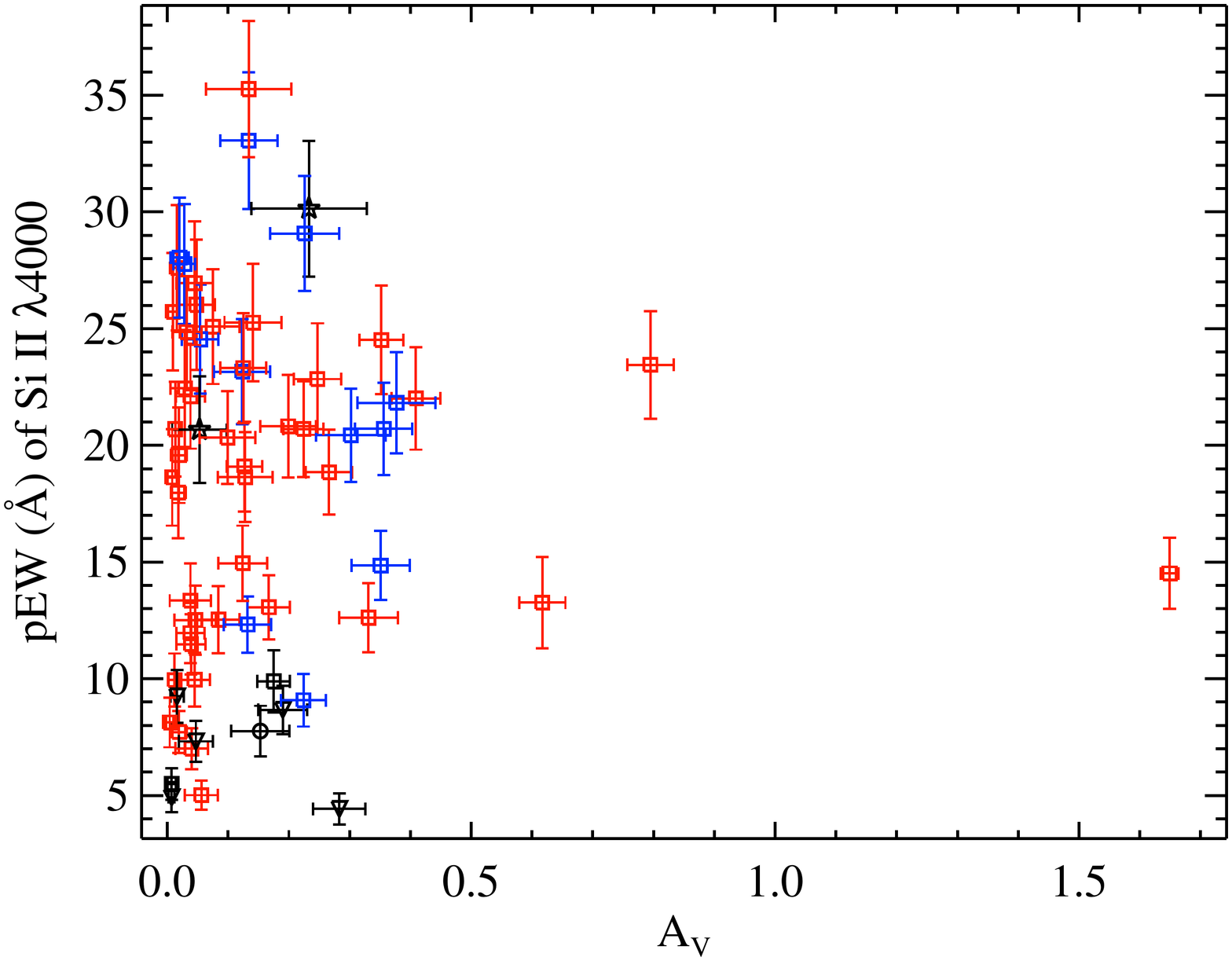}
\end{array}$
\caption[pEW of Si~II $\lambda$4000 versus $\Delta$ and
$A_V$]{The pEW of the \ion{Si}{II} $\lambda$4000 feature versus 
  MLCS2k2 light-curve width parameter $\Delta$ ({\it top}) and MLCS2k2 reddening
  parameter $A_V$ ({\it bottom}). Colours and shapes of data points are the same as in 
  Figure~\ref{f:v_bv_si6355}. In the top plot, the solid line is the
  linear least-squares fit and the dashed lines are the standard error
  of the fit.}\label{f:ew_MLCS_si4000} 
\end{figure}

While the above investigation focused on the SALT2 light-curve fitter,
we can investigate correlations between the pEW of the \ion{Si}{II}
$\lambda$4000 feature and photometric parameters from
MLCS2k2. Figure~\ref{f:ew_MLCS_si4000} shows the 63 BSNIP SNe which
have MLCS2k2 fits and measured pEW values for the \ion{Si}{II}
$\lambda$4000 feature within 5~d of maximum brightness. The pEWs are
plotted against $\Delta$ and $A_V$.

The correlation coefficient between pEW of \ion{Si}{II}
$\lambda$4000 and $\Delta$ is 0.74, which is larger than previously
observed \citep{Nordin11a} and implies that these parameters are
highly correlated (at a significance $>3\sigma$). This is expected
based on the high degree of 
correlation between the pEW of \ion{Si}{II} $\lambda$4000 and $x_1$ since
both $\Delta$ and $x_1$ are measurements of the width of SN~Ia light
curves. In Figures~\ref{f:ew_MLCS_si4000} and \ref{f:ew_SALT_si4000}
the Ia-99aa objects lie at the extreme low-pEW end of the
relationship, though perhaps they are slightly systematically below
the trend in the top panel of Figure~\ref{f:ew_MLCS_si4000}. On the
other hand, the Ia-91bg objects in the top panel of
Figure~\ref{f:ew_MLCS_si4000} fall significantly below the linear
trend.

As in \citet{Nordin11a}, there is no significant correlation between
the pEW of \ion{Si}{II} $\lambda$4000 and $A_V$, even when objects with
$A_V > 0.5$ mag are removed (correlation coefficients of $< 0.13$ in
both cases using the BSNIP data). In all of the plots presented in
this section, HV and Ia-norm objects 
overlap significantly. This is expected since in BSNIP~II it was shown
that the pEW of the \ion{Si}{II} $\lambda$4000 feature is extremely
similar for these two subclasses. We also note that the plot of
$\Delta m_{15}(B)$ versus the pEW of \ion{Si}{II} $\lambda$4000 looks
nearly identical to the top panel of Figure~\ref{f:ew_MLCS_si4000} and
has a larger correlation coefficient of 0.85 (again, significant at
$>3\sigma$).

\subsubsection{\ion{Fe}{II} and \ion{Mg}{II}}\label{sss:ew_fe}

\begin{figure}
\centering
\includegraphics[width=3.45in]{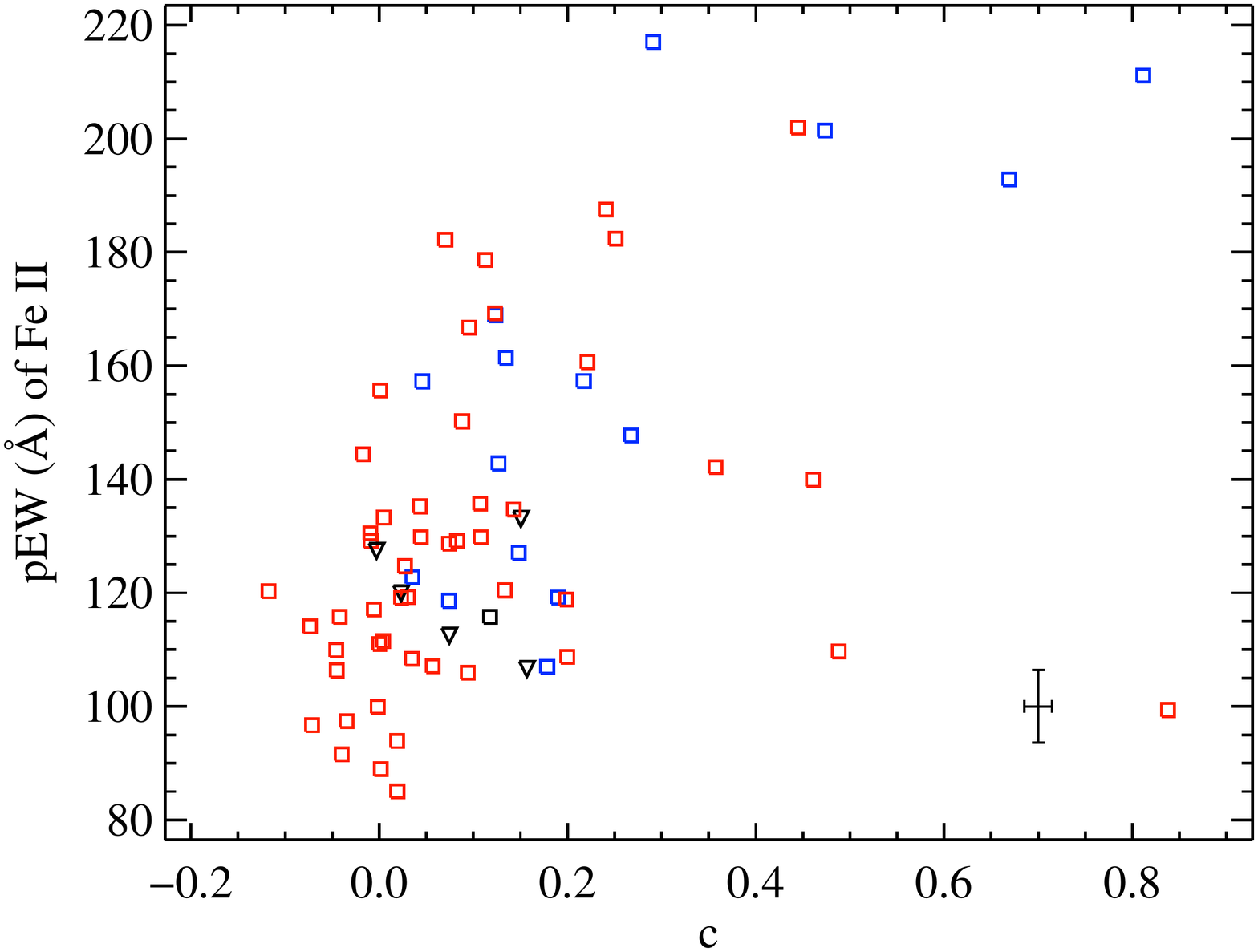}
\caption[pEW of Fe~II versus $c$]{The pEW of the \ion{Fe}{II}
  complex versus the SALT2 colour parameter $c$. Colours and
  shapes of data points are the same as in
  Figure~\ref{f:v_bv_si6355}. The median uncertainty in both
  directions is shown in the lower-right corner.}\label{f:ew_c_fe}
\end{figure}

\citet{Nordin11a} found that the $\Delta$pEW of \ion{Fe}{II} within
3~d of maximum brightness is well correlated with SALT colour. In
Figure~\ref{f:ew_c_fe} we present the 63 SNe in the BSNIP sample 
which have SALT2 fits and a measurement of the pEW of the \ion{Fe}{II}
complex within 5~d of maximum brightness. The
two observables plotted have a correlation coefficient of
0.49 (with significance $>3\sigma$), increasing slightly for objects
with $c < 0.5$. If, 
however, only spectra within 3~d of maximum brightness are 
used, the correlation increases slightly but the sample size decreases
by nearly one-quarter. The strength of this correlation is slightly
higher than that found by \citet{Nordin11a}, though we point out that
they used SALT colour while we use SALT2 colour, and they used
$\Delta$pEW (i.e., the difference between the measured pEW and the
average pEW evolution) while we use the actual measured pEW. Ia-99aa
objects appear to have typical values of both the pEW of \ion{Fe}{II} and
$c$, though there are a very small number of objects of this spectral
subtype. On the other hand, HV SNe seem to be both redder and have
larger pEWs, but as before, there is significant overlap with Ia-norm
objects as well.

\begin{figure}
\centering
\includegraphics[width=3.45in]{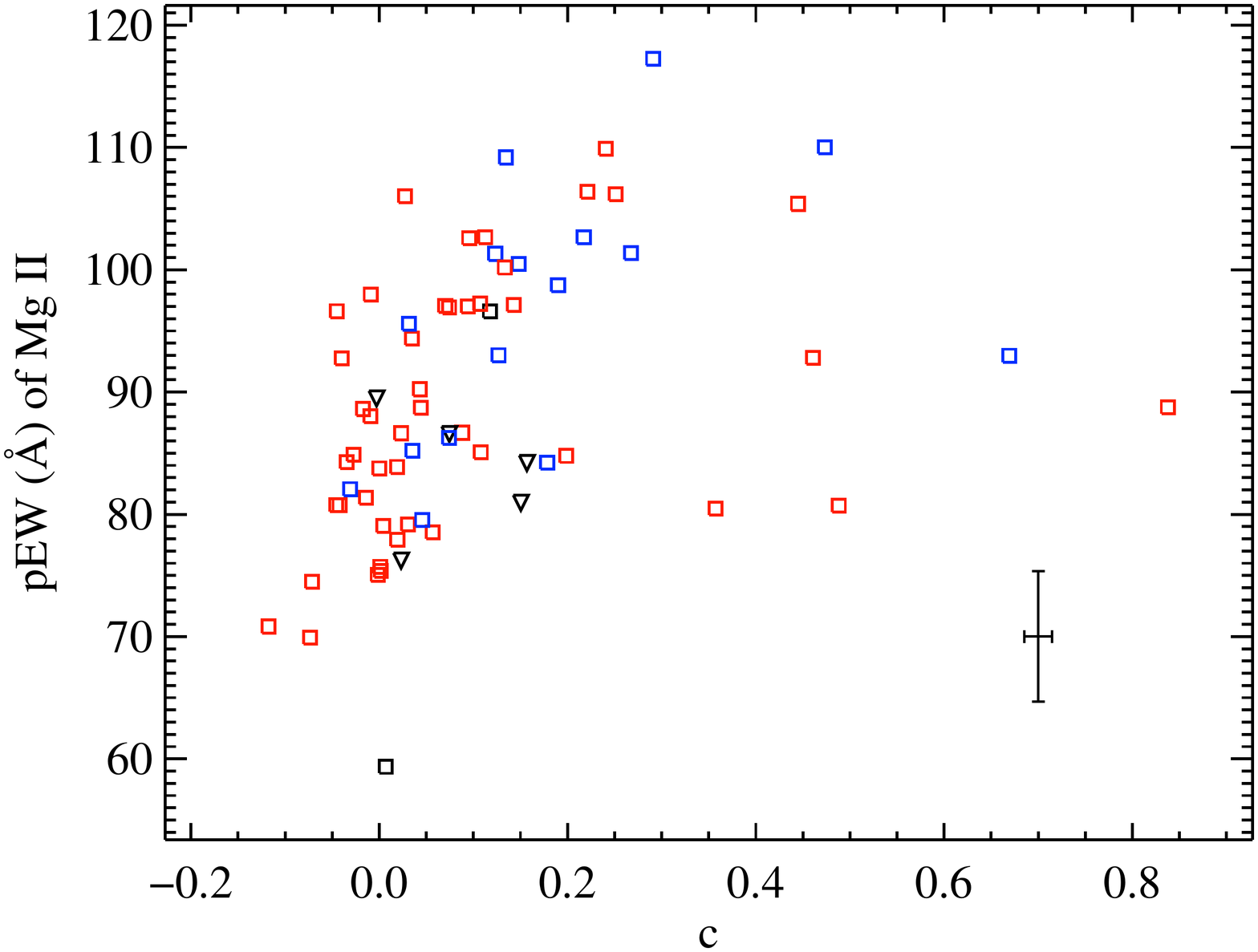}
\caption[pEW of Mg~II versus $c$]{The pEW of the \ion{Mg}{II}
  complex versus the SALT2 colour parameter $c$. Colours and
  shapes of data points are the same as in
  Figure~\ref{f:v_bv_si6355}. The median uncertainty in both
  directions is shown in the lower-right corner.}\label{f:ew_c_mg} 
\end{figure}

The pEW of the \ion{Mg}{II} complex is also correlated with $c$, 
as seen in Figure~\ref{f:ew_c_mg}. In that figure, 64 SNe within 5~d of
maximum brightness are plotted, and they have a correlation
coefficient of 0.44 (for objects with $c < 0.5$) with the
significance of the correlation \about3$\sigma$). Once again, when
using only spectra within 3~d of 
maximum brightness the correlation becomes slightly stronger at the
expense of significant decrease in the sample size. This correlation
has been observed previously, though at slightly lower significance
and with about one-third the number of low-$z$ SNe
\citep{Walker11}. The data presented in Figure~\ref{f:ew_c_mg} match
well with what was shown by \citet{Walker11} for low-$z$ objects, but
their high-$z$ sample spans a much larger range of \ion{Mg}{II} pEW
values and shows nearly no correlation with $c$. There seems to be
effectively no difference between the various spectroscopic subtypes
in this parameter space.

The significant correlations between $c$ and both the pEWs of \ion{Mg}{II}
and \ion{Fe}{II} are promising. Measuring a pEW of a broad feature in
a single spectrum near maximum brightness is much simpler than
obtaining photometric data for a full light curve that needs to then
be modeled by a light-curve fitter (such as SALT2). Furthermore, since
interstellar reddening cannot affect pEWs significantly, it seems that
the {\it intrinsic} colour of the SN is correlated with both the pEWs
of \ion{Mg}{II} and \ion{Fe}{II}.

As with all other spectral measurements discussed in this work, we
constructed Hubble diagrams using the pEW of 
\ion{Mg}{II} and \ion{Fe}{II} and models of the form shown in
Equations~\ref{eq:m1}--\ref{eq:m4}. All but one of these models
performed worse than the standard $\left(x_1,c\right)$ model. The
model including $x_1$, $c$, and the pEW of \ion{Mg}{II} was only as 
accurate as the standard model ($\Delta_{x_1,c} = -0.004 \pm
0.004$). However, using the pEW of
\ion{Mg}{II} or \ion{Fe}{II} as a replacement for $c$ or in addition
to $c$ (along with $x_1$) is a tantalising possibility that should be
explored using future datasets.

Furthermore, we attempt two Hubble diagrams using no light-curve
information whatsoever. One uses only the pEWs of the \ion{Si}{II}
$\lambda$4000 feature and the \ion{Mg}{II} complex, and the other uses
only the pEWs of the \ion{Si}{II} $\lambda$4000 feature and the
\ion{Fe}{II} complex. The idea is that the pEW of \ion{Si}{II}
$\lambda$4000 is a good proxy for $x_1$, and the pEWs of the
\ion{Mg}{II} and \ion{Fe}{II} features are reasonably good proxies for
$c$. These Hubble diagrams included only a subset of the data used in
the flux-ratio study (see Section~\ref{ss:ratios}) since they needed
to have well-measured pEWs. While both the \ion{Mg}{II} and
\ion{Fe}{II} models had quite low WMRS values (0.274 and
0.297, respectively), they were not as low as the WRMS values
using the standard $\left(x_1,c\right)$ model (0.118 and 0.121,
respectively).

\subsubsection{\ion{S}{II} ``W''}\label{sss:ew_sw}

As discussed in Section~\ref{ss:depth}, the depth of the {\it bluer
  absorption} of the \ion{S}{II} ``W'' feature relative to the
pseudo-continuum was shown by \citet{Blondin11} to decrease the
scatter of Hubble residuals by about 10~per~cent. In that section we
showed that the relative depths of the {\it redder absorption} of the
\ion{S}{II} ``W'' in the BSNIP data were marginally correlated with
$x_1$, opposite to what was seen by \citet{Blondin11}. However, both
studies agree that $c$ and the colour- and width-corrected Hubble
residual are uncorrelated with the relative depth of the
\ion{S}{II} ``W.''

As discussed in BSNIP~II, the relative depth of a spectral
feature relies on a spline fit to the spectra and can fairly easily be
contaminated by local noise. The pEW, however, is less prone to this
type of contamination, relies only on the definition of the
pseudo-continuum (and not any additional fit to the data), and often
contains the same information as the relative depth. For these reasons
the pEW values were used in favor of the $a$ values in the analysis
performed in BSNIP~II. Furthermore, both \citet{Blondin11} and
BSNIP~II measure the pEW of the {\it entire} \ion{S}{II} ``W''
feature, 
and thus pEW values are a more fair comparison between the two studies
than are the $a$ values.



We find that $c$ is uncorrelated with the pEW of the 
\ion{S}{II} ``W'' (correlation coefficient of 0.15 for objects
with $c < 0.5$). In contrast with the relative depth of the \ion{S}{II}
``W,'' however, $x_1$ is {\it also} uncorrelated with the pEW of the
\ion{S}{II} ``W'' (correlation coefficient $-0.16$). This is
significantly weaker than the correlation between $x_1$ and the
relative depth of this feature found in
Section~\ref{ss:depth}. Equations~\ref{eq:m1}--\ref{eq:m4} were used 
to create Hubble diagrams involving the pEW of the \ion{S}{II} ``W,''
but none of these models led to an improvement in the WRMS.

\subsubsection{\ion{Si}{II} $\lambda$5972}\label{sss:ew_si5972}

In BSNIP~II it was shown that the pEW of the \ion{Si}{II}
$\lambda$5972 feature correlated well with the spectral luminosity
indicator $\Re$(\ion{Si}{II}) (see Section~\ref{ss:si_ratio} for more
information on this parameter). Thus, one might expect this pEW to be
an accurate luminosity indicator as well, and in fact evidence for a
correlation between the pEW of \ion{Si}{II} $\lambda$5972 and both 
$x_1$ and $\Delta m_{15}(B)$ has been seen in previous work
\citep{Nordin11a,Hachinger06}.

\begin{figure}
\centering
\includegraphics[width=3.5in,angle=180]{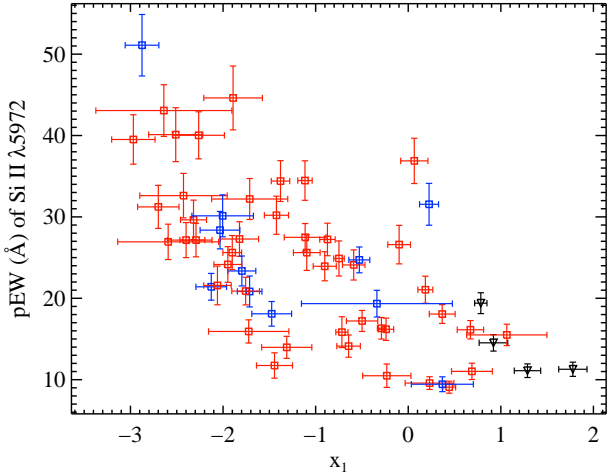}
\caption[pEW of Si~II $\lambda$5972 versus $x_1$]{The pEW of
  \ion{Si}{II} $\lambda$5972 versus the SALT2 light-curve width
  parameter $x_1$. Colours and shapes of data points are the same as in
  Figure~\ref{f:v_bv_si6355}.}\label{f:ew_x1_si5972} 
\end{figure}

Figure~\ref{f:ew_x1_si5972} shows the 55 SNe which have a SALT2 fit as
well as a measured pEW for the \ion{Si}{II} $\lambda$5972
feature. We find a correlation
coefficient of $-0.66$ with a significance of $>3\sigma$, which is
stronger than what was found by 
\citet{Nordin11a}. As in the relationship between $x_1$ and the pEW of
\ion{Si}{II} $\lambda$4000, the Ia-99aa objects appear to follow the
relation.

\begin{figure}
\centering
\includegraphics[width=3.5in]{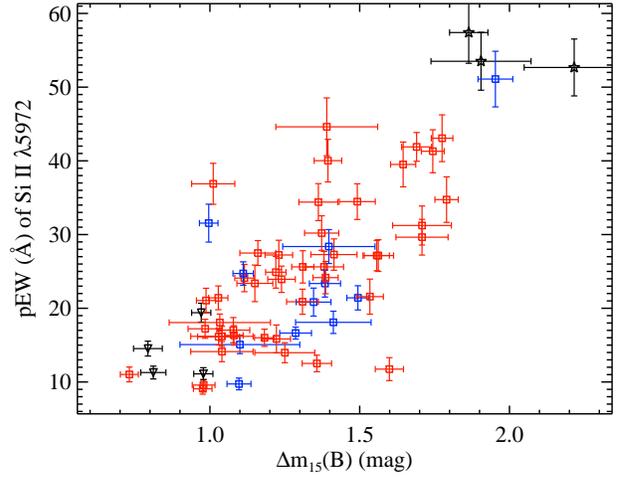}
\caption[pEW of Si~II $\lambda$5972 versus $\Delta
m_{15}(B)$]{The pEW of the \ion{Si}{II} $\lambda$5972 feature versus
  $\Delta m_{15}(B)$. Colours and shapes of data points are the same as in
  Figure~\ref{f:v_bv_si6355}.}\label{f:ew_dm15_si5972} 
\end{figure}

Similarly, a strong linear correlation has been observed between the
pEW of \ion{Si}{II} $\lambda$5972 and $\Delta m_{15}(B)$
\citep{Hachinger06}. This relationship is also found in the BSNIP
data, as shown in Figure~\ref{f:ew_dm15_si5972}: there are
62 SNe, the parameters have a 
correlation coefficient of 0.76 (significant at $>3\sigma$), and as
with the relationship with 
$x_1$, Ia-99aa objects occupy the bottom of the
correlation while the Ia-norm and HV objects are highly
overlapping.

\citet{Hachinger06} denote the Benetti type of each
object on their plot of the pEW of \ion{Si}{II} $\lambda$5972 versus
$\Delta m_{15}(B)$ and note that FAINT objects are found at the top of
the correlation while LVG SNe are found at the bottom (with most HVG objects
occupying the middle of the trend). The BSNIP data show a similar
behaviour for the FAINT objects (again, much like the Ia-91bg objects
in Figure~\ref{f:ew_dm15_si5972}); however, our data show no
differentiation between the LVG and HVG objects in this
parameter space. 

Finally, we note that when plotting the pEW of the
\ion{Si}{II} $\lambda$5972 feature against the MLCS2k2 $\Delta$
parameter, nearly the exact same results are seen as those in
Figure~\ref{f:ew_dm15_si5972}. As with the \ion{S}{II} ``W,'' no distance model utilising the pEW of
\ion{Si}{II} $\lambda$5972 led to an improvement in the Hubble
residuals.

\subsubsection{\ion{Si}{II} $\lambda$6355}\label{sss:ew_si6355}

Much like the pEW of \ion{Si}{II} $\lambda$5972, the pEW of
\ion{Si}{II} $\lambda$6355 has been seen to correlate marginally well
with $x_1$ and to separate various spectral subtypes when compared to
$\Delta m_{15}(B)$ \citep[][respectively]{Nordin11a,Hachinger06}.

\begin{figure}
\centering$
\begin{array}{c}
\includegraphics[width=3.4in]{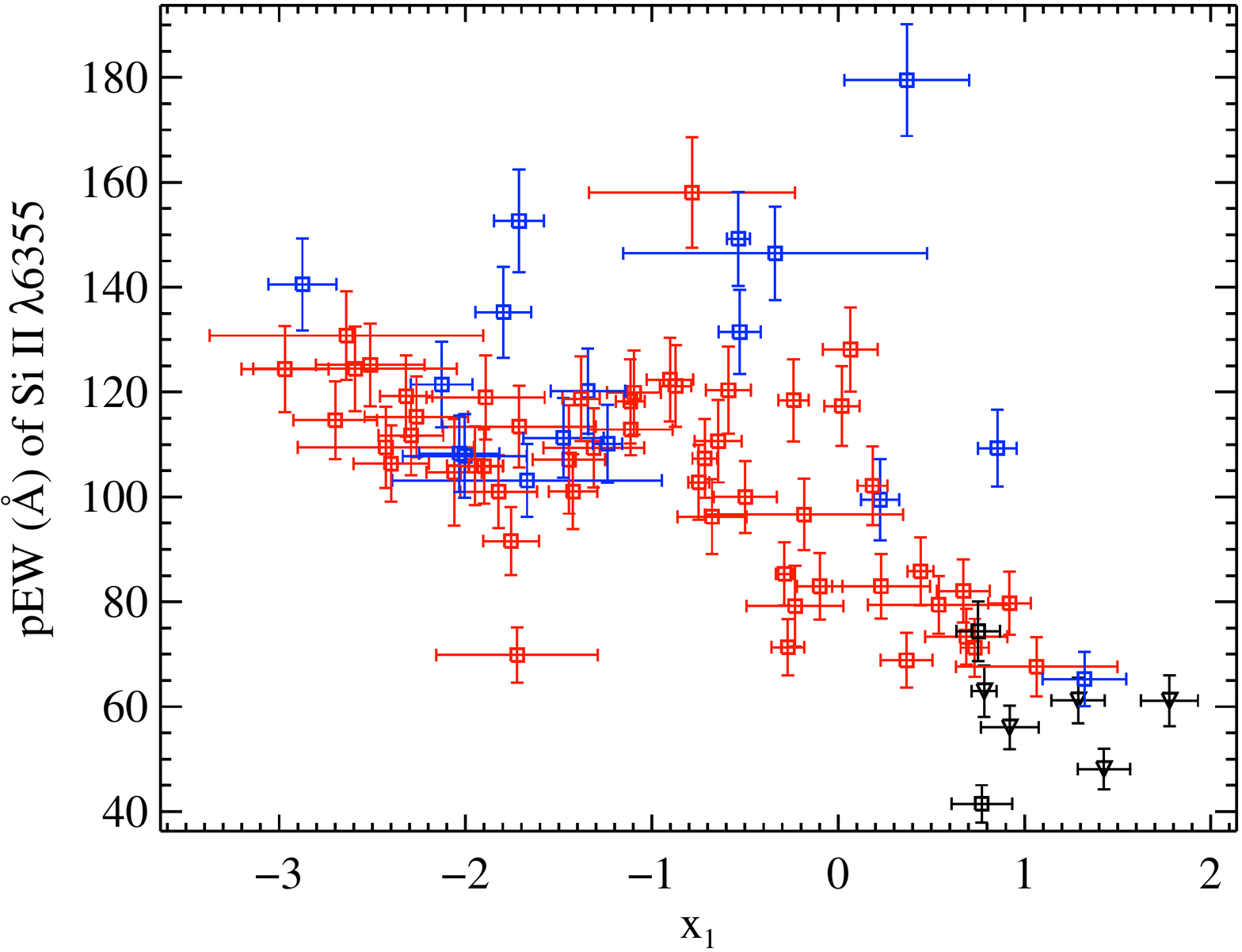} \\
\includegraphics[width=3.4in]{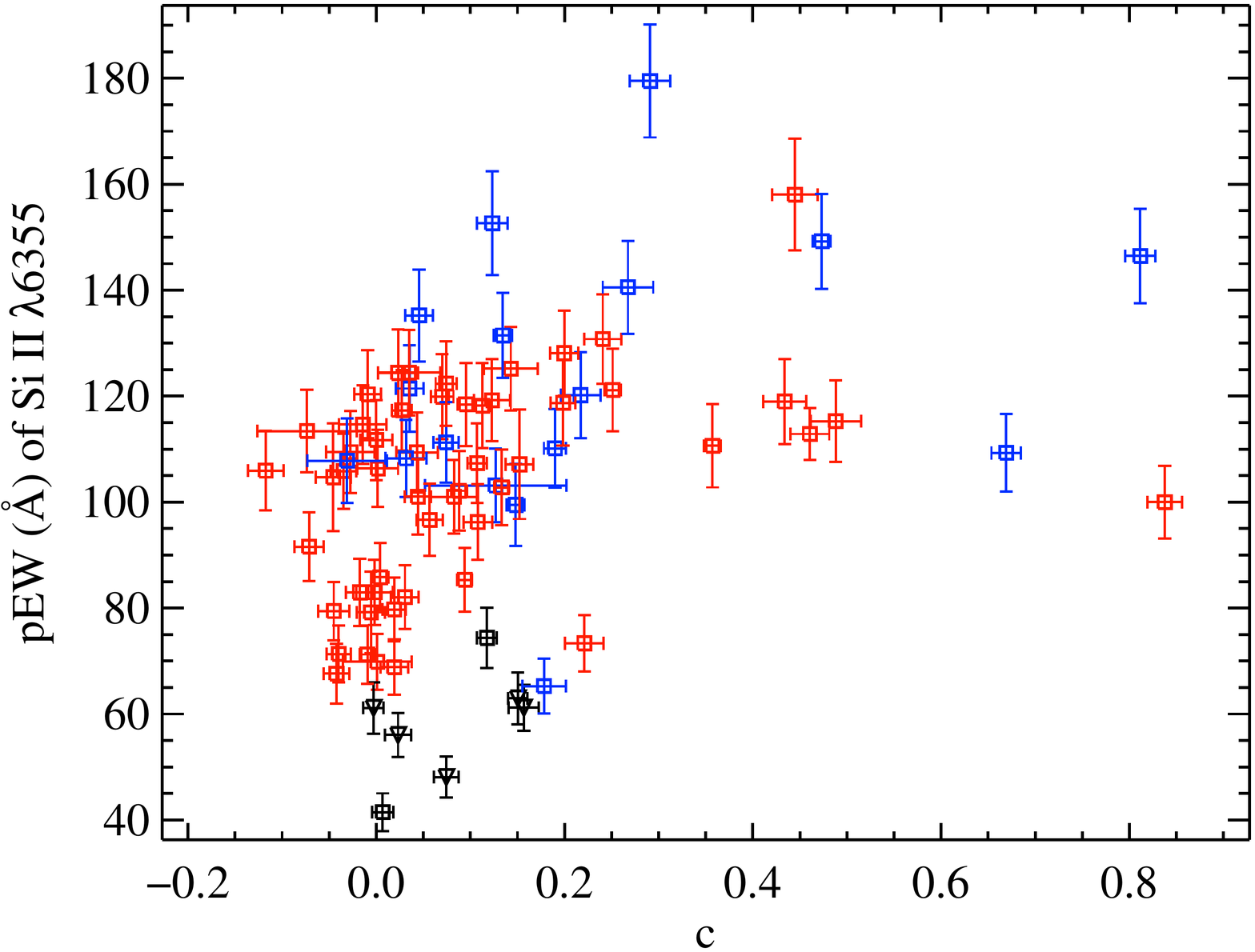} \\
\includegraphics[width=3.4in]{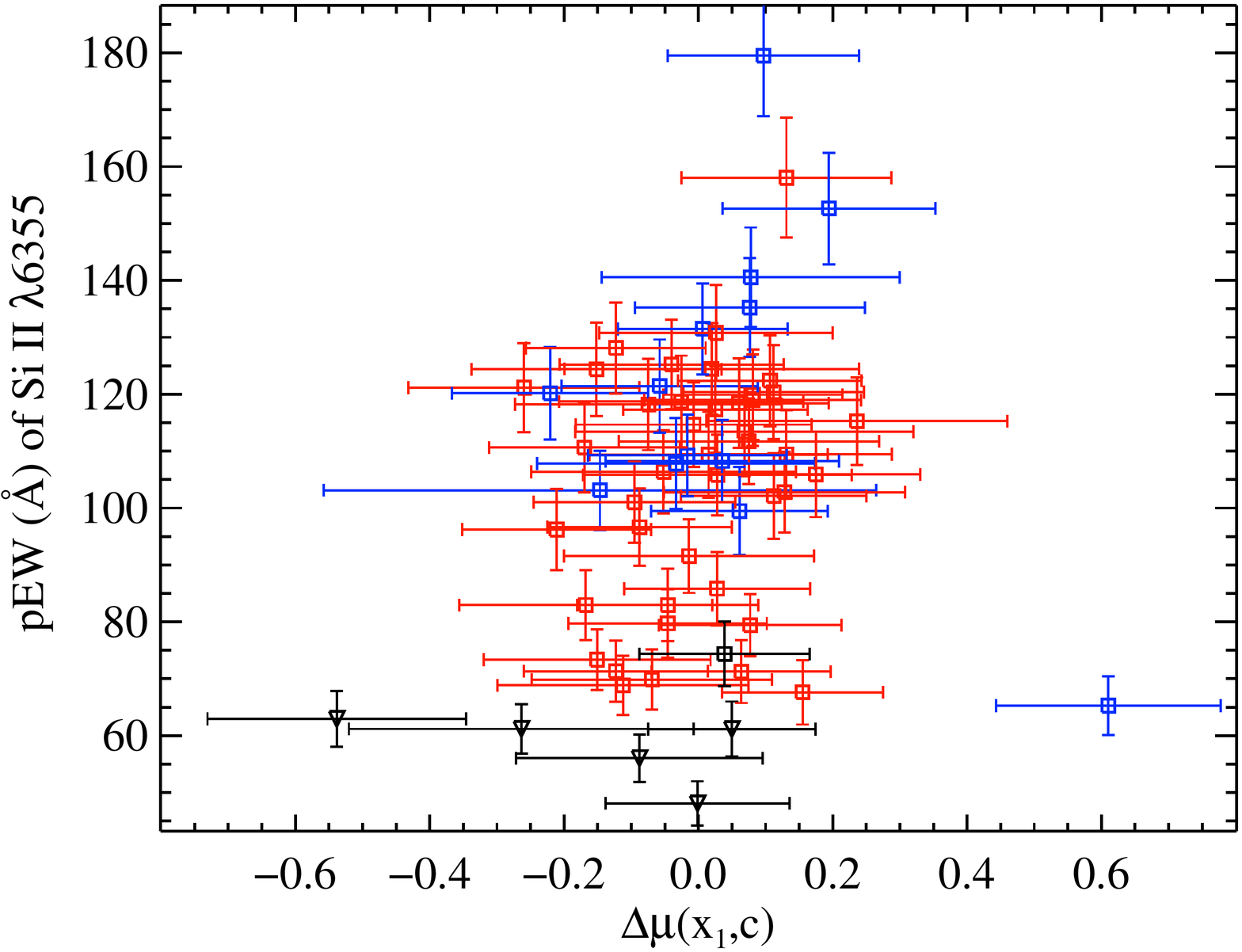} 
\end{array}$
\caption[pEW of Si~II $\lambda$6355 versus $x_1$, $c$, and
Hubble residual]{The pEW of the \ion{Si}{II} $\lambda$6355 feature versus
  SALT2 light-curve width parameter $x_1$ ({\it top}), SALT2 colour
  parameter $c$ ({\it middle}), and Hubble residuals corrected for light-curve width
  and colour ({\it bottom}). Colours and shapes of data points are the same as in
  Figure~\ref{f:v_bv_si6355}.}\label{f:ew_SALT_si6355}
\end{figure}

In Figure~\ref{f:ew_SALT_si6355} the 66 SNe with SALT2 fits and pEW
values for the \ion{Si}{II} $\lambda$6355 feature are shown. The pEW
values are plotted against $x_1$, $c$, and Hubble residuals corrected
for light-curve width and colour (for objects that are part of the
Hubble diagram). A correlation
coefficient of $-0.58$ is calculated for the top panel, which is
significant at $>3\sigma$. This is 
consistent with what was observed in the data studied by
\citet{Nordin11a}. As in the previous two relationships between $x_1$
and the pEW of \ion{Si}{II} features, the Ia-99aa objects follow the
linear relation.

We find that $c$ is somewhat correlated with the pEW of \ion{Si}{II} $\lambda$6355
(correlation coefficient of 0.43 and significance of \about3$\sigma$)
for objects with $c < 0.5$. This pEW 
is even less correlated with Hubble residuals corrected for $x_1$ and
$c$ (correlation coefficient 0.23). A distance model which includes
$x_1$, $c$, and the pEW of the \ion{Si}{II} $\lambda$6355 feature leads to
a 4~per~cent decrease in WRMS, a 6~per~cent decrease in $\sigma_{\rm
  pred}$, and is significant at the 1.2$\sigma$ level. So while this
is technically an improvement over the standard $\left(x_1,c\right)$
model, it may not actually be very helpful.

\begin{figure}
\centering
\includegraphics[width=3.5in]{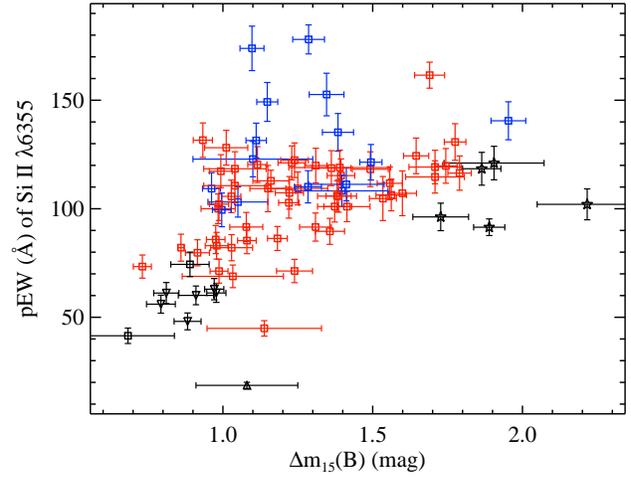}
\caption[pEW of Si~II $\lambda$6355 versus $\Delta
m_{15}(B)$]{The pEW of \ion{Si}{II} $\lambda$6355 versus $\Delta
  m_{15}(B)$. Colours and shapes of data points are the same as in
  Figure~\ref{f:v_bv_si6355}.}\label{f:ew_dm15_si6355} 
\end{figure}

Plotting the pEW of \ion{Si}{II} $\lambda$6355 versus $\Delta
m_{15}(B)$, \citet{Hachinger06} are able to separate FAINT, LVG, and
HVG objects relatively accurately. This is also seen, though at a
lower significance, in the BNSIP data. In
Figure~\ref{f:ew_dm15_si6355} we plot 80 SNe; the
parameters have a correlation coefficient of 0.45 (again with a
significance of \about3$\sigma$), but the 
Ia-91bg, Ia-99aa, and the lone Ia-91T objects are all reasonably well
separated from the bulk of the SNe. There is even some evidence for a 
difference between HV and Ia-norm objects in this parameter space. As
mentioned above, \citet{Hachinger06} denote the Benetti type of
each object on their plot of the pEW of \ion{Si}{II} $\lambda$6355 versus
$\Delta m_{15}(B)$ and state that the three subtypes are well
separated. Again the BSNIP data support this conclusion, but at a
weaker significance. FAINT objects are found in the same part of
parameter space as the Ia-91bg objects, while HVG and HV SNe tend to
occupy a different part of parameter space compared with the LVG and
Ia-norm/91T/99aa objects. However, there is quite a lot of overlap
among all of the non-FAINT (and non-Ia-91bg) SNe. 

\begin{figure}
\centering$
\begin{array}{c}
\includegraphics[width=3.4in]{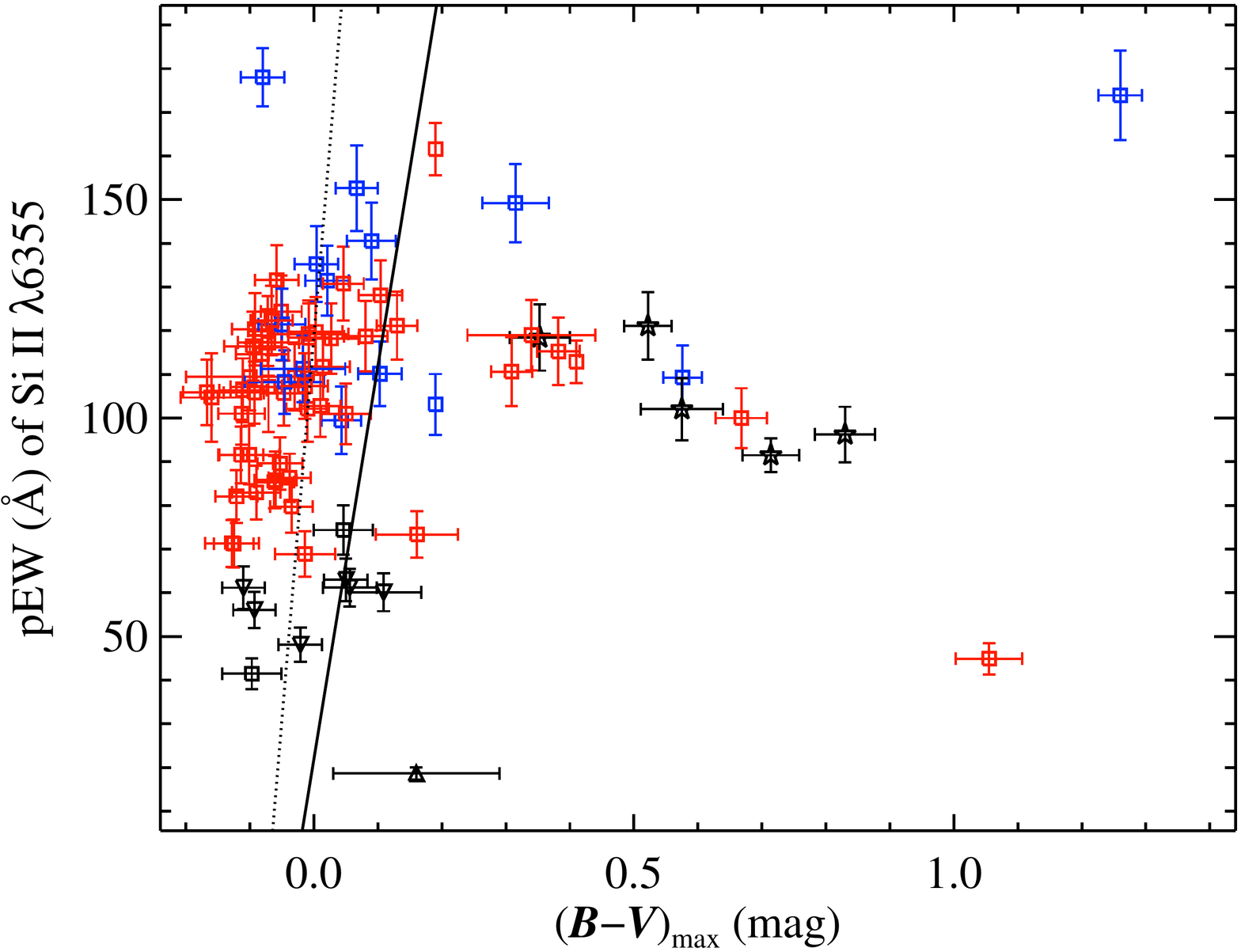} \\
\includegraphics[width=3.4in]{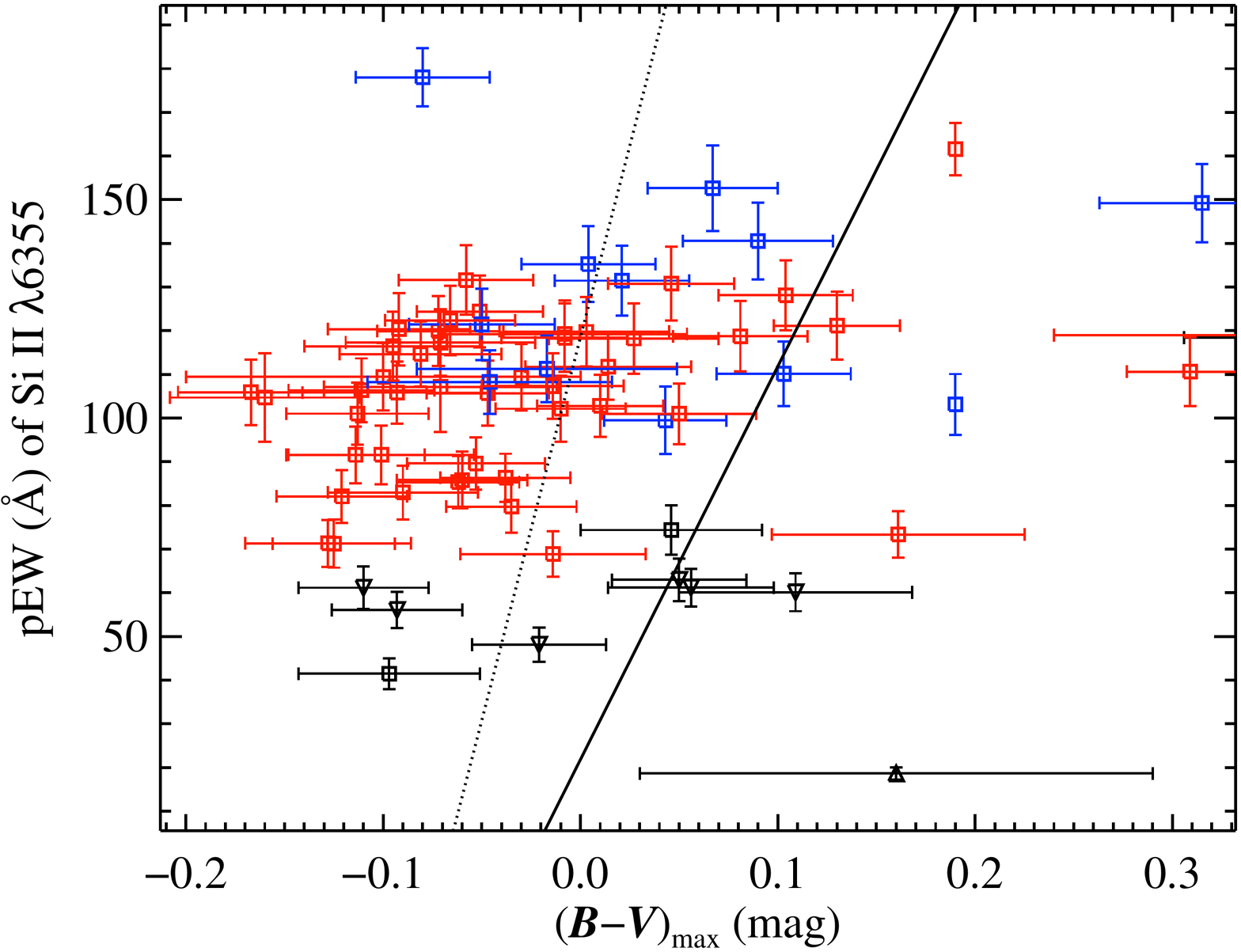} 
\end{array}$
\caption[pEW of Si~II $\lambda$6355 versus
$\left(B-V\right)_\textrm{max}$]{The pEW of the \ion{Si}{II}
  $\lambda$6355 feature versus \bvmax\ ({\it top}) and a close-up view
  of objects with $\left(B-V\right)_\textrm{max} < 0.319$~mag
  ({\it bottom}).  Colours and shapes of data points are the same as in
  Figure~\ref{f:v_bv_si6355}. The solid line
  is the fit to all of the data while the dotted line is the fit only
  to objects with $(B-V)_{\rm max} < 0.319$~mag.}\label{f:ew_bv_si6355}
\end{figure}

Figure~\ref{f:ew_bv_si6355} presents the pEW of the \ion{Si}{II}
$\lambda$6355 feature plotted against
$\left(B-V\right)_\textrm{max}$. The top panel shows all 78 SNe from
the BSNIP dataset for which both of these values have been measured
and the bottom panel displays a close-up view of objects with
$\left(B-V\right)_\textrm{max} < 0.319$~mag. The linear least-squares
fit to all of the points is represented by the solid line, and the fit to SNe
with $\left(B-V\right)_\textrm{max} < 0.319$~mag is shown by the
dotted line.

The correlations for the full sample and the less reddened
sample are weak (coefficients 0.12 and 0.16,
respectively). Qualitatively, this matches the work of
\citet{Foley11:velb}, even though they plot {\it intrinsic} colours
and we plot {\it observed} colours. The pEW of the \ion{Si}{II}
$\lambda$6355 
feature is less correlated with observed \bvmax\ than the
velocity near maximum brightness of that same spectral feature.

\subsubsection{\ion{Ca}{II} and \ion{O}{I}}\label{sss:ew_cair}

Much like \citet{Hachinger06}, we searched for possible correlations
between the pEW of each spectral feature investigated and various
photometric parameters. Many of the strongest and most interesting of
these possible correlations have been discussed in the preceding
sections. For \ion{Ca}{II}~H\&K as well as the \ion{O}{I} triplet, no
pairs of pEWs and photometric parameters were found to have 
correlation coefficients $> 0.4$.

However, the pEW of the \ion{Ca}{II} near-IR triplet is found to
correlate with $\Delta$ (with a correlation coefficient of
0.66 and significance of $>3\sigma$) and with $c$ (with a correlation
coefficient of 0.50 for SNe with $c < 0.5$, though the significance of
this correlation is only at the \about2$\sigma$ level). 
One of these correlations is marginal; however, this spectral region
has been studied very little in the past. A strength of the
BSNIP data is that the average wavelength coverage
(3300--10,400~\AA, BSNIP~I) is significantly wider than that of most
other SN~Ia spectral datasets. For example, one of the largest previously
published SN~Ia spectral datasets had an average wavelength coverage
of 3700--7400~\AA\ \citep{Matheson08}. Thus, the \ion{Ca}{II}~H\&K
feature, the \ion{O}{I} triplet, and the \ion{Ca}{II} near-IR triplet
have been ignored almost entirely in past spectral analyses like the
one presented here.


We once again constructed Hubble diagrams from
Equations~\ref{eq:m1}--\ref{eq:m4} and using the pEW values of
\ion{Ca}{II}~H\&K, the \ion{O}{I} triplet, and the \ion{Ca}{II}
near-IR triplet. All but two models were significantly worse at
measuring distances than the standard $\left(x_1,c\right)$
model. The $(x_1,c,$ \ion{Ca}{II}~H\&K pEW$)$ and
$(x_1,c,$ \ion{O}{I} triplet pEW$)$ models both slightly
decreased the WRMS, but at almost imperceptible levels
($\Delta_{x_1,c} = -0.0023 \pm 0.0096$ and $\Delta_{x_1,c} = -0.0063
\pm 0.0148$, respectively).

\subsection{The Si~II Ratio}\label{ss:si_ratio}

Historically, one of the first spectral luminosity indicators investigated 
was the \ion{Si}{II} ratio, $\Re$(\ion{Si}{II}), defined by \citet{Nugent95}
as the ratio of the depth of the \ion{Si}{II} $\lambda$5972 feature to
the depth of the \ion{Si}{II} $\lambda$6355
feature. \citet{Hachinger06} redefined the \ion{Si}{II} ratio as the
pEW of \ion{Si}{II} $\lambda$5972 divided by the pEW of \ion{Si}{II}
$\lambda$6355. In BSNIP~II it was shown that these are nearly
equivalent definitions, so in order to be consistent with that work we
define the \ion{Si}{II} ratio for the present study to be 
\begin{equation}
\Re\left(\textrm{\ion{Si}{II}}\right) \equiv \frac{\textrm{pEW}\left(\textrm{\ion{Si}{II} $\lambda$5972}\right)}{\textrm{pEW}\left(\textrm{\ion{Si}{II} $\lambda$6355}\right)}.
\end{equation}

\begin{figure}
\centering
\includegraphics[width=3.5in]{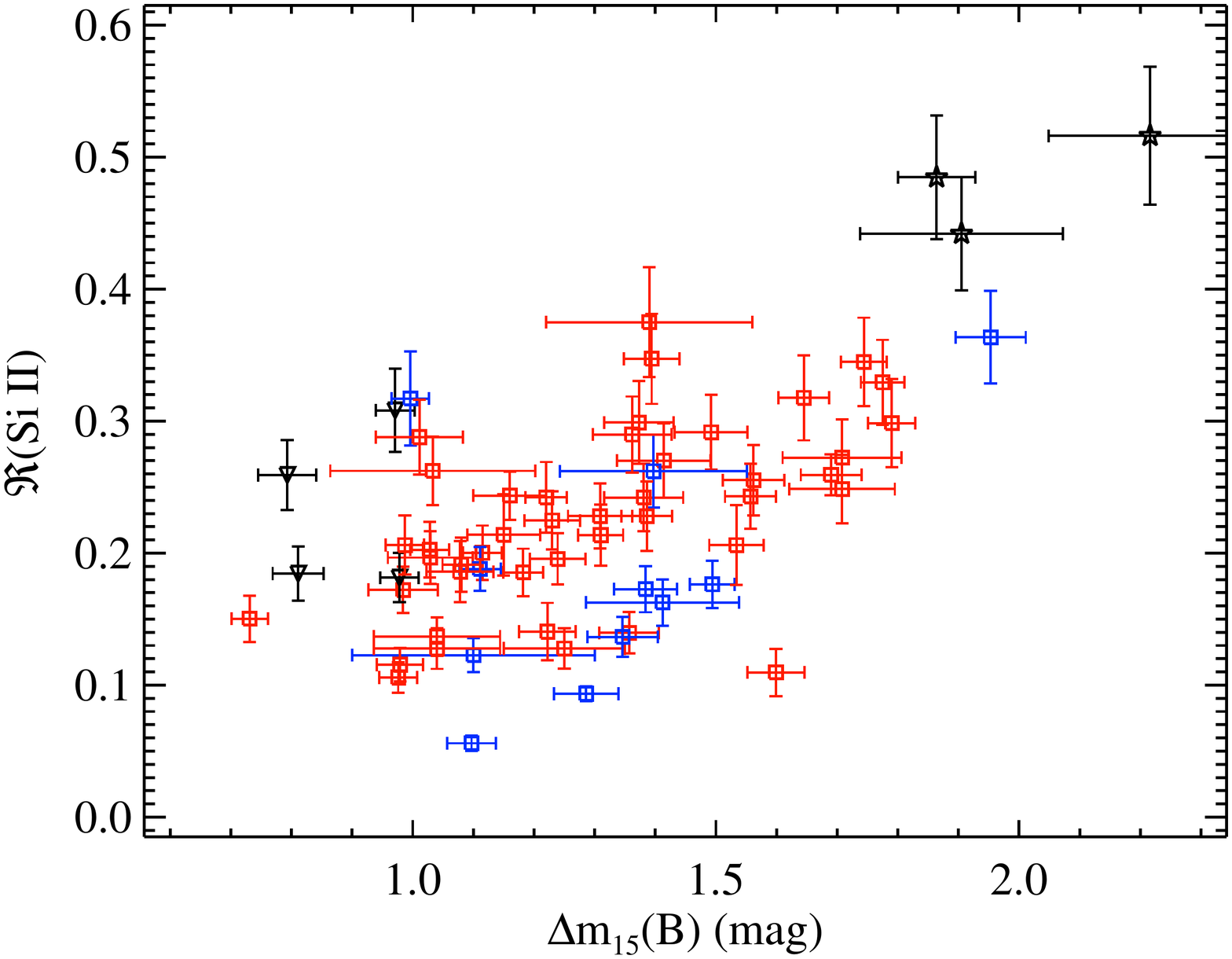}
\caption[The Si~II ratio versus $\Delta m_{15}(B)$]{The \ion{Si}{II} ratio
  versus $\Delta m_{15}(B)$. Colours and shapes of
  data points are the same as in
  Figure~\ref{f:v_bv_si6355}.}\label{f:R_Si_dm15}  
\end{figure}

The \ion{Si}{II} ratio has been shown to correlate with maximum
absolute $B$-band magnitude and $\Delta m_{15}(B)$, which is why it has
been used as a spectral luminosity indicator
\citep[e.g.,][]{Nugent95,Benetti05,Hachinger06}.
Figure~\ref{f:R_Si_dm15} shows 62 SNe~Ia with both $\Delta m_{15}(B)$
and $\Re$(\ion{Si}{II}). The data are correlated with a 
correlation coefficient of 0.62, which is significant at the
$>3\sigma$ level. Ia-91bg objects appear at the 
upper right of the plot and form a continuous relationship with the
Ia-norm objects. Ia-99aa objects appear to lie above the main trend
(though there are only a handful of these SNe in the figure), while HV
objects 
lie below the main trend. When removing the Ia-99aa objects, the
correlation increases slightly. 


If we instead tag each data point in Figure~\ref{f:R_Si_dm15} by its
Benetti type, we find that the FAINT objects are found in the
upper right of the main trend \citep[similar to the Ia-91bg objects
and as seen in previous studies, e.g.,][]{Benetti05,Hachinger06}. The 
LVG and HVG objects are found in the lower-left portion of the plot
with a significant level of overlap between the two subclasses. This
differs from previous work, where there have been claims
that LVG objects have larger \ion{Si}{II} ratios and lie above the
main trend \citep{Benetti05,Hachinger06}, though these studies and
the BSNIP data both observe larger scatter in $\Re$(\ion{Si}{II})
values in the lower $\Delta m_{15}(B)$ objects. When removing the LVG
objects, the correlation is effectively unchanged. Comparing
$\Re$(\ion{Si}{II}) to the MLCS2k2 $\Delta$ parameter yields
results similar to those seen in Figure~\ref{f:R_Si_dm15}.

The BSNIP distribution of $\Delta m_{15}(B)$ values, while not evenly
distributed, is more continuous than in previous studies similar to the
present one. For example, the data in \citet{Hachinger06} contained
only one 
object with $\Delta m_{15}(B)$ between 1.5 and 1.7~mag, while the
BSNIP data have 6 objects in that range. A more continuous distribution
of $\Delta m_{15}(B)$ values, combined with the spectroscopic
subclasses presented in Figure~\ref{f:R_Si_dm15}, complicates the
relatively simplistic view that underpins the basic Phillips
relation. For $0.95 \la \Delta m_{15}(B) \la 1.0$~mag there are
Ia-99aa, Ia-norm, {\it and} HV objects. On the other end of the
$\Delta m_{15}(B)$ distribution, between \about1.75 and \about1.95~mag
there are Ia-91bg, Ia-norm, {\it and again} HV objects. Thus, for a
given light-curve width (or decline rate), there exist SNe~Ia of
significantly different subclasses.

As discussed in BSNIP~II, objects tagged by SNID as Ia-91bg or Ia-99aa
are the most spectroscopically peculiar objects and probably only
represent the extreme ends of a continuous distribution of spectra. If
true, this means that the most spectroscopically peculiar objects {\it
  may not} have the most extreme light curves. The reverse may also be
true, namely that the SNe with the most extreme light curves may not
be the most spectroscopically peculiar. This is further supported by
the relatively wide scatter in the main trend of
Figure~\ref{f:R_Si_dm15}. At any value of $\Delta m_{15}(B)$ (with a
significant number of objects) there is a broad range in
$\Re$(\ion{Si}{II}) values.

In BSNIP~II, it was pointed out that the relative strength of the two
\ion{Si}{II} features that go into calculating $\Re$(\ion{Si}{II}) is
fairly robust at differentiating between the various ``SNID types''
and ``Wang types.'' Thus, from a spectrum, either using SNID or the pEWs of 
\ion{Si}{II} features, one may declare an object to be Ia-91bg or
Ia-99aa, whereas based on the light curve of the same object it might
be considered relatively normal. The significant amount of scatter in
the correlation between the \ion{Si}{II} ratio and $\Delta m_{15}(B)$
also cautions one against simply measuring $\Re$(\ion{Si}{II}) from a 
single spectrum and then using that value and a fit to the data in
Figure~\ref{f:R_Si_dm15} to calculate a $\Delta m_{15}(B)$ value.



\begin{figure}
\centering$
\begin{array}{c}
\includegraphics[width=3.4in]{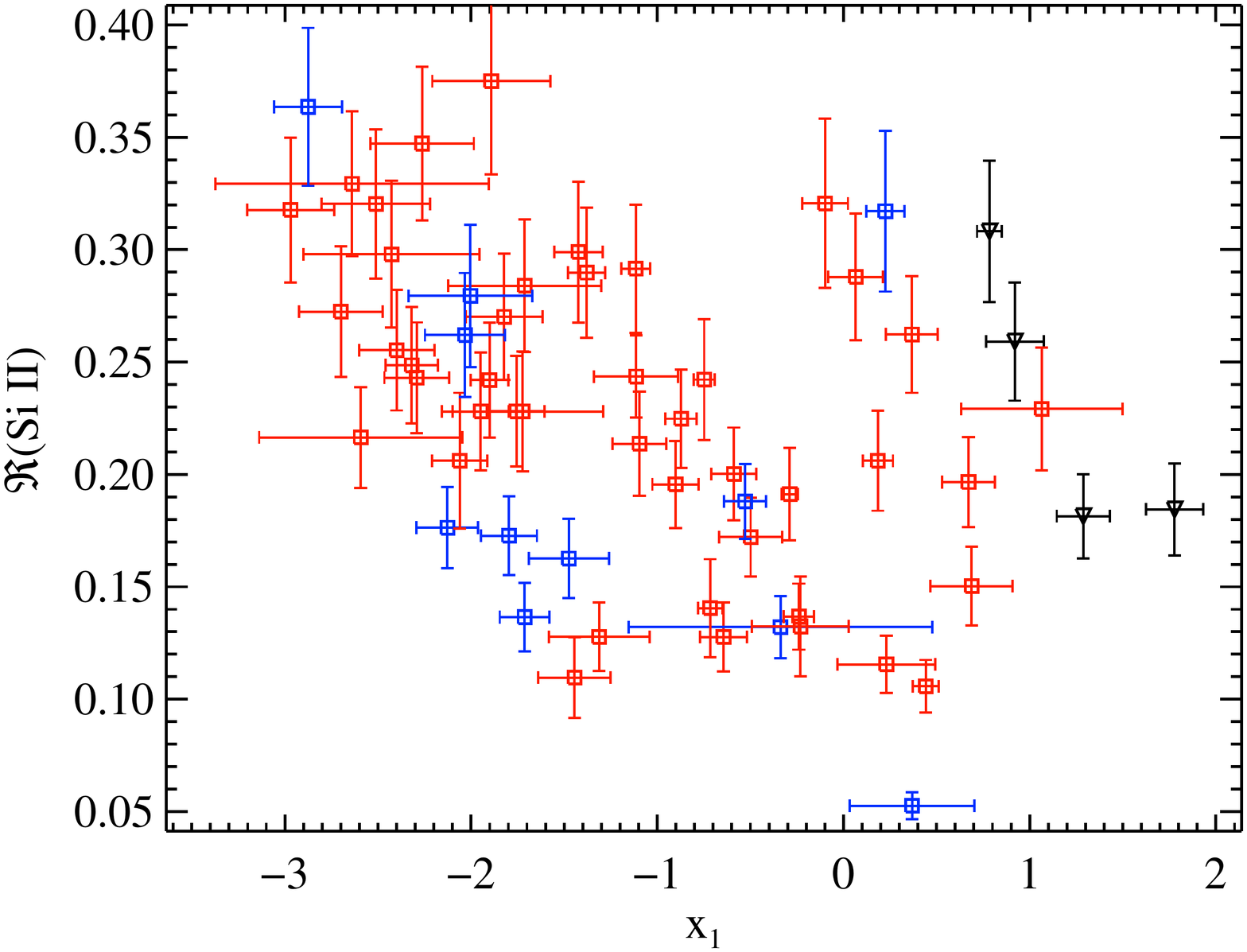} \\
\includegraphics[width=3.4in]{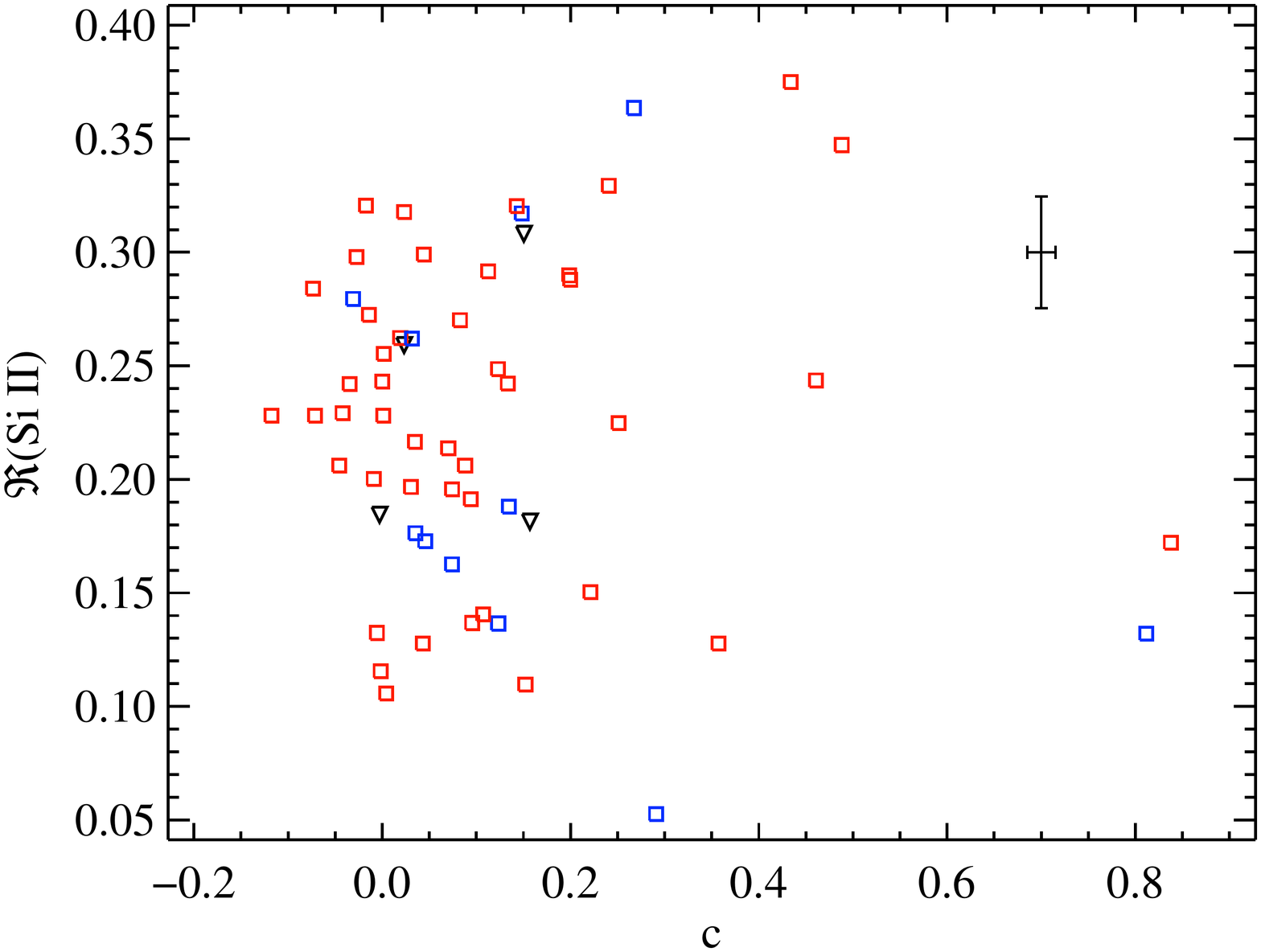} \\
\includegraphics[width=3.4in]{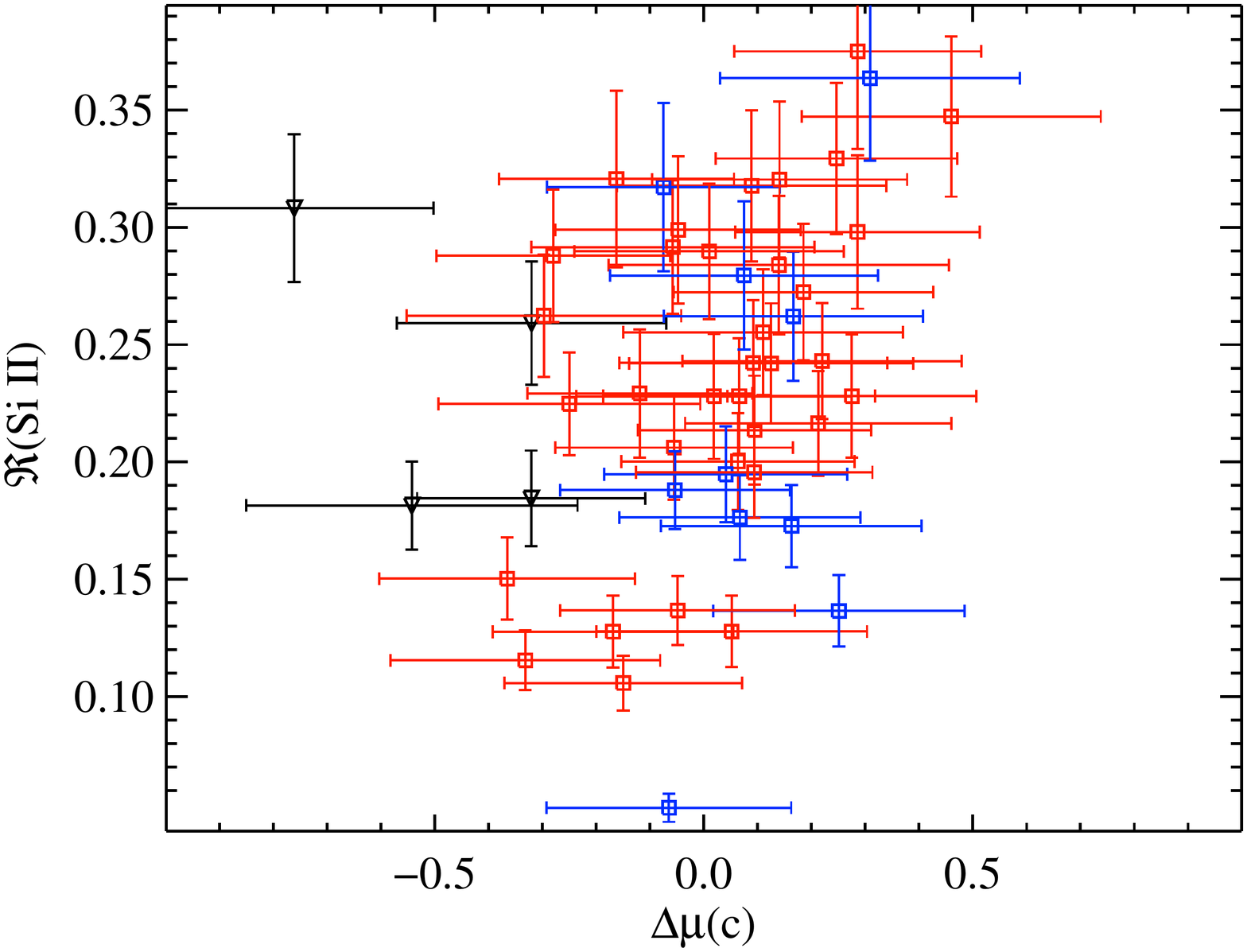} 
\end{array}$
\caption[The Si~II ratio versus $x_1$, $c$, and Hubble
residual]{The \ion{Si}{II} ratio versus 
  SALT2 light-curve width parameter $x_1$ ({\it top}), SALT2 colour
  parameter $c$ ({\it middle}), and Hubble residuals corrected for colour
  only ({\it bottom}). Colours and shapes of data points are the same as in
  Figure~\ref{f:v_bv_si6355}. In the middle plot, the
  median uncertainty in both directions is shown in the upper-right
  corner.}\label{f:SALT_R_Si}
\end{figure}

In Figure~\ref{f:SALT_R_Si} we show the 51 BSNIP SNe which have SALT2 
fits and \ion{Si}{II} ratios within 5~d of maximum
brightness. $\Re$(\ion{Si}{II}) is plotted against $x_1$, $c$, and
Hubble residuals corrected for colour only (for SNe which are used in
the Hubble diagram).

From the BSNIP data we find that the \ion{Si}{II} ratio is only
marginally correlated with $x_1$ (top plot of
Figure~\ref{f:SALT_R_Si}) with a correlation coefficient of
$-0.40$ and significance of \about2$\sigma$. This is a weaker
relationship than what has been found before 
\citep{Blondin11}. The Ia-99aa objects appear to be above the main
trend and most of the HV objects seem to be below it.
The middle plot of Figure~\ref{f:SALT_R_Si} shows no evidence for a
correlation between $\Re$(\ion{Si}{II}) and $c$, with a 
correlation coefficient of 0.14 for objects with $c < 0.5$.

The bottom plot of Figure~\ref{f:SALT_R_Si} shows a
low-significance correlation between the \ion{Si}{II} ratio and Hubble
residuals 
corrected for colour only (coefficient of 0.34), and it is again
significantly weaker than what has been found before
\citep{Blondin11}. In fact, \citet{Blondin11} go so far as to say that
$\Re$(\ion{Si}{II}) acts as a replacement for $x_1$, but the BSNIP
data do not support such a claim.
Using the current sample, the best
model which includes the \ion{Si}{II} ratio also includes both $x_1$
{\it and} $c$, but it is only about as accurate as the standard
$\left(x_1,c\right)$ model ($\Delta_{x_1,c} = 0.0156 \pm 0.0114$). 

Interestingly, \citet{Blondin11} found that the subluminous (but
Ia-norm) SN~2000k is a 2$\sigma$ outlier in their plot of
$\Re$(\ion{Si}{II}) versus $c$, but part of the main
correlation of $\Re$(\ion{Si}{II}) versus $x_1$. This object is in the
BSNIP dataset and, while we agree that it is subluminous and
spectroscopically normal, it is not a significant outlier in any of
the three plots in Figure~\ref{f:SALT_R_Si}.

\subsection{The Ca~II Ratio}\label{ss:ca_ratio}

The \ion{Ca}{II} ratio was defined by \citet{Nugent95} as the ratio of
the flux at the red edge of the \ion{Ca}{II}~H\&K feature to the flux
at the blue edge of that feature. In the notation from BSNIP~II this
is 
\begin{equation}
\Re\left(\textrm{\ion{Ca}{II}}\right) \equiv \frac{F_r\left(\textrm{\ion{Ca}{II}~H\&K}\right)}{F_b\left(\textrm{\ion{Ca}{II}~H\&K}\right)}.
\end{equation}
Like the \ion{Si}{II} ratio, it has been found to correlate with
maximum absolute $B$-band magnitude \citep{Nugent95}.

\begin{figure}
\centering
\includegraphics[width=3.5in]{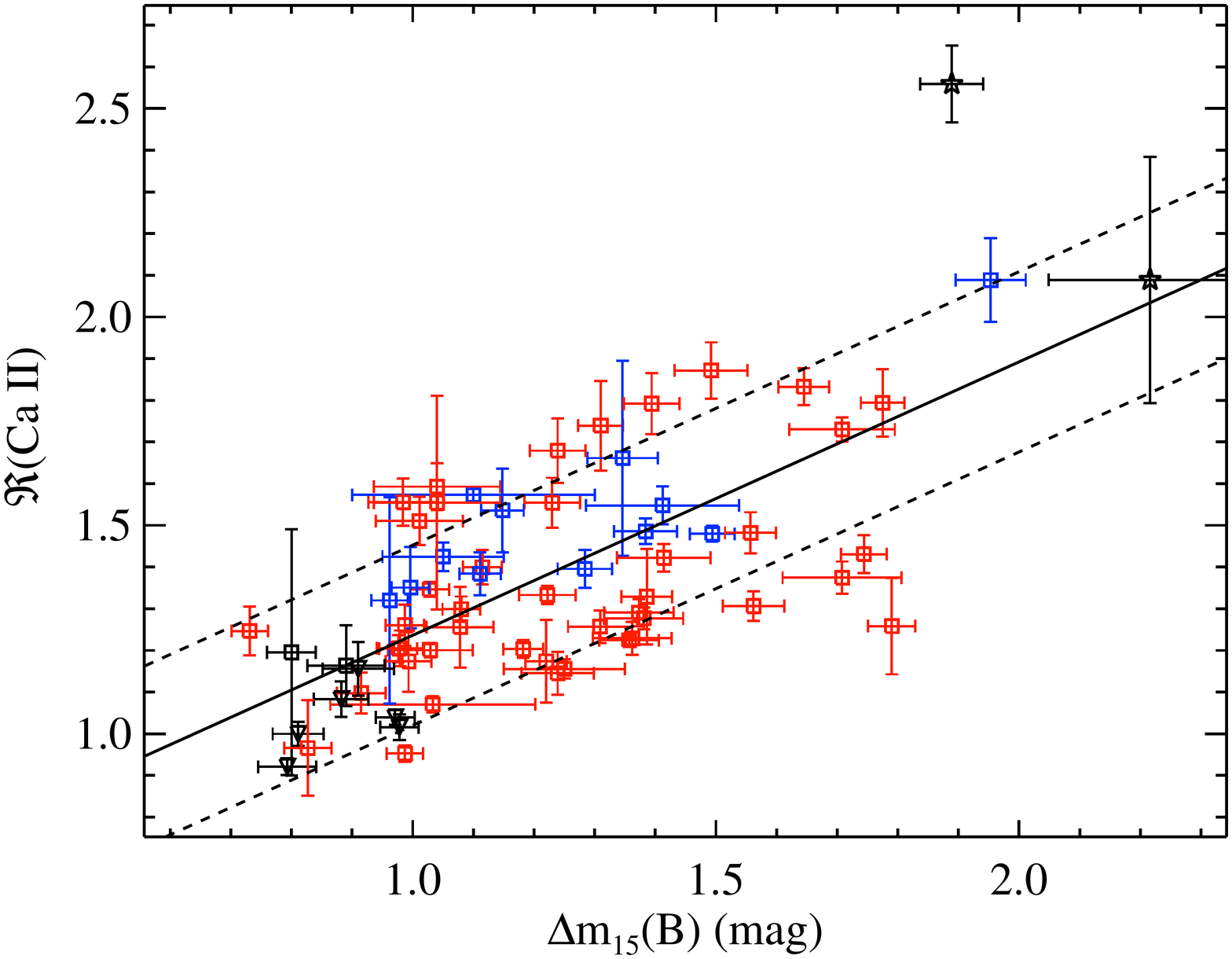}
\caption[The Ca~II ratio versus $\Delta m_{15}(B)$]{The \ion{Ca}{II} ratio
  versus $\Delta m_{15}(B)$. Colours and shapes of data points are the
  same as in Figure~\ref{f:v_bv_si6355}. The solid line
  is the linear least-squares fit and the dashed lines are the
  standard error of the fit.}\label{f:R_Ca_dm15}  
\end{figure}

In BSNIP~II it was shown that the \ion{Ca}{II} ratio and the
\ion{Si}{II} ratio are uncorrelated, even though both of them have
been used as spectral luminosity indicators. In
Section~\ref{ss:si_ratio} it was shown that $\Re$(\ion{Si}{II}) is
correlated with $\Delta m_{15}(B)$. Figure~\ref{f:R_Ca_dm15} illustrates
that $\Re$(\ion{Ca}{II}) is correlated with $\Delta m_{15}(B)$ as well
(correlation coefficient 0.70, significance $>3\sigma$). The plot 
contains 65 SNe; the least-squares linear fit to the data
is shown as the solid line in Figure~\ref{f:R_Ca_dm15}, and the dashed
lines are the standard error.

Interestingly, the HV objects seem to occupy a relatively narrow
region of parameter space that is surrounded on all sides by mainly
Ia-norm SNe. Ia-99aa objects mostly make up the lowest end of the linear
trend, while one of the two Ia-91bg objects in the plot perhaps does not
follow the main relationship. Unsurprisingly, comparing $\Delta$ to 
$\Re$(\ion{Ca}{II}) results in the same trends seen in
Figure~\ref{f:R_Ca_dm15}.

\begin{figure}
\centering$
\begin{array}{c}
\includegraphics[width=3.4in]{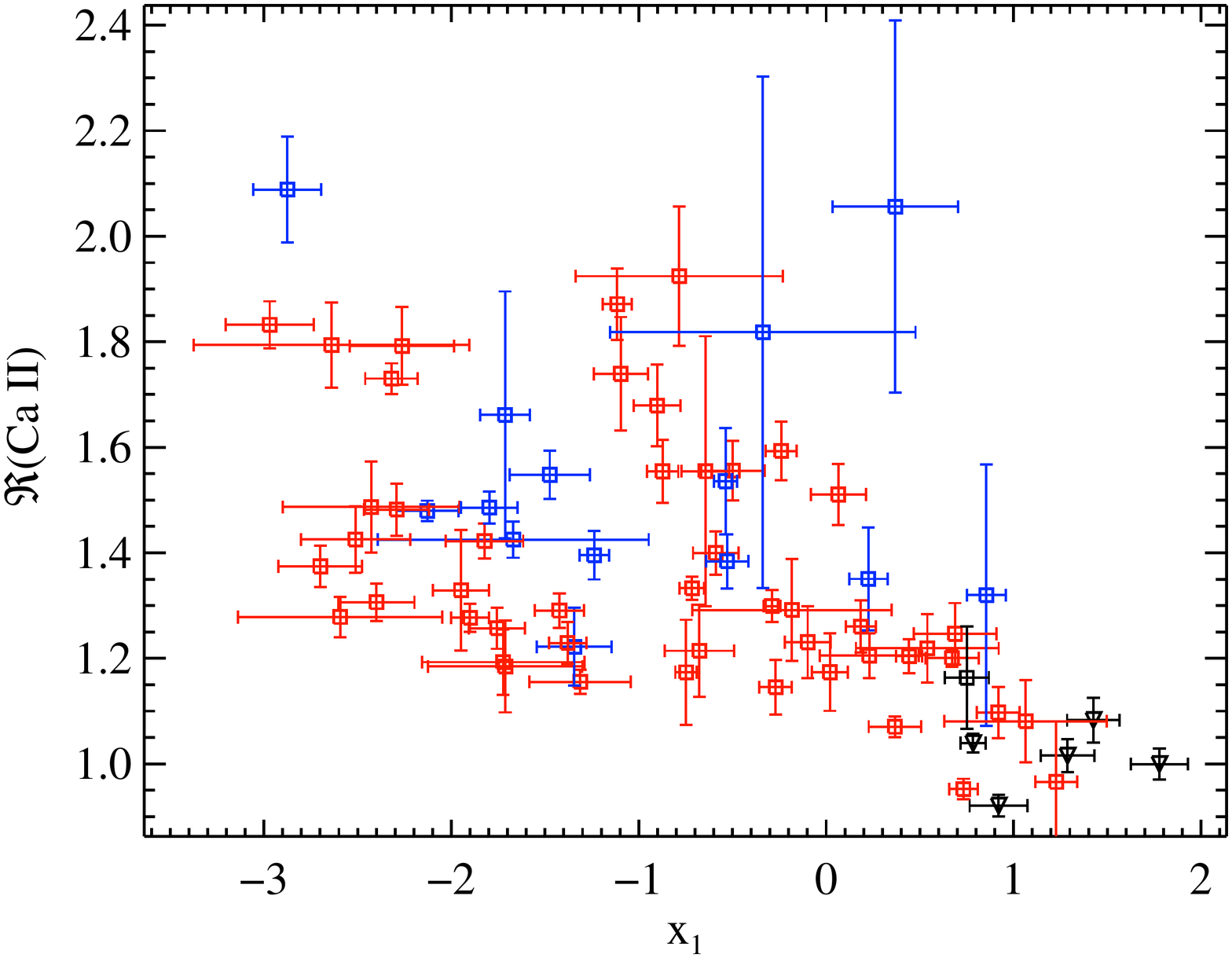} \\
\includegraphics[width=3.4in]{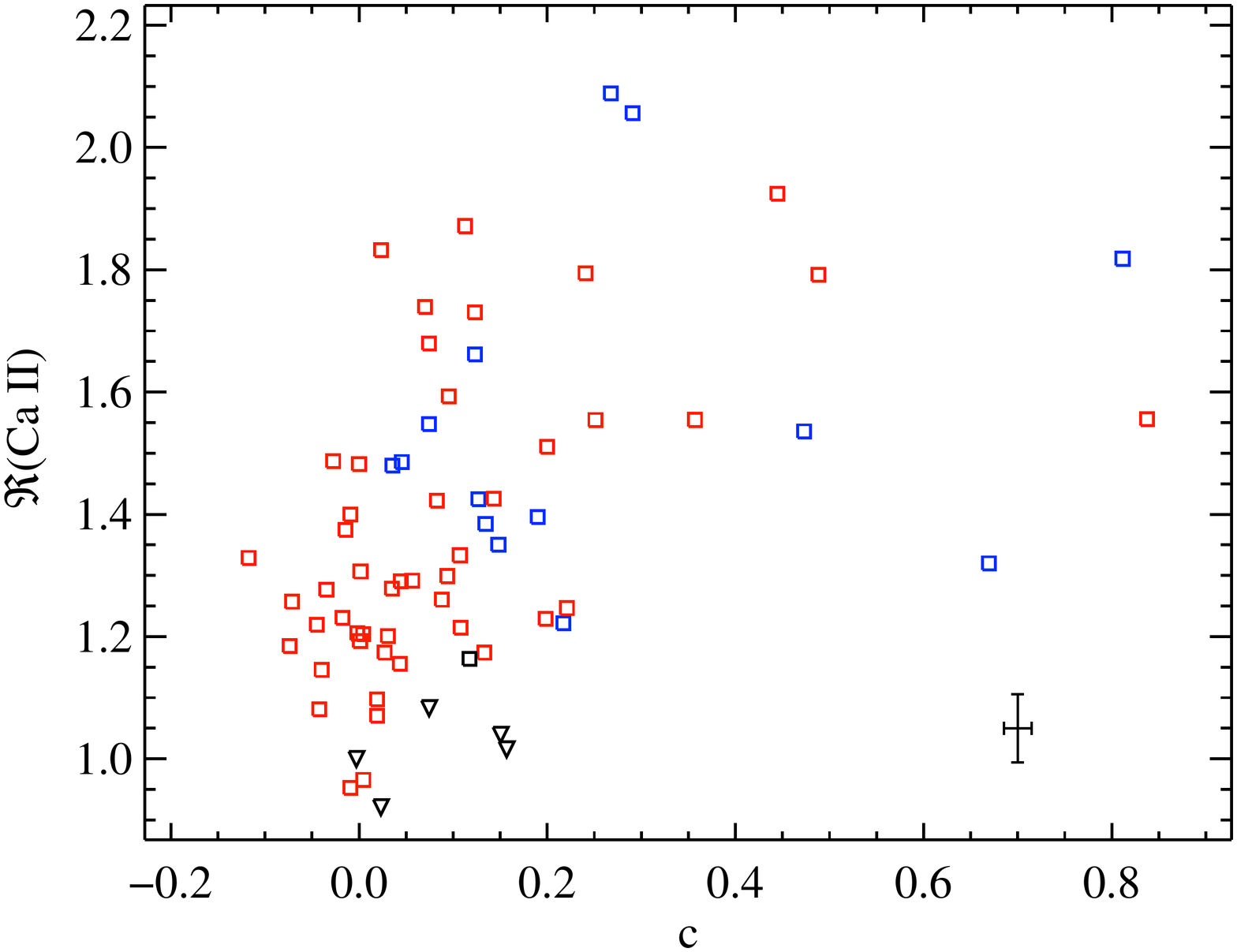} \\
\includegraphics[width=3.4in]{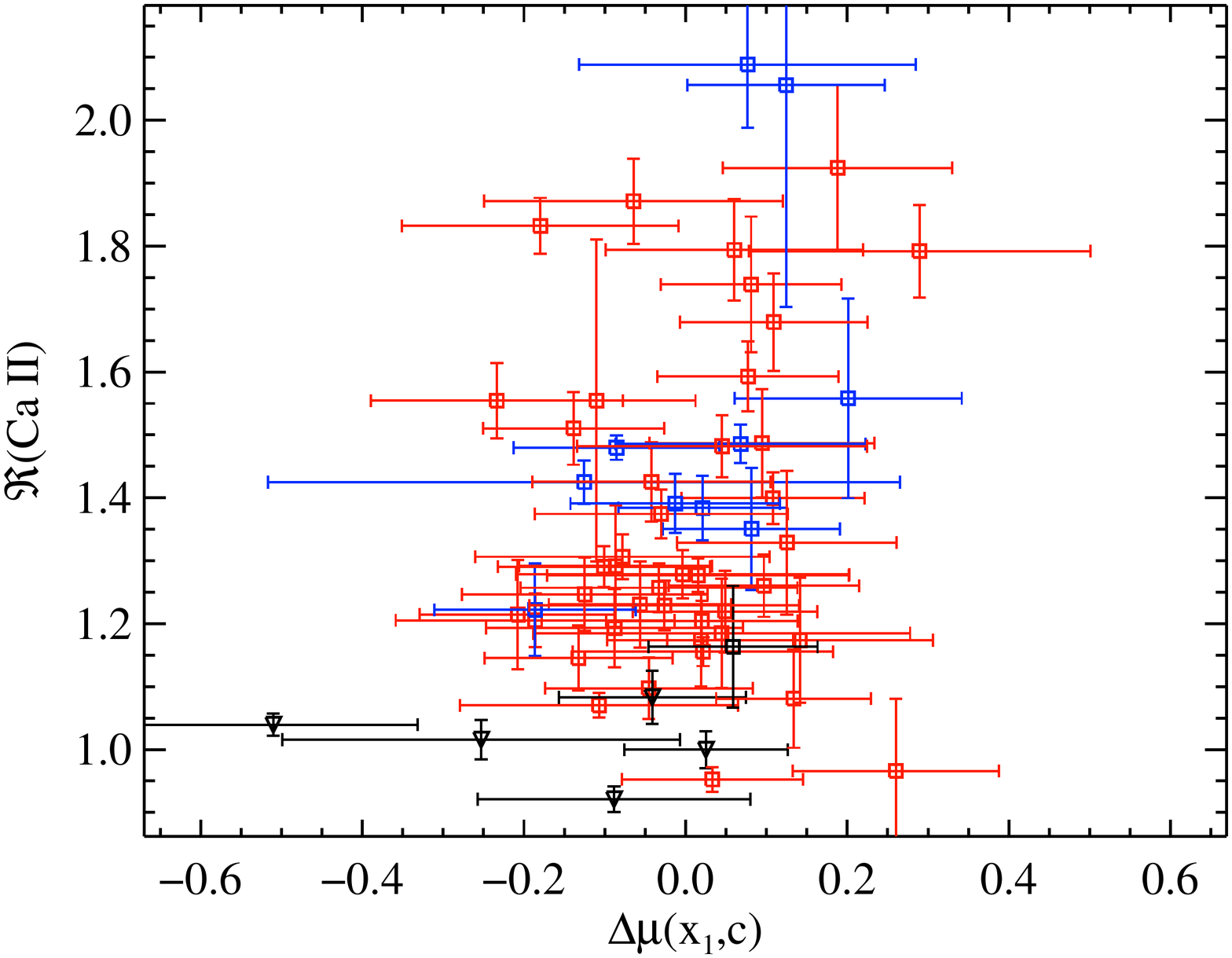} 
\end{array}$
\caption[The Ca~II ratio versus $x_1$, $c$, and Hubble
residual]{The \ion{Ca}{II} ratio versus 
  SALT2 light-curve width parameter $x_1$ ({\it top}), SALT2 colour
  parameter $c$ ({\it middle}), and Hubble residuals corrected for light-curve width
  and colour ({\it bottom}). Colours and shapes of data points are the same as in
  Figure~\ref{f:v_bv_si6355}. In the middle plot,
  the median uncertainty in both directions is shown in the
  lower-right corner.}\label{f:SALT_R_Ca}
\end{figure}

Figure~\ref{f:SALT_R_Ca} displays the 64 BSNIP SNe which have SALT2 
fits as well as \ion{Ca}{II} ratios within 5~d of maximum
brightness. $\Re$(\ion{Ca}{II}) is plotted against $x_1$, $c$, and
Hubble residuals corrected for light-curve width and colour (for
objects used when constructing the Hubble diagram).

The \ion{Ca}{II} ratio appears to be well
anticorrelated with $x_1$ (correlation coefficient $-0.53$ with
significance $>3\sigma$) and 
correlated with $c$ (coefficient of 0.46, again with significance
$>3\sigma$). The bottom plot of 
Figure~\ref{f:SALT_R_Ca} shows that the $x_1$ and $c$ corrected Hubble
residuals and $\Re$(\ion{Ca}{II}) are not well correlated with
each other (coefficient of 0.24). However, a model that uses $x_1$,
$c$, {\it and} the \ion{Ca}{II} ratio decreases the WRMS by \about6~per~cent and
the $\sigma_{\rm pred}$ by \about33~per~cent, although the 
significance of this
improvement is only at the 1.1$\sigma$ level ($\Delta_{x_1,c} =
-0.0207 \pm 0.0191$).

\subsection{The ``SiS'' Ratio}\label{ss:sis_ratio}

Analogous to the \ion{Ca}{II} ratio, the ``SiS ratio'' was introduced
by \citet{Bongard06} as the ratio of the flux at the red edge of the
\ion{S}{II} ``W'' feature to the flux at the red edge of the
\ion{Si}{II} $\lambda$6355 feature. In the notation used in BSNIP~II
this is 
\begin{equation}
\Re\left(\textrm{SiS}\right) \equiv \frac{F_r\left(\textrm{\ion{S}{II} ``W''}\right)}{F_r\left(\textrm{\ion{Si}{II} $\lambda$6355}\right)}.
\end{equation}
In a sample of 8 SNe, $\Re$(SiS) has been seen to correlate with
maximum absolute $B$-band magnitude in the same way as
$\Re$(\ion{Ca}{II}) \citep{Bongard06}. 

\begin{figure}
\centering
\includegraphics[width=3.5in]{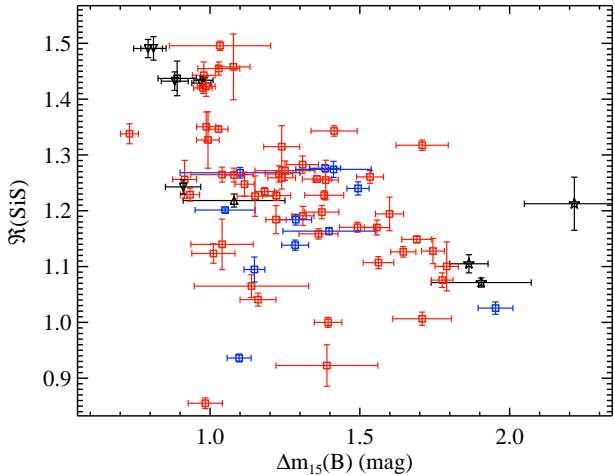}
\caption[The SiS ratio versus $\Delta m_{15}(B)$]{The SiS ratio
  versus $\Delta m_{15}(B)$. Colours and shapes of data points are the
  same as in Figure~\ref{f:v_bv_si6355}. 
  The correlation
  coefficient is $-0.54$.}\label{f:R_SiS_dm15}  
\end{figure}

The SiS ratio and the \ion{Si}{II} ratio were found to be only
marginally correlated in BSNIP~II, and Figure~\ref{f:R_SiS_dm15} (which
contains 72 SNe) shows that $\Re$(SiS) is anticorrelated with $\Delta
m_{15}(B)$ (correlation coefficient $-0.50$ and significance
\about3$\sigma$).  
Note that this relationship is in the opposite sense of the
one between the \ion{Ca}{II} ratio and $\Delta m_{15}(B)$. Here
Ia-99aa/91T objects lie above the main
relationship while the Ia-norm and HV SNe are well mixed. Once again, 
comparing $\Delta$ or $x_1$ to $\Re$(SiS) yields similar results to
what is seen in Figure~\ref{f:R_SiS_dm15}.

\begin{figure}
\centering
\includegraphics[width=3.5in]{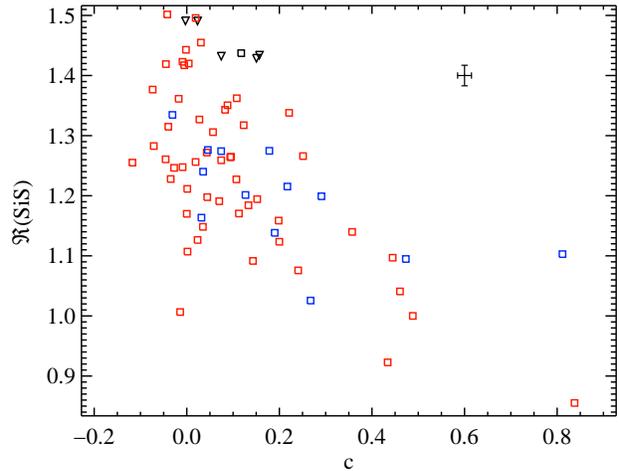}
\caption[The SiS ratio versus $c$]{The SiS ratio
  versus $c$. Colours and shapes of data points are the
  same as in Figure~\ref{f:v_bv_si6355}. The median
  uncertainty in both directions is 
  shown in the upper-right corner.}\label{f:R_SiS_c}  
\end{figure}

The SiS ratio appears to be well correlated with $c$, when including
the most reddened objects. In Figure~\ref{f:R_SiS_c} are 71 SNe, and the
correlation coefficient is $-0.61$ (significant at the $>3\sigma$
level). However, if one removes 
the most highly reddened objects with $c > 0.5$, the correlation
weakens slightly to $-0.56$ (still with significance $>3\sigma$). No
distance model involving the SiS ratio 
is more accurate than the $\left(x_1,c\right)$ model. However, when
$\Re$(SiS) is combined with just $c$ or both $x_1$ and $c$, the
accuracy is on par with the standard $\left(x_1,c\right)$ model
($\Delta_{x_1,c} = 0.0209 \pm 0.0223$ and $\Delta_{x_1,c} = 0.0074 \pm
0.0105$, respectively). We also find (in Section~\ref{ss:ratios}) that
out of 17,822 flux ratios combined with $c$, the most accurate
distances are calculated using flux ratios that are effectively the
SiS ratio, and that it is nearly as accurate as using the standard 
$\left(x_1,c\right)$ model. Finally, we note that \citet{Blondin11}
found that $\Re$(SiS) performs significantly worse than the usual
$\left(x_1,c\right)$ model.

\subsection{The ``SSi'' Ratio}\label{ss:ssi_ratio}

Yet another possible spectroscopic luminosity indicator is the ratio
of the pEW of the \ion{S}{II} ``W'' to that of the \ion{Si}{II}
$\lambda$5972 feature \citep{Hachinger06}. This SSi ratio is defined
in BSNIP~II as
\begin{equation}
\Re\left(\textrm{S,Si}\right) \equiv \frac{\textrm{pEW}\left(\textrm{\ion{S}{II} ``W''}\right)}{\textrm{pEW}\left(\textrm{\ion{Si}{II} $\lambda$5972}\right)}.
\end{equation}
\citet{Hachinger06} found that the SSi ratio is linearly
anticorrelated with $\Delta m_{15}$ (which is opposite to the
relationship between $\Re$(\ion{Si}{II}) and $\Delta m_{15}$). The
analysis in BSNIP~II seemed to confirm this observation by showing
that the SSi ratio was strongly anticorrelated (nonlinearly) with
the \ion{Si}{II} ratio. 

\begin{figure}
\centering
\includegraphics[width=3.5in]{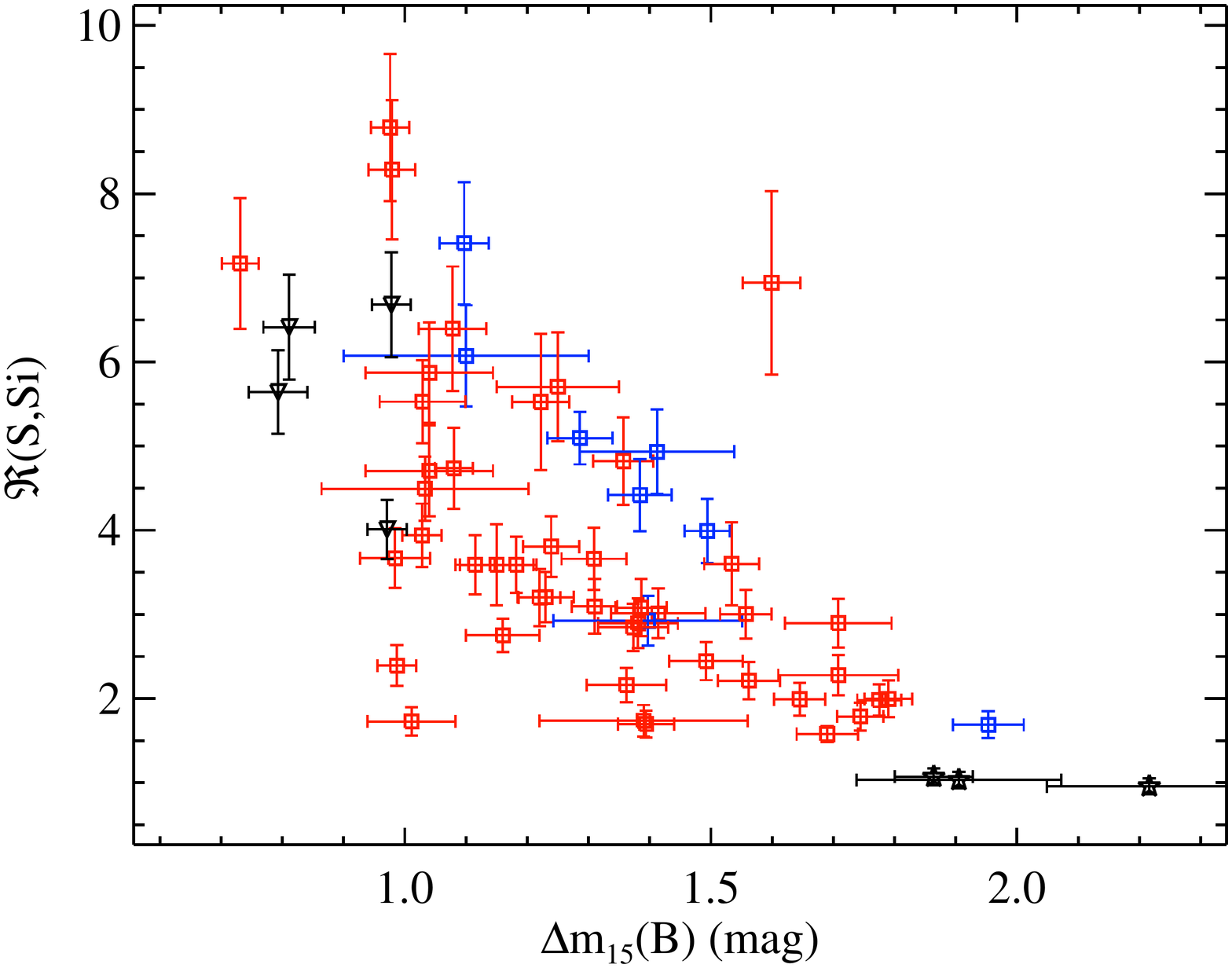}
\caption[The SSi ratio versus $\Delta m_{15}(B)$]{The SSi ratio
  versus $\Delta m_{15}(B)$. Colours and shapes of data points are the
  same as in Figure~\ref{f:v_bv_si6355}.}\label{f:R_SSi_dm15}  
\end{figure}

$\Re$(S,Si) is plotted against $\Delta m_{15}(B)$ for 59 SNe in
Figure~\ref{f:R_SSi_dm15}. The results of \citet{Hachinger06} and the
speculation in BSNIP~II are confirmed: the SSi ratio is strongly
anticorrelated with $\Delta m_{15}(B)$ (correlation coefficient of
$-0.67$ with significance $>3\sigma$). Here, the Ia-91bg and Ia-99aa 
objects follow the main 
trend and are found at the lower and upper ends of the correlation, 
respectively. There are only a few HV objects in
Figure~\ref{f:R_SSi_dm15}, but there is some evidence that they have
larger than average $\Re$(S,Si) values (which was also seen in Fig.~16
of BSNIP~II).

Plots of $\Delta$ and $x_1$ versus $\Re$(S,Si) display trends like
that of Figure~\ref{f:R_SSi_dm15}. However, the Ia-99aa objects fall
off of the main 
correlation in both of these parameter spaces. In both cases these SNe
have lower $\Re$(S,Si) values than one would expect from the main
correlation.

\subsection{The ``SiFe'' Ratio}\label{ss:sife_ratio}

Analogous to the SSi ratio, the ``SiFe ratio'' was defined as the
ratio of the pEW
of the \ion{Si}{II} $\lambda$5972 feature to that of the \ion{Fe}{II}
complex, and it was shown to be an accurate spectroscopic luminosity
indicator 
\citep{Hachinger06}. In BSNIP~II, $\Re$(Si,Fe) was defined as
\begin{equation}
\Re\left(\textrm{Si,Fe}\right) \equiv \frac{\textrm{pEW}\left(\textrm{\ion{Si}{II} $\lambda$5972}\right)}{\textrm{pEW}\left(\textrm{\ion{Fe}{II}}\right)},
\end{equation}
and found to be strongly correlated with the \ion{Si}{II} ratio.

\begin{figure}
\centering
\includegraphics[width=3.5in]{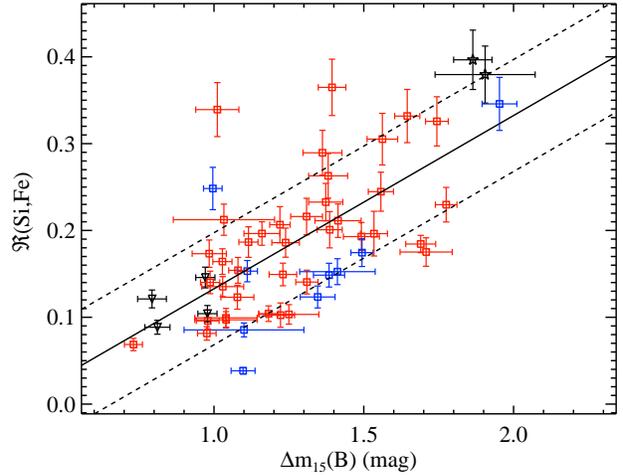}
\caption[The SiFe ratio versus $\Delta m_{15}(B)$]{The SiFe ratio
  versus $\Delta m_{15}(B)$. Colours and shapes of data points are the
  same as in Figure~\ref{f:v_bv_si6355}. The solid line
  is the linear least-squares fit and the dashed lines are the
  standard error of the fit.}\label{f:R_SiFe_dm15}  
\end{figure}

We plot $\Re$(Si,Fe) versus $\Delta m_{15}(B)$ for 53 SNe in
Figure~\ref{f:R_SiFe_dm15}. The results of \citet{Hachinger06} and the
speculation in BSNIP~II are again confirmed: the SiFe ratio is
strongly (linearly) correlated with $\Delta m_{15}(B)$, with a 
correlation coefficient of 0.68 (and with significance
$>3\sigma$). The solid line in the figure is the 
linear least-squares fit and the dashed lines are the standard error
of the fit.

Similar to $\Re$(\ion{Si}{II}), Ia-99aa objects are found at the
lowest end of the linear trend while the Ia-91bg objects in the
plot appear to be above the main relationship.  In
Figure~\ref{f:R_SiFe_dm15} there are only a few HV SNe, but they
appear to have smaller than average $\Re$(Si,Fe) values (which was also
seen in Fig.~17 of BSNIP~II). When comparing $\Delta$ and $x_1$ to
$\Re$(Si,Fe), the basic trend seen in Figure~\ref{f:R_SiFe_dm15} is
recovered, but with larger scatter (even though the correlation
coefficients are similar).

\subsection{Arbitrary Flux Ratios}\label{ss:ratios}

\citet{Bailey09} found that by using ratios of fluxes from a single,
binned SN~Ia spectrum they could decrease the scatter in their Hubble 
diagrams. These ratios are defined as
$\mathcal{R}(\lambda_y/\lambda_x) \equiv F(\lambda_y)/F(\lambda_x)$, 
where $\lambda_y$ and $\lambda_x$ are the rest-frame central
wavelengths of given bins.\footnote{This is similar to the definition
  of \citet{Bailey09}, but is the reciprocal of the definition used
  by \citet{Blondin11}. However, this only really matters for the
  plots of $\lambda_y$ versus $\lambda_x$ in
  Figure~\ref{f:2d}. Thus, each panel in Figure~7 of \citet{Blondin11}
  is the transpose of the panels in Figure~\ref{f:2d}. When using
  either definition, note that the first wavelength listed for a given 
  $\mathcal{R}$ is 
  the numerator in the actual ratio of fluxes.} The spectra are
forced to cover a wavelength range of exactly 3500--8500~\AA\ and are
binned into 134 equal-sized (in $\ln \lambda$) bins (corresponding to
2000~\kms\ per bin). The data are also 
deredshifted and dereddened using the redshift and reddening values
presented in Table~1 of BSNIP~I and assuming that the extinction
follows the \citet{Cardelli89} extinction law modified by
\citet{ODonnell94}. As in \citet{Bailey09} and \citet{Blondin11}, a
colour-corrected version of this flux ratio
($\mathcal{R}^c(\lambda_y/\lambda_x)$) is also used, and it is defined
as the ratio of fluxes as measured from a spectrum that has been
corrected for SALT2 $c$ using the colour law from \citet{Guy07}. We use
these colour-corrected flux ratios when testing models that also adopt
the SALT2 colour parameter (i.e., Equations~\ref{eq:m3} and
\ref{eq:m4}).

As with the rest of the current study, we only investigate spectra
within 5~d of maximum brightness since it was shown in BSNIP~II that
the spectra do not evolve significantly during these epochs.  
Also, as mentioned above, we do not use only spectra within 2.5~d of
maximum \citep[as has been done previously,][]{Bailey09,Blondin11}
because the significance of our results would be weakened due to the
smaller number of objects. Since the average spectrum in BSNIP extends
to 3300~\AA\ (BSNIP~I), we perform the current flux-ratio analysis
with the requirement that all spectra cover a wavelength range of
3300--8500~\AA. No ratios involving wavelengths below 3500~\AA\ are
found to decrease the WRMS significantly. We also vary the binning of
the spectra used in the flux-ratio analysis and investigate data with
bin sizes of 4000, 8000, and 10,000~\kms. We find that as the spectra
are binned more, the WRMS values increase for all ratios. This can be
explained by the idea that larger bins will ``blend'' wavelength bins
of flux ratios that decrease the Hubble residuals with ones that do not
and will thus add ``noise'' into flux ratios.

Since we are utilising SALT2 fits and Hubble diagrams, we again
require that SNe have $z_{\rm helio} > 0.01$, $c < 0.50$, and reduced
$\chi^2 < 2$. \citet{Blondin11} also require that the absolute
difference between $B-V$ colour at maximum brightness derived from the spectrum
and derived from the photometry be less than 0.1~mag. This is used as
a proxy for  
their relative spectrophotometric accuracy. In BSNIP~I it was shown
(in Table~3) that the relative spectrophotometric accuracy is often $<
0.1$~mag for the BSNIP data. In fact, $B-V$ colour is only inaccurate
at the 0.1~mag level for the oldest ($t > 20$~d) and noisiest (S/N $<
20$) BSNIP spectra. Therefore, the spectra investigated here should
all be spectrophotometrically accurate enough for the flux-ratio
analysis. 

Of the data studied here, 62 objects have flux ratios calculated for
the entire wavelength range mentioned above and reliable SALT2
fits that pass our Hubble diagram criteria (see
Section~\ref{ss:hubble}). We randomly divide our sample into 9 groups
of 7~SNe and 1 group with 6~SNe when doing 10-fold CV.

\subsubsection{Flux-Ratio Results}

The ``best'' flux ratios for each model are chosen
to be the ones with the lowest WRMS values. Ranking by other
parameters, such as the intrinsic prediction error \citep[as used
in][]{Blondin11}, yields different values for the best-performing flux
ratios. However, since our main goal is to minimise the scatter in the
Hubble diagram, and since the WRMS has a relatively straightforward
interpretation, we rank the best flux ratios by their WRMS values.
As discussed in Section~\ref{ss:hubble}, for each model involving a
flux ratio 
(Equations~\ref{eq:m1}--\ref{eq:m4}) we calculate, in addition to the 
WRMS, the intrinsic prediction error ($\sigma_{\rm pred}$), the
intrinsic correlation ($\rho_{x_1,c}$) of the residuals with residuals
using the $\left(x_1,c\right)$ model (Equation~\ref{eq:m5}), and the
difference ($\Delta_{x_1,c}$) in intrinsic prediction error with
respect to the $\left(x_1,c\right)$ model and its significance. These
parameters, along 
with the wavelengths, of the top 10 ratios for each model which
includes a flux ratio (Equations~\ref{eq:m1}--\ref{eq:m4}), are shown
in Table~\ref{t:top}.

\begin{table*}
\begin{center}
\caption{Top 10 Flux Ratios for Each Model}\label{t:top}
\begin{tabular}{lcccccc}
\hline\hline
Rank &  $\lambda_y$ &  $\lambda_x$ & WRMS (mag) & $\sigma_{\rm pred}$ (mag) & $\rho_{x_1,c}$  & $\Delta_{x_1,c}$  \\
\hline

\multicolumn{7}{c}{$\mathcal{R}$} \\
\hline
 1 & 7770 & 3750 & $0.218 \pm 0.027$ & $0.179 \pm 0.023$ & $-0.11$ & \phantom{$-$}$ 0.095 \pm 0.030$ (3.2$\sigma$) \\
 2 & 7670 & 3750 & $0.223 \pm 0.028$ & $0.189 \pm 0.023$ & \phantom{$-$}$ 0.09$ & \phantom{$-$}$ 0.104 \pm 0.030$ (3.5$\sigma$) \\
 3 & 7720 & 3750 & $0.223 \pm 0.027$ & $0.188 \pm 0.023$ & $-0.05$ & \phantom{$-$}$ 0.103 \pm 0.030$ (3.4$\sigma$) \\
 4 & 7930 & 3750 & $0.226 \pm 0.030$ & $0.187 \pm 0.023$ & $-0.10$ & \phantom{$-$}$ 0.104 \pm 0.031$ (3.4$\sigma$) \\
 5 & 6990 & 3750 & $0.227 \pm 0.030$ & $0.193 \pm 0.023$ & $-0.17$ & \phantom{$-$}$ 0.111 \pm 0.031$ (3.6$\sigma$) \\
 6 & 7880 & 3750 & $0.227 \pm 0.027$ & $0.190 \pm 0.023$ & $-0.06$ & \phantom{$-$}$ 0.106 \pm 0.030$ (3.5$\sigma$) \\
 7 & 7670 & 3780 & $0.230 \pm 0.030$ & $0.199 \pm 0.024$ & \phantom{$-$}$ 0.08$ & \phantom{$-$}$ 0.114 \pm 0.030$ (3.8$\sigma$) \\
 8 & 6990 & 3780 & $0.231 \pm 0.029$ & $0.202 \pm 0.023$ & $-0.21$ & \phantom{$-$}$ 0.121 \pm 0.031$ (3.9$\sigma$) \\
 9 & 6900 & 3750 & $0.233 \pm 0.029$ & $0.198 \pm 0.024$ & $-0.18$ & \phantom{$-$}$ 0.116 \pm 0.031$ (3.7$\sigma$) \\
10 & 7040 & 3780 & $0.234 \pm 0.029$ & $0.208 \pm 0.023$ & $-0.15$ & \phantom{$-$}$ 0.127 \pm 0.031$ (4.1$\sigma$) \\
\hline

\multicolumn{7}{c}{$\left(x_1,\mathcal{R}\right)$} \\
\hline
 1 & 6990 & 3750 & $0.199 \pm 0.022$ & $0.160 \pm 0.022$ & $-0.19$ & \phantom{$-$}$ 0.080 \pm 0.031$ (2.6$\sigma$) \\
 2 & 7770 & 3750 & $0.200 \pm 0.024$ & $0.160 \pm 0.022$ & $-0.09$ & \phantom{$-$}$ 0.077 \pm 0.030$ (2.6$\sigma$) \\
 3 & 6720 & 3750 & $0.203 \pm 0.024$ & $0.166 \pm 0.022$ & $-0.28$ & \phantom{$-$}$ 0.088 \pm 0.031$ (2.8$\sigma$) \\
 4 & 6900 & 3750 & $0.204 \pm 0.024$ & $0.164 \pm 0.022$ & $-0.18$ & \phantom{$-$}$ 0.083 \pm 0.031$ (2.7$\sigma$) \\
 5 & 6760 & 3750 & $0.206 \pm 0.027$ & $0.165 \pm 0.022$ & $-0.22$ & \phantom{$-$}$ 0.085 \pm 0.031$ (2.8$\sigma$) \\
 6 & 6950 & 3750 & $0.206 \pm 0.024$ & $0.166 \pm 0.023$ & $-0.19$ & \phantom{$-$}$ 0.085 \pm 0.031$ (2.7$\sigma$) \\
 7 & 6850 & 3750 & $0.206 \pm 0.024$ & $0.166 \pm 0.022$ & $-0.22$ & \phantom{$-$}$ 0.085 \pm 0.031$ (2.8$\sigma$) \\
 8 & 7930 & 3750 & $0.209 \pm 0.027$ & $0.167 \pm 0.022$ & $-0.12$ & \phantom{$-$}$ 0.085 \pm 0.030$ (2.8$\sigma$) \\
 9 & 6590 & 3750 & $0.209 \pm 0.028$ & $0.166 \pm 0.023$ & $-0.23$ & \phantom{$-$}$ 0.092 \pm 0.032$ (2.9$\sigma$) \\
10 & 6990 & 3780 & $0.209 \pm 0.027$ & $0.177 \pm 0.022$ & $-0.19$ & \phantom{$-$}$ 0.097 \pm 0.030$ (3.2$\sigma$) \\
\hline

\multicolumn{7}{c}{$\left(c,\mathcal{R}^c\right)$} \\
\hline
 1 & 5580 & 6330 & $0.146 \pm 0.021$ & $0.083 \pm 0.022$ & \phantom{$-$}$ 0.43$ & $-0.004 \pm 0.018$ (0.2$\sigma$) \\
 2 & 3980 & 4140 & $0.147 \pm 0.017$ & $0.076 \pm 0.025$ & \phantom{$-$}$ 0.56$ & $-0.002 \pm 0.017$ (0.1$\sigma$) \\
 3 & 6330 & 5580 & $0.149 \pm 0.021$ & $0.086 \pm 0.022$ & \phantom{$-$}$ 0.46$ & \phantom{$-$}$ 0.002 \pm 0.018$ (0.1$\sigma$) \\
 4 & 5730 & 6370 & $0.149 \pm 0.019$ & $0.087 \pm 0.022$ & \phantom{$-$}$ 0.41$ & \phantom{$-$}$ 0.002 \pm 0.019$ (0.1$\sigma$) \\
 5 & 6370 & 5580 & $0.150 \pm 0.019$ & $0.094 \pm 0.021$ & \phantom{$-$}$ 0.42$ & \phantom{$-$}$ 0.009 \pm 0.018$ (0.5$\sigma$) \\
 6 & 5580 & 6290 & $0.150 \pm 0.022$ & $0.076 \pm 0.025$ & \phantom{$-$}$ 0.39$ & $-0.012 \pm 0.021$ (0.6$\sigma$) \\
 7 & 5730 & 6330 & $0.150 \pm 0.019$ & $0.087 \pm 0.023$ & \phantom{$-$}$ 0.43$ & $-0.000 \pm 0.019$ (0.0$\sigma$) \\
 8 & 5580 & 6420 & $0.151 \pm 0.021$ & $0.093 \pm 0.021$ & \phantom{$-$}$ 0.43$ & \phantom{$-$}$ 0.003 \pm 0.019$ (0.2$\sigma$) \\
 9 & 5690 & 6330 & $0.151 \pm 0.021$ & $0.085 \pm 0.022$ & \phantom{$-$}$ 0.18$ & \phantom{$-$}$ 0.009 \pm 0.022$ (0.4$\sigma$) \\
10 & 6370 & 5730 & $0.151 \pm 0.018$ & $0.083 \pm 0.024$ & \phantom{$-$}$ 0.64$ & $-0.002 \pm 0.019$ (0.1$\sigma$) \\
\hline

\multicolumn{7}{c}{$\left(x_1,c,\mathcal{R}^c\right)$} \\
\hline
 1 & 3780 & 4580 & $0.130 \pm 0.017$ & $0.050 \pm 0.029$ & \phantom{$-$}$ 0.81$ & $-0.020 \pm 0.012$ (1.7$\sigma$) \\
 2 & 3610 & 4890 & $0.132 \pm 0.015$ & $0.056 \pm 0.028$ & \phantom{$-$}$ 0.71$ & $-0.016 \pm 0.014$ (1.2$\sigma$) \\
 3 & 5360 & 6900 & $0.132 \pm 0.017$ & $0.062 \pm 0.025$ & \phantom{$-$}$ 0.84$ & $-0.014 \pm 0.010$ (1.4$\sigma$) \\
 4 & 4760 & 3800 & $0.132 \pm 0.020$ & $0.065 \pm 0.025$ & \phantom{$-$}$ 0.84$ & $-0.012 \pm 0.010$ (1.2$\sigma$) \\
 5 & 7980 & 8140 & $0.133 \pm 0.015$ & $0.062 \pm 0.025$ & \phantom{$-$}$ 0.69$ & $-0.021 \pm 0.015$ (1.4$\sigma$) \\
 6 & 7720 & 8200 & $0.133 \pm 0.015$ & $0.061 \pm 0.026$ & \phantom{$-$}$ 0.94$ & $-0.013 \pm 0.008$ (1.6$\sigma$) \\
 7 & 6080 & 6210 & $0.133 \pm 0.020$ & $0.054 \pm 0.029$ & \phantom{$-$}$ 0.79$ & $-0.019 \pm 0.013$ (1.5$\sigma$) \\
 8 & 3610 & 4170 & $0.133 \pm 0.015$ & $0.064 \pm 0.025$ & \phantom{$-$}$ 0.81$ & $-0.009 \pm 0.010$ (0.9$\sigma$) \\
 9 & 4280 & 3780 & $0.133 \pm 0.015$ & $0.062 \pm 0.026$ & \phantom{$-$}$ 0.80$ & $-0.011 \pm 0.011$ (1.0$\sigma$) \\
10 & 3610 & 4230 & $0.133 \pm 0.017$ & $0.062 \pm 0.024$ & \phantom{$-$}$ 0.73$ & $-0.016 \pm 0.013$ (1.2$\sigma$) \\
\hline

\multicolumn{7}{c}{$\left(x_1,c\right)$} \\
\hline
$\cdots$ & $\cdots$ & $\cdots$ & $0.144 \pm 0.019$ & $0.076 \pm 0.023$ & \phantom{$-$}$\cdots$ & \phantom{$-$}$\cdots$ \\
\hline\hline
\end{tabular}
\end{center}
\end{table*}

\begin{figure}
\centering
\includegraphics[width=3.45in]{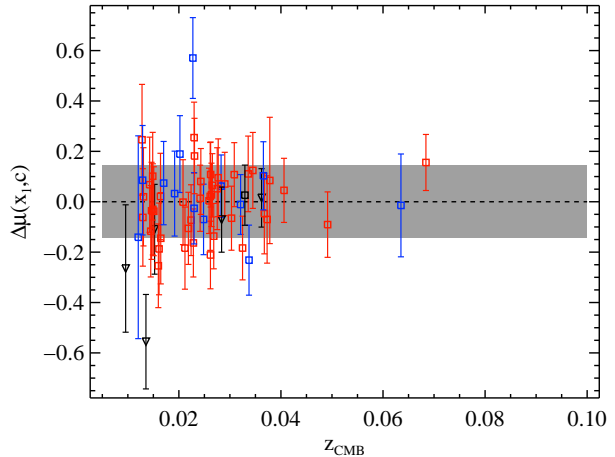}
\caption[Residuals versus $z_{\rm CMB}$ for the $\left(x_1,c\right)$
model]{Hubble diagram residuals versus $z_{\rm CMB}$ for the standard
  $\left(x_1,c\right)$ model (Equation~\ref{eq:m5}). The grey band is
  the WRMS for the model.  Colours and shapes of data points are the
  same as in Figure~\ref{f:v_bv_si6355}.}\label{f:x1_c_z}
\end{figure}

Also displayed in Table~\ref{t:top} is the WRMS and $\sigma_{\rm
  pred}$ of our benchmark $\left(x_1,c\right)$
model. Figure~\ref{f:x1_c_z} shows the Hubble diagram residuals for
this model versus redshift for the 62 SNe~Ia mentioned above. The grey
band indicates the WRMS for the model.

\begin{figure*}
\centering$
\begin{array}{cc}
\hspace{.25in}\mathcal{R} & \hspace{.25in}\mathcal{R} \\
\includegraphics[width=2.7in]{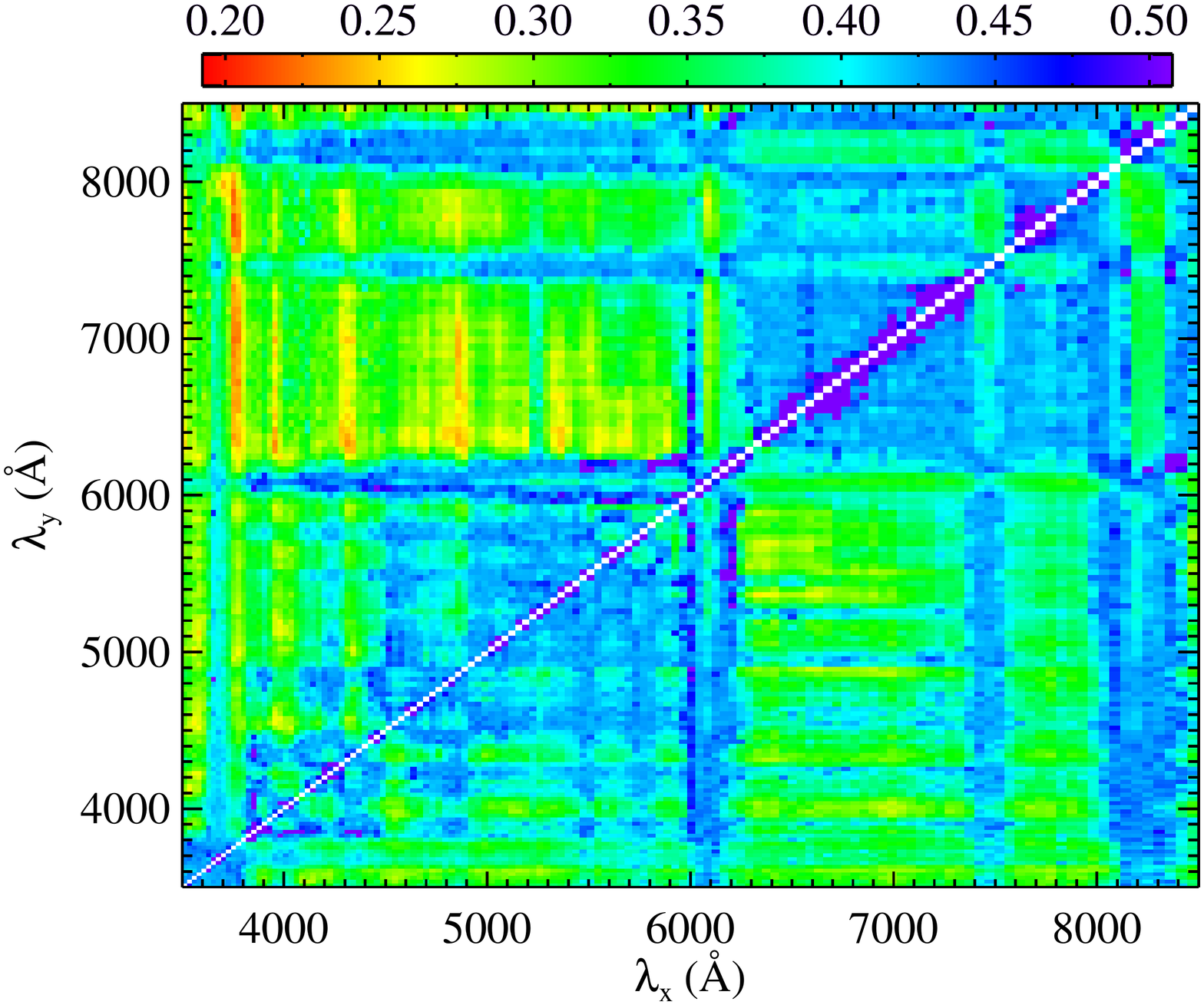} &
\includegraphics[width=2.7in]{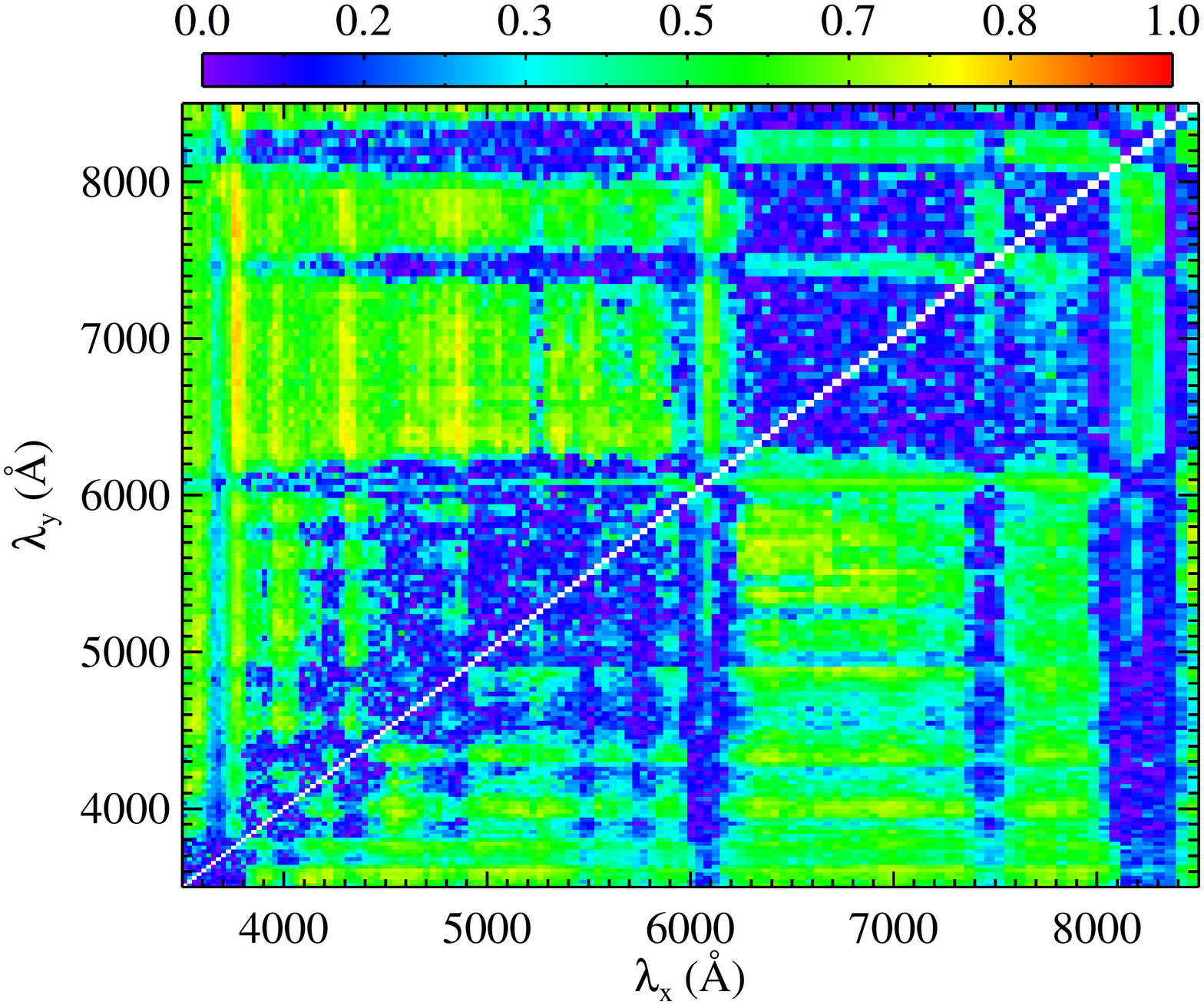} \\
\hspace{.25in}\left(x_1,\mathcal{R}\right) & \hspace{.25in}\left(x_1,\mathcal{R}\right) \\
\includegraphics[width=2.7in]{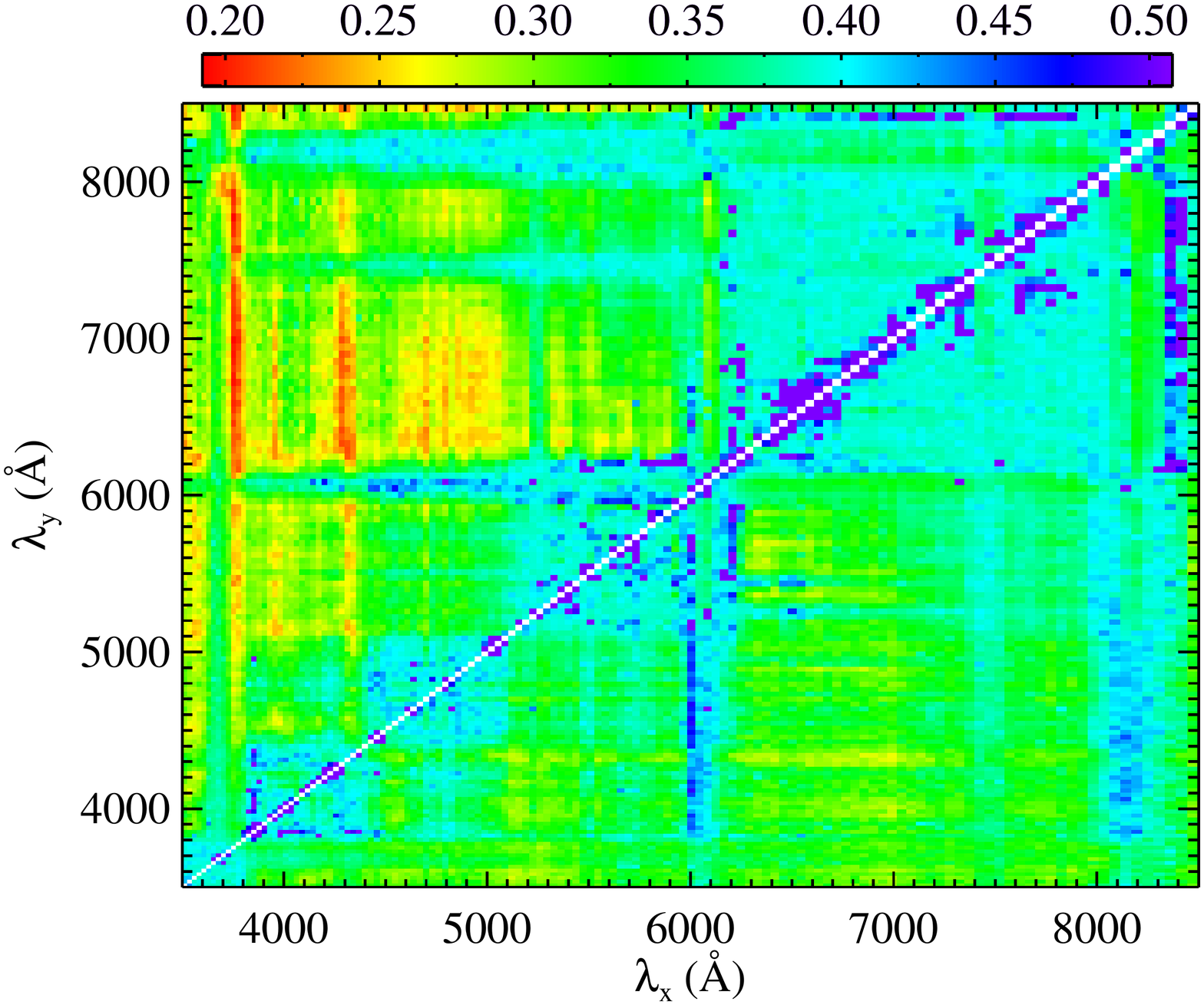} &
\includegraphics[width=2.7in]{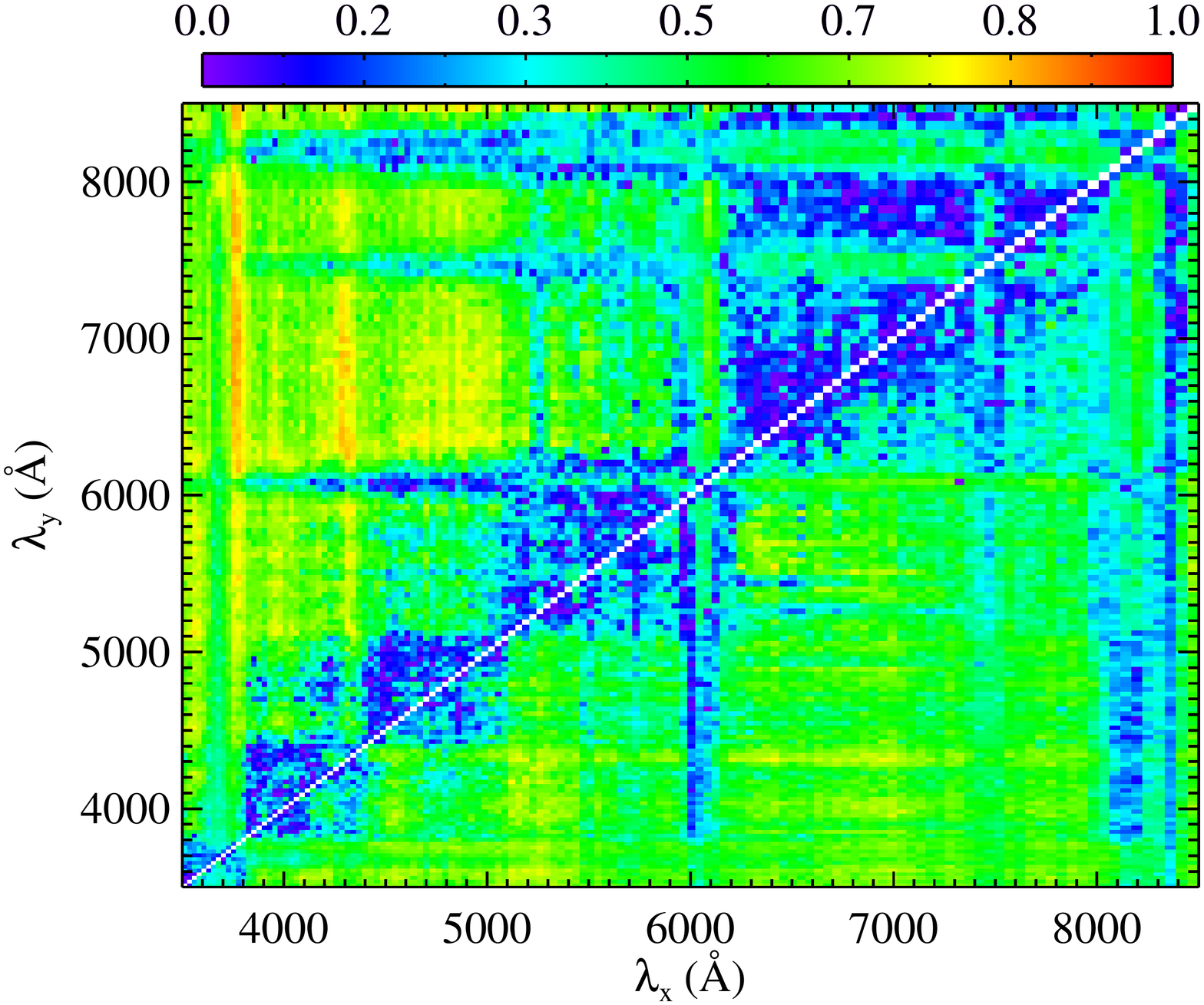} \\
\hspace{.25in}\left(c,\mathcal{R}^c\right) & \hspace{.25in}\left(c,\mathcal{R}^c\right) \\
\includegraphics[width=2.7in]{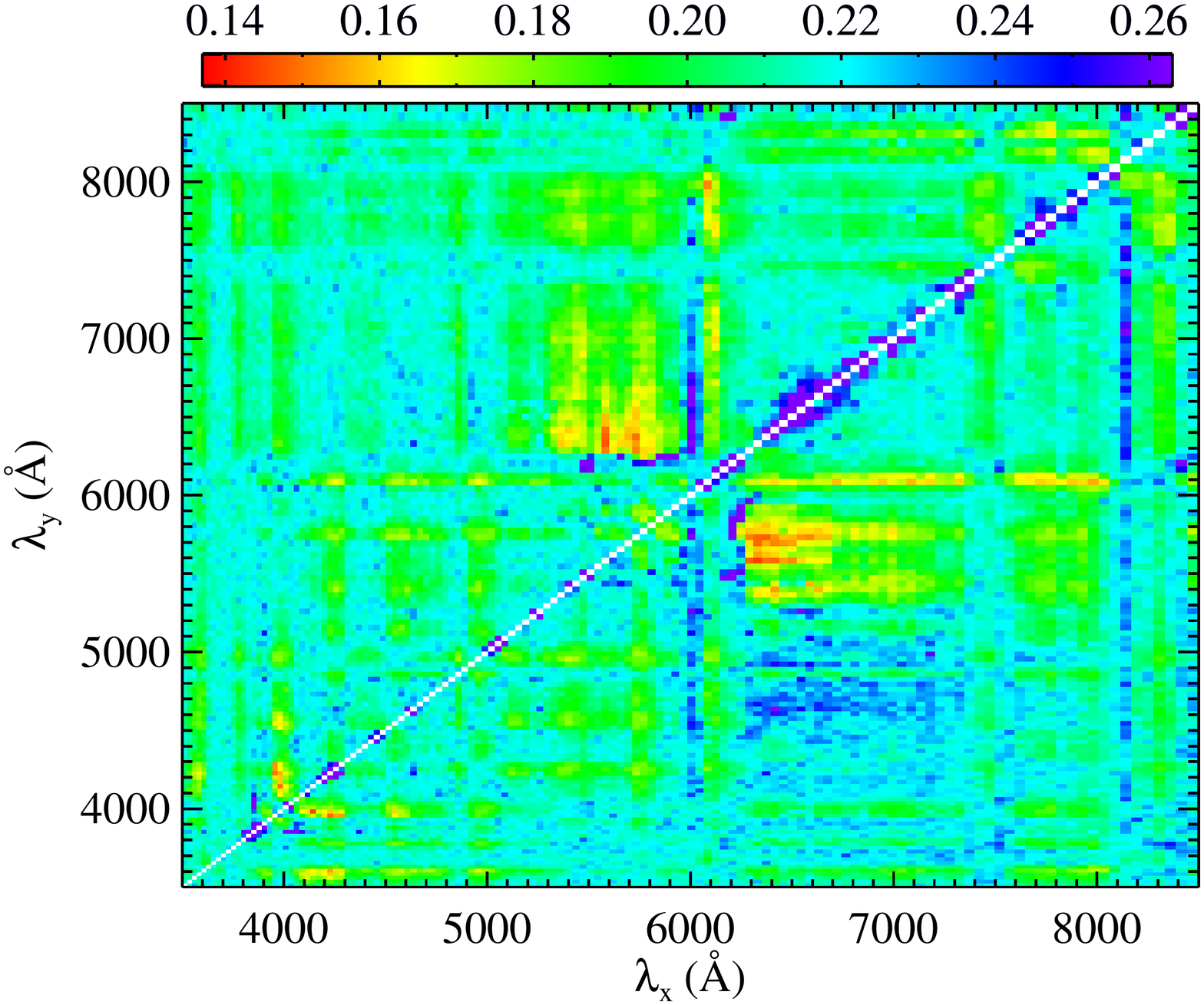} &
\includegraphics[width=2.7in]{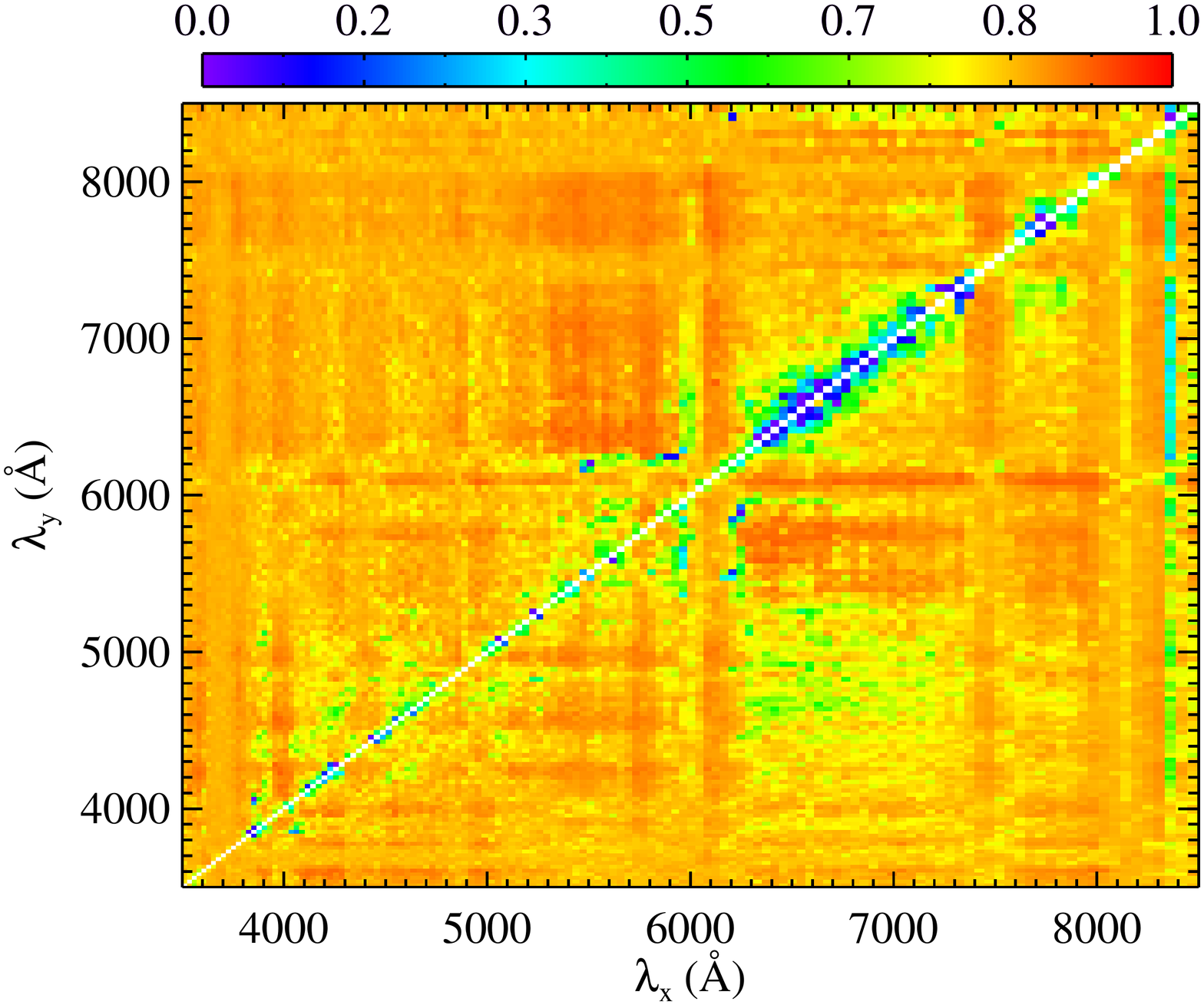} \\
\hspace{.25in}\left(x_1,c,\mathcal{R}^c\right) & \hspace{.25in}\left(x_1,c,\mathcal{R}^c\right) \\
\includegraphics[width=2.7in]{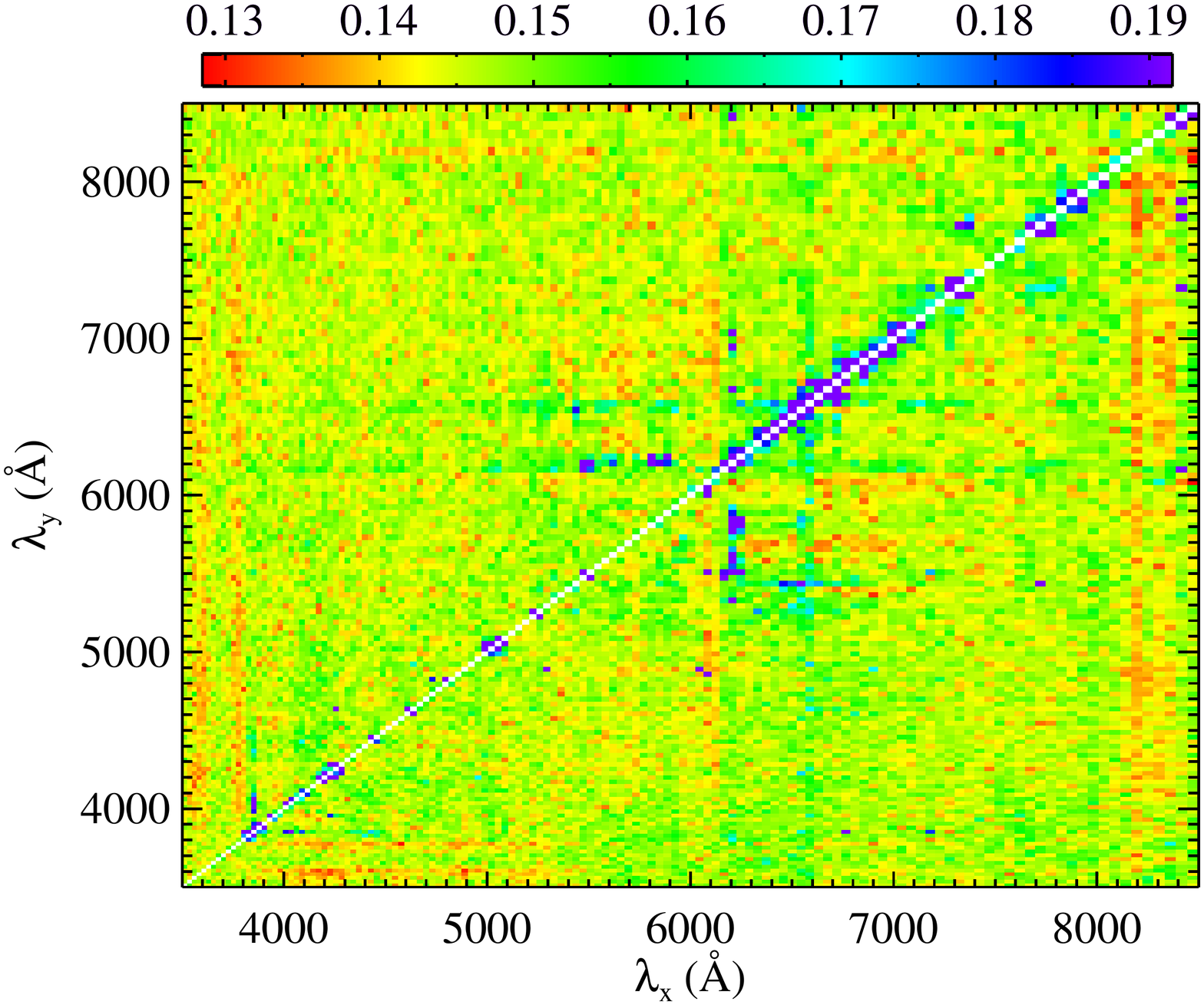} &
\includegraphics[width=2.7in]{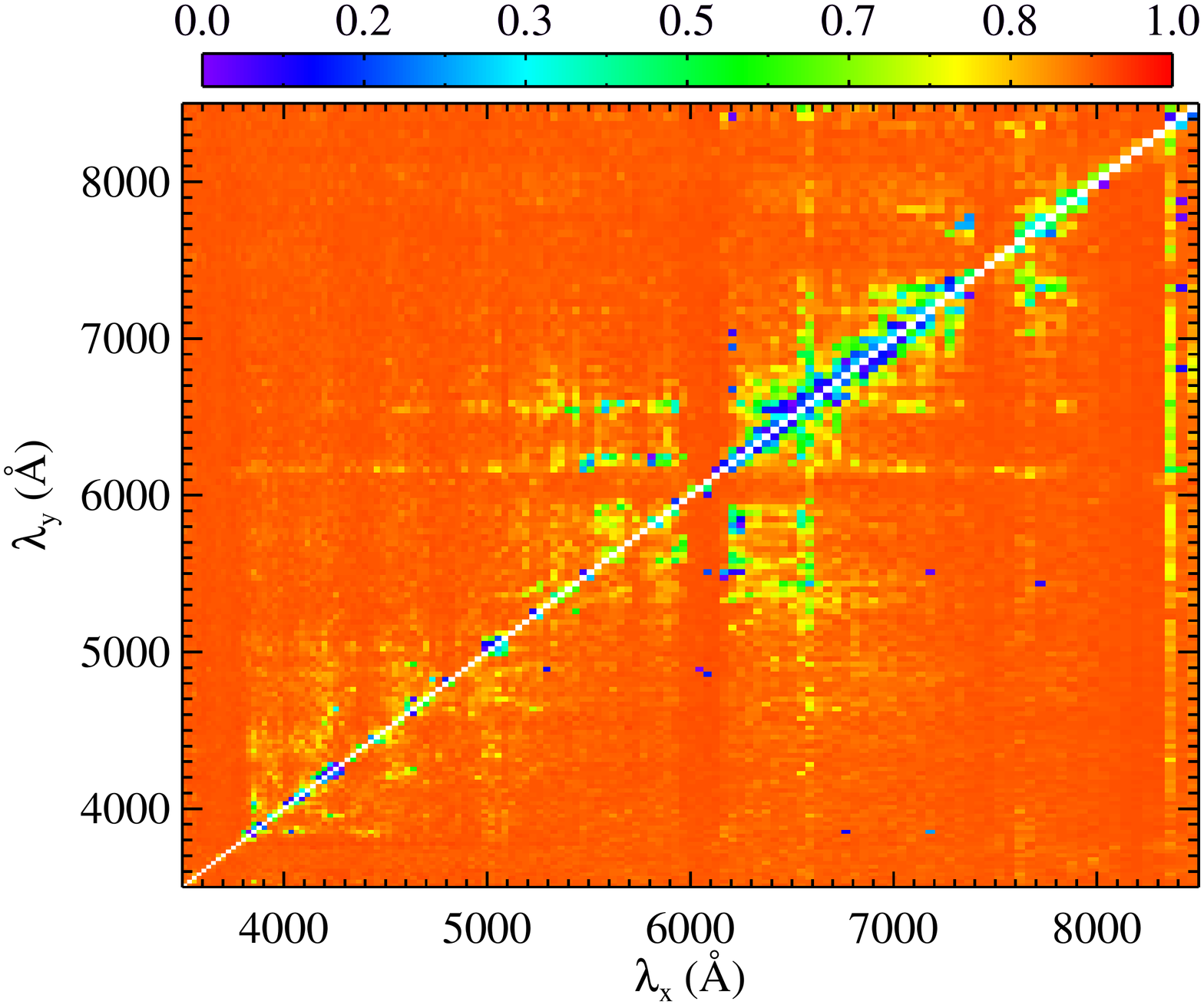} \\
\end{array}$
\caption{A map of the WRMS ({\it left column}) and the absolute 
  Pearson correlation coefficient of the correction term with the
  uncorrected Hubble residuals ({\it right column}) for all 17,822
  flux ratios used in all four models involving a flux ratio
  ($\mathcal{R}$ only, $\left(x_1,\mathcal{R}\right)$,
  $\left(c,\mathcal{R}^c\right)$, and
  $\left(x_1,c,\mathcal{R}^c\right)$, {\it top} to {\it bottom},
  respectively). All ratios with WRMS values $\ge 2\sigma$ above the
  mean have the same colour.}\label{f:2d} 
\end{figure*}

Figure~\ref{f:2d} shows the WRMS (left column) and absolute Pearson
correlation coefficient of the correction term (either $\gamma
\mathcal{R}$ or $\gamma \mathcal{R}^c$) with the uncorrected Hubble
residuals (right column) for all 17,822 ($=134 \times 133$) flux
ratios in all four models involving a flux ratio: $\mathcal{R}$,
$\left(x_1,\mathcal{R}\right)$, $\left(c,\mathcal{R}^c\right)$, and
$\left(x_1,c,\mathcal{R}^c\right)$ (top to bottom, respectively). All
ratios with WRMS values $\ge 2\sigma$ above the mean are displayed
using the same colour.

The left column of Figure~\ref{f:2d} is a proxy for the overall
scatter in the model while the right column indicates how
much ``new'' information is gained by adding in the correction term
($\gamma \mathcal{R}$ or $\gamma \mathcal{R}^c$). In general, a model
may have a low WRMS (or $\sigma_{\rm pred}$), meaning that the model
is fitting the data well, but since the data have uncertainties
associated with them, the model might be {\it overfitting} the data
and actually end up fitting noise. 

One way to discern whether this is the case is to see how well
the correction terms correlate with the uncorrected Hubble
residuals. As described in Section~\ref{sss:vels},  if the terms that
do not contain $x_c$ or $c$ (i.e., $\gamma \mathcal{R}$ or $\gamma
\mathcal{R}^c$) are well correlated with the uncorrected Hubble
residuals, then the measured observable is fitting information that
actually exists in the data (as opposed to noise). However, a large
correlation does not necessarily imply a good model. This is obvious
(for example) in the top row of Figure~\ref{f:2d}, where most flux
ratios have large WRMS values, including ones that also have quite
high correlations between the correction term and the uncorrected
Hubble residuals.

\subsubsection{Model 1: \scriptr\ Only}\label{sss:model1}

Using only a flux ratio (Equation~\ref{eq:m1}) leads to no improvement
over the usual $\left(x_1,c\right)$ model (Equation~\ref{eq:m5}). In
fact, this model seems to perform significantly worse, as can be seen
by the relatively large $\Delta_{x_1,c}$ values. The WRMS and
$\sigma_{\rm pred}$ of the ``best'' ratios are quite a bit larger than
those of the $\left(x_1,c\right)$ model. This differs from the
conclusion of 
previous work, which found that models using a flux ratio alone could perform
as well as the $\left(x_1,c\right)$ model \citep{Bailey09,Blondin11}.
The best-performing ratio in the \scriptr-only model,
$\mathcal{R}\left(7770/3750\right)$, is not correlated with $x_1$, but is
highly correlated with $c$ (correlation coefficient $0.81$). This
implies that $\mathcal{R}\left(7770/3750\right)$ is effectively a
colour indicator.




The best ratio using this model seen by \citet{Blondin11},
$\mathcal{R}\left(6630/4400\right)$, was found in their study to be
similarly correlated with SALT2 colour and to improve the Hubble
diagram residuals over using the $\left(x_1,c\right)$ model (albeit with
a low 
significance). We tested all of our flux-ratio models using a randomly
selected subset of 26 SNe from the BSNIP sample in order to match the
number of objects used by \citet{Blondin11}. These models all yielded
WRMS values similar to what was found by \citet{Blondin11}, though the
best-performing flux ratios were significantly different and the
significance of the improvement was decreased (mostly due to the fact
that the $\left(x_1,c\right)$ model performed much worse when using
only 26 objects). Much like our analysis which only used spectra
within 2.5~d of maximum brightness, the decrease in sample size was
chiefly responsible for the weakened significance of Hubble residual
improvement. 

\citet{Bailey09} also saw an overall improvement when
using their top ratio for this model,
$\mathcal{R}\left(6420/4430\right)$. The WRMS for
both of these ratios using the BSNIP data is \about0.35~mag, which is
much larger than the WRMS values seen for our best ratios. When the
BSNIP data are fit using $K = 2$, the WRMS of the best-performing
ratio does decrease as compared to our $K = 10$ run. However, the
best-performing ratios involve wavelengths that are randomly scattered
in wavelength space which may imply that these results are less
reliable than the more statistically rigorous run using $K = 10$. 

\subsubsection{Model 2: \scriptr\ and $x_1$}

Combining a flux ratio with the SALT2 stretch parameter $x_1$
(Equation~\ref{eq:m2}) also leads to no improvement over the
$\left(x_1,c\right)$ model. The WRMS and $\sigma_{\rm pred}$ values
are smaller than when using the \scriptr-only model, but they
are still significantly larger than those from the standard
$\left(x_1,c\right)$ model. We again point out that \citet{Blondin11}
found that the $\left(x_1,\mathcal{R}\right)$ model did better than
the $\left(x_1,c\right)$ model.



The best flux ratio using the $\left(x_1,\mathcal{R}\right)$ model,
$\mathcal{R}\left(6990/3750\right)$, is (like the \scriptr-only model)
not correlated with $x_1$, but strongly correlated with $c$
(coefficient of $0.83$). The ratio $\mathcal{R}\left(6990/3750\right)$
is therefore a proxy for $c$, and so it is unsurprising that this is
the top-ranked ratio in a model employing only a ratio and $x_1$.
Using this model \citet{Blondin11} again found that the ratio
$\mathcal{R}\left(6630/4400\right)$ was best and, as mentioned above,
it was similarly correlated with SALT2 $c$.
Nearly all of the top ten ratios for the \scriptr-only and
$\left(x_1,\mathcal{R}\right)$ models have very similar numerator and
denominator wavelengths, and the difference in wavelength between the
two fluxes is significant (3000--4000~\AA). This again supports the
idea presented above that the top-ranked flux ratios for these two
models are effectively proxies for colour.


\subsubsection{Model 3: \scriptrc\ and $c$}

Some of the top-ranked flux ratios with the SALT2 colour parameter $c$
(Equation~\ref{eq:m3}) are consistent with the results when using the
$\left(x_1,c\right)$ model. This lack of improvement is once again at
odds with what was seen by \citet{Blondin11}. However, as mentioned in 
Section~\ref{sss:model1}, the apparent improvement in \citet{Blondin11} 
is likely due to their smaller sample size and
the relatively poor performance of the $\left(x_1,c\right)$ model.



Many ratios appear to have large correlations between the correction
terms and uncorrected residuals, but the lowest WRMS values are
tightly clustered in wavelength space. This is also apparent in
Table~\ref{t:top}, where all but one of the top ten ratios for the
$\left(c,\mathcal{R}^c\right)$ model involve wavelengths near
5600~\AA\ and 6300~\AA. Flux ratios which include these wavelengths
are effectively the same as $\Re$(SiS), the SiS ratio
(Section~\ref{ss:sis_ratio}). 

The SiS ratio was shown above to be anticorrelated with $\Delta
m_{15}(B)$ (as well as correlated with $x_1$). The best flux ratio
using the $\left(c,\mathcal{R}^c\right)$ model,
$\mathcal{R}^c\left(5580/6330\right)$, is strongly correlated with
$x_1$ (correlation coefficient 0.83) and effectively uncorrelated
with $c$ (correlation coefficient $-0.18$). Therefore, the
top-ranked ratio for this model can be thought of as equivalent to
$\Re$(SiS) and/or $x_1$. Furthermore, the lack of correlation between
$\mathcal{R}^c\left(5580/6330\right)$ and $c$ implies that dereddening
the data using the SALT2 $c$ and colour law is working as intended.

With the $\left(c,\mathcal{R}^c\right)$ model, \citet{Blondin11}
showed that their top-ranked ratio was
$\mathcal{R}^c\left(6420/5290\right)$. This is similar to the
reciprocal of the best ratio found with the BSNIP data; thus, it is not
surprising that they find a strong {\it anticorrelation} between
$\mathcal{R}^c\left(6420/5290\right)$ and $x_1$. Their ratio also
leads to a larger decrease in WRMS
\citep[\about15~per~cent,][]{Blondin11}. \citet{Bailey09} again showed an
improvement when using their top ratio for this model,
$\mathcal{R}^c\left(6420/5190\right)$, which is also very close to the
best ratio of \citet{Blondin11}, as well as the reciprocal of the best
ratio presented here. Despite this similarity in wavelength space, the
top-ranked ratios for this model from \citet{Bailey09} and
\citet{Blondin11} yield WRMS values of \about0.21~mag when using the
BSNIP data.

\subsubsection{Model 4: \scriptrc, $x_1$, and $c$}

\begin{figure}
\centering$
\begin{array}{c}
\includegraphics[width=3.4in]{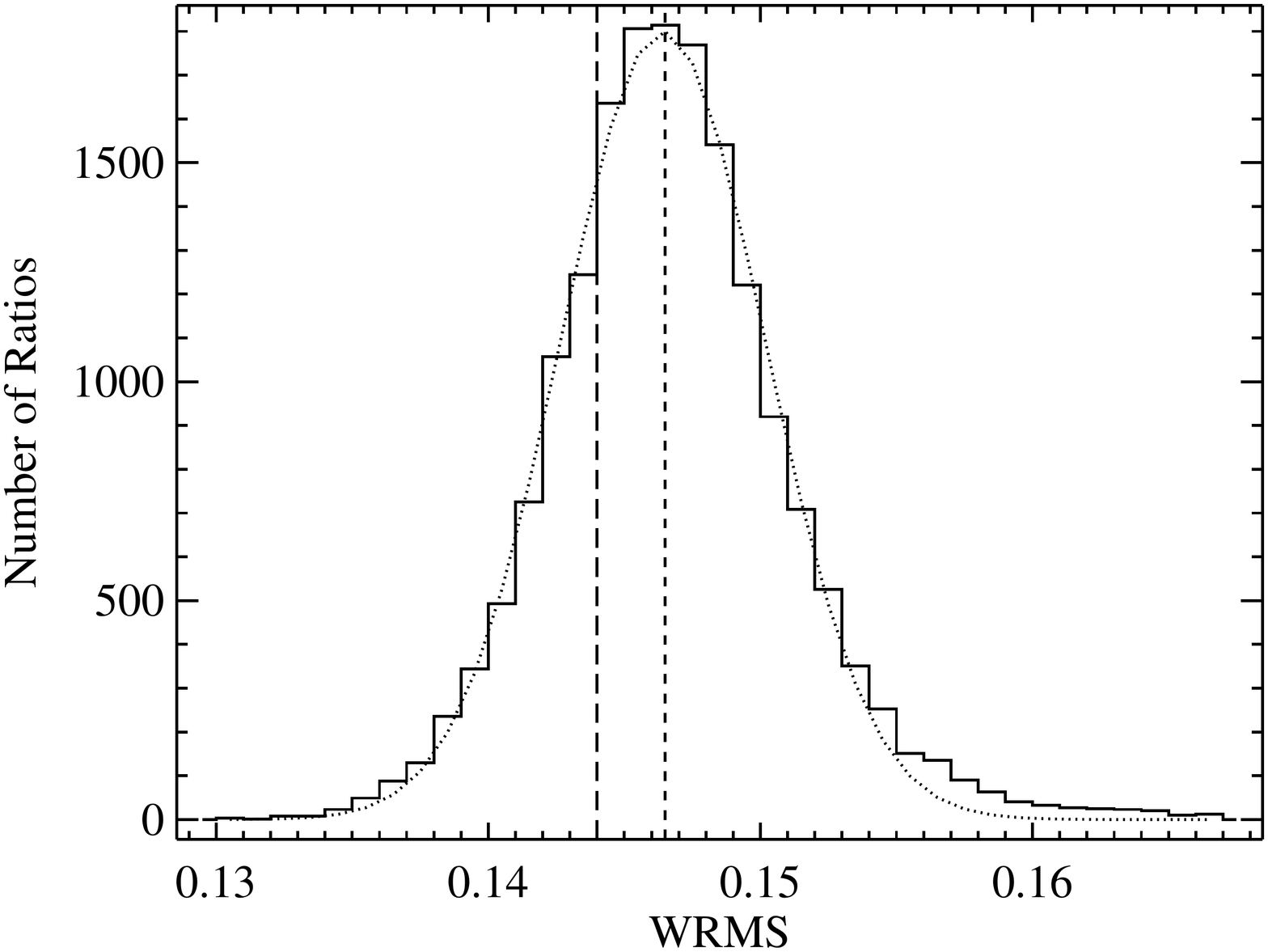} \\
\includegraphics[width=3.4in]{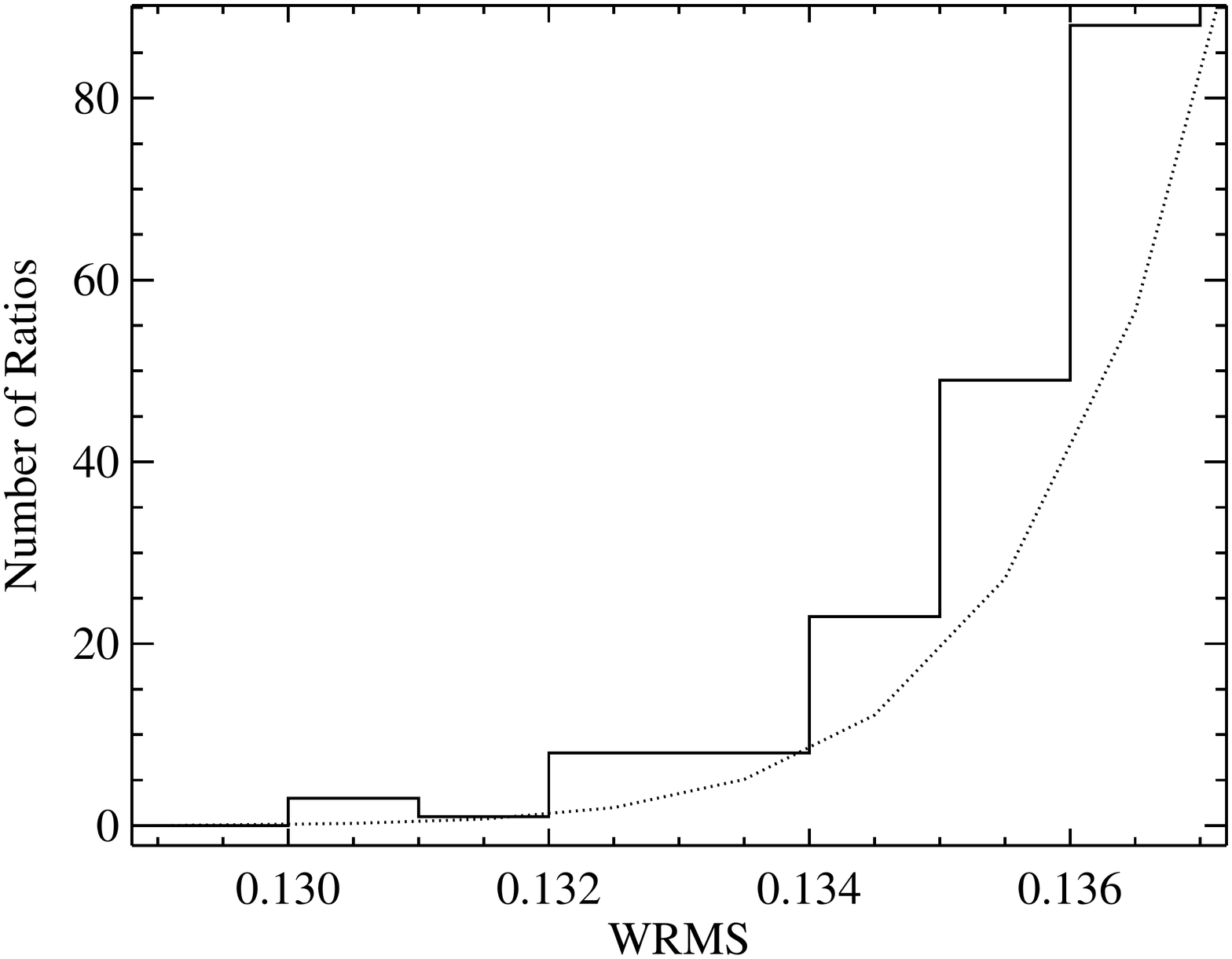}
\end{array}$
\caption{A histogram of the WRMS values for the
  $\left(x_1,c,\mathcal{R}^c\right)$ model (solid line) and the
  expected Gaussian distribution (dotted line) for ratios with WRMS
  values that are $< 5\sigma$ away from the peak of the distribution
  ({\it top}). The short-dashed and long-dashed lines are the peak of
  the WRMS distribution and the WRMS value of the standard
  $\left(x_1,c\right)$ model, respectively. A close-up view of the smallest
  WRMS values ({\it bottom}).}\label{f:wrms_hist} 
\end{figure}

Quite significant improvements over the standard $\left(x_1,c\right)$
model are seen when using the top-ranked flux ratios along with both
$x_1$ and $c$ (Equation~\ref{eq:m4}). In fact, the top ten ratios lead
to improvements at about the 1--2$\sigma$ level (see
Table~\ref{t:top}). Furthermore, a good fraction of the ratios lead to
some improvement over the standard model. Figure~\ref{f:wrms_hist}
shows a histogram of the WRMS values for the
$\left(x_1,c,\mathcal{R}^c\right)$ model (solid line) and the expected
Gaussian distribution (dotted line). The short-dashed and
long-dashed lines are the peak of the WRMS distribution and the WRMS
value of the standard $\left(x_1,c\right)$ model, respectively. The
distribution is roughly Gaussian when one ignores the high-WRMS tail
(i.e., $\textrm{WRMS} \ga 0.17$, or ratios with WRMS values that are
$> 5\sigma$ away from the peak of the distribution). The bottom panel
of Figure~\ref{f:wrms_hist} shows a close-up view of the smallest WRMS
values. The fact that the best performing flux ratios (i.e., the ones
yielding the smallest WRMS values) lie quite a bit above the Gaussian
expectation seems to indicate that these are in fact statistically
significant decreases in the WRMS, as indicated by the last column of 
Table~\ref{t:top}. 

\begin{figure}
\centering
\includegraphics[width=3.45in]{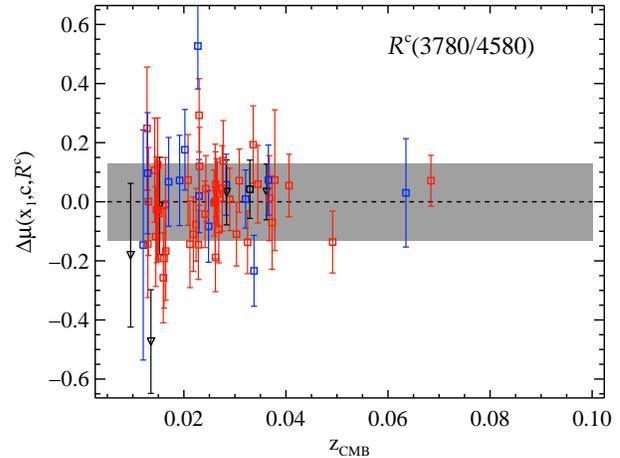}
\caption[Residuals versus $z_{\rm CMB}$ for the
$\left(x_1,c,\mathcal{R}^c\right)$ model]{Hubble diagram residuals 
versus $z_{\rm CMB}$ for the best flux ratio,
$\mathcal{R}^c\left(3780/4580\right)$, using the
$\left(x_1,c,\mathcal{R}^c\right)$ model (Equation~\ref{eq:m4}). The grey
band is the WRMS for the model.  Colours and shapes of data points are the
  same as in Figure~\ref{f:v_bv_si6355}.}\label{f:r_x1_c_z}
\end{figure}

The $\left(x_1,c,\mathcal{R}^c\right)$ model and
its best ratio, $\mathcal{R}^c\left(3780/4580\right)$, decrease the
WRMS by \about10~per~cent and $\sigma_{\rm pred}$ by \about34~per~cent
from the $\left(x_1,c\right)$ model. The Hubble diagram for the
top-ranked flux ratio using the $\left(x_1,c,\mathcal{R}^c\right)$
model is shown in Figure~\ref{f:r_x1_c_z}, along with the WRMS for the
best model (the grey band).

The decrease in WRMS using this model is smaller than previously seen,
but the decrease in $\sigma_{\rm  pred}$ is larger and the overall
improvement is at a higher significance; \citet{Blondin11} found a
20~per~cent decrease in WRMS with a 1.6$\sigma$
significance. The fact that the current dataset has nearly
three times as many objects as the sample studied by \citet{Blondin11}
is 
likely the reason why the $\left(x_1,c,\mathcal{R}^c\right)$ model
yields more significant improvements in WRMS values.

Most ratios have low values of WRMS and extremely large correlations
between the correction terms and uncorrected residuals, implying that
the $\left(x_1,c,\mathcal{R}^c\right)$ model is performing better
overall than any of the other models investigated. This is consistent
with \citet{Blondin11}, although they found no flux
ratio to have a strong correlation between the correction terms and
the uncorrected Hubble residuals. The wavelengths of six of the top
ten ratios for this model are near \about3750~\AA\ and \about4550~\AA\
with wavelength baselines of \about800~\AA. These approximately
correspond to the midpoint of the \ion{Ca}{II}~H\&K feature and the
border between the \ion{Mg}{II} and \ion{Fe}{II} complexes. 

\begin{figure}
\centering$
\begin{array}{c}
\includegraphics[width=3.5in]{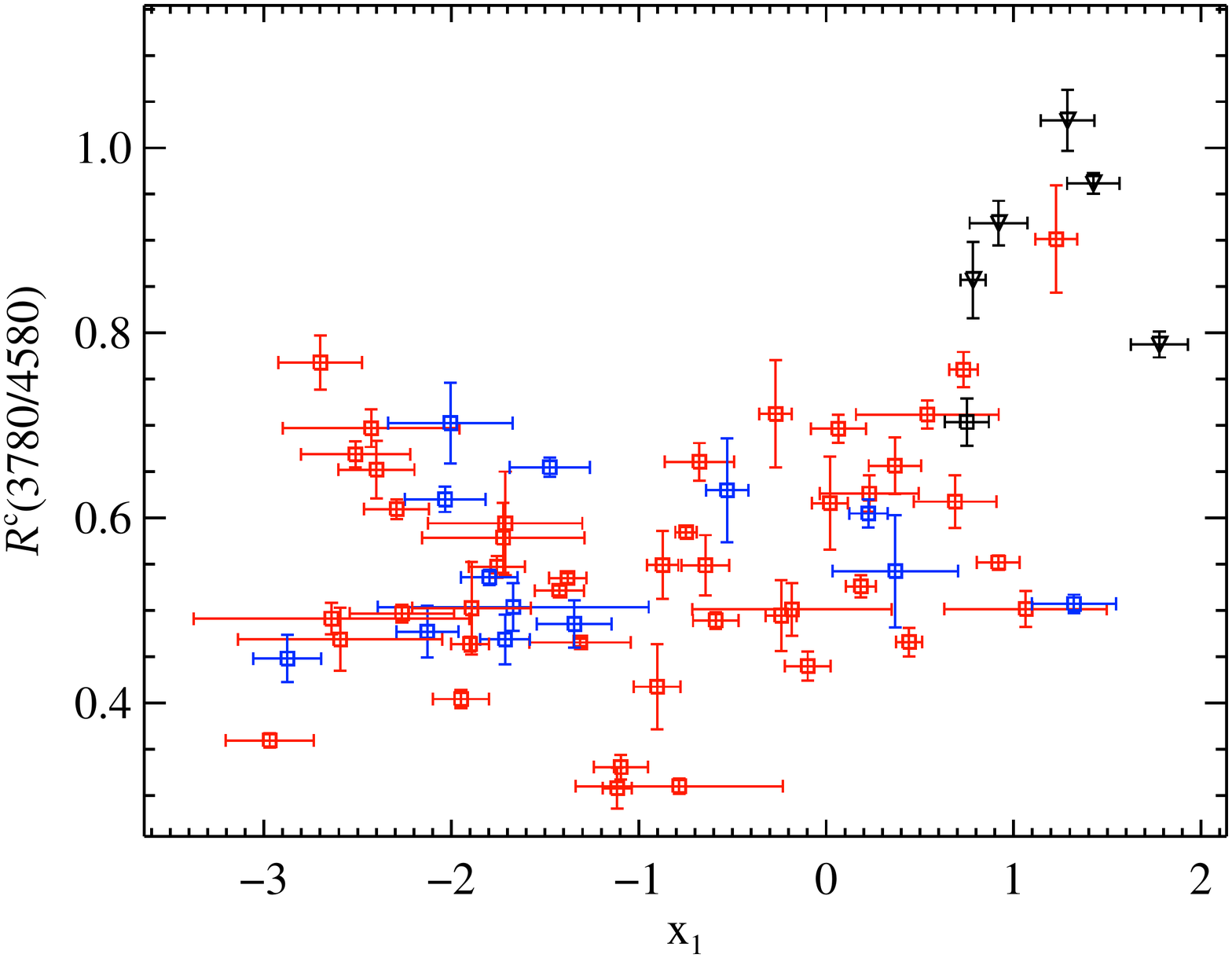} \\
\includegraphics[width=3.5in]{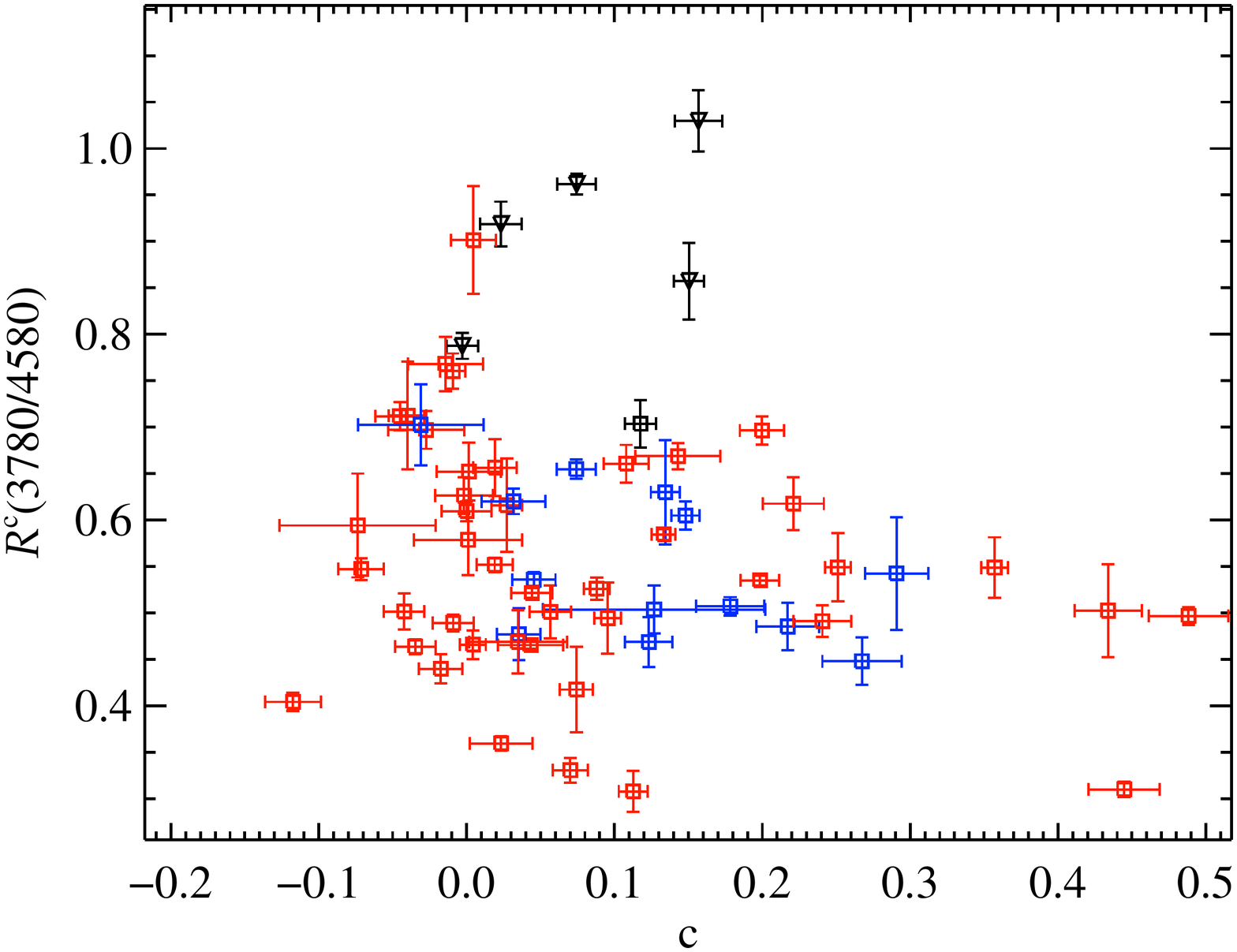}
\end{array}$
\caption[The best flux ratio using the $\left(x_1,c,\mathcal{R}^c\right)$
model versus $x_1$ and $c$]{The top-ranked flux ratio,
  $\mathcal{R}^c\left(3780/4580\right)$, using the
  $\left(x_1,c,\mathcal{R}^c\right)$ model 
  (Equation~\ref{eq:m4}) versus the SALT2 parameters $x_1$ and
  $c$.  Colours and shapes of data points are the
  same as in Figure~\ref{f:v_bv_si6355}.}\label{f:salt_r_x1_c} 
\end{figure}

Figure~\ref{f:salt_r_x1_c} shows the best flux ratio using the
$\left(x_1,c,\mathcal{R}^c\right)$ model,
$\mathcal{R}^c\left(3780/4580\right)$, versus the SALT2 parameters
$x_1$ and $c$. There is a slight correlation of this ratio with
$x_1$ (correlation coefficient 0.46) and effectively no correlation
with $c$. 
Since this ratio is essentially uncorrelated with both SALT2
parameters and it decreases the WRMS at the \about2$\sigma$ level, it
yields useful information about each SN beyond light-curve stretch and
colour. It is intriguing that this new information is found at the
blue end of the optical range, since there is evidence 
that spectral features in this region do not correlate with
light-curve parameters, yet contain information related to SN~Ia
luminosity \citep[e.g.,][]{Foley08:uv}.

Note that \citet{Blondin11} found that flux ratios with wavelengths
near \about5300~\AA\ and baselines of $< 400$~\AA\ gave the best
results for the $\left(x_1,c,\mathcal{R}^c\right)$ model. Using the
best ratio from their study, $\mathcal{R}^c\left(5690/5360\right)$,
the BSNIP data yield a WRMS of \about0.15~mag, which is not as good
as our top ten ratios.


\section{Conclusions}\label{s:conclusions}

This is the third paper in the BSNIP series and presents a comparison
between spectral feature measurements and photometric properties of
108 low-redshift ($z < 0.1$) SNe~Ia within 5~d of maximum
brightness. The spectral data all come from BSNIP~I, and the
photometric data come mainly from the LOSS sample and are published by
\citet{Ganeshalingam10:phot_paper}. The details of the spectral
measurements can be found in BSNIP~II, and the light-curve fits and
photometric parameters are in Ganeshalingam \etal (in preparation). A
combination of light-curve parameters (specifically the SALT2 stretch
and colour parameters $x_1$ and $c$) and spectral measurements are
used to calculate distances to SNe~Ia. We then compare the residuals
from these models to the standard model which only uses light-curve
stretch and colour. Future BSNIP papers will incorporate host-galaxy
properties and SN spectra at later epochs into the analysis presented
here.

\subsection{Summary of Investigated Correlations}\label{ss:summary}

The velocity gradient \citep{Benetti05} is compared to the light-curve
width, and it is shown that, as in BSNIP~II, the various classifications
based on the value of $\dot{v}$ overlap significantly. Similarly,
velocities at maximum brightness ($v_0$) are compared 
to photometric observables and classifications based on velocity
gradient, and there is a large amount of overlap in all of these
parameters as well. In earlier work, HV and HVG objects have been used
almost interchangeably, as have normal velocity and LVG objects
\citep[e.g.,][]{Hachinger06,Pignata08,Wang09}. However, the analyses
of BSNIP~II and this work show that these associations are not as
distinct as previously thought. 

The measured velocities of
the \ion{Si}{II} $\lambda$6355 and \ion{Si}{II} $\lambda$5972
features near maximum brightness are uncorrelated with observed
\bvmax. Furthermore, the HV objects and normal-velocity objects have
similar distributions of observed \bvmax\ values, though the HV
objects tend to have slightly larger observed \bvmax. 
When distances to SNe~Ia are 
computed using light-curve width ($x_1$) and colour ($c$) parameters,
no significant improvement in the accuracy of the distances
is found when the velocity of the \ion{Si}{II} $\lambda$6355
feature is added. This is contrary to what was seen by
\citet{Blondin11}. Furthermore, models involving $x_1$, $c$, and the
velocity of the \ion{Ca}{II}~H\&K feature fare as poorly. However, the
velocity of the \ion{S}{II} ``W,'' when used in conjunction with $x_1$
and $c$, leads to a decrease in the WRMS of the distances at the
1.8$\sigma$ level. Despite what was seen in the analysis of
\citet{Blondin11}, the use of relative depths of none of the features
analysed here leads to an improvement in the Hubble residuals.

The pEW of the \ion{Si}{II} $\lambda$4000 feature is strongly
anticorrelated with $x_1$, which confirms many previous studies
\citep{Arsenijevic08,Walker11,Blondin11,Nordin11a,Chotard11}. Furthermore, 
when using a model that includes $x_1$, $c$, and the pEW of this
feature, the residuals are as low as when using the standard
$\left(x_1,c\right)$ model. The pEWs of the \ion{Mg}{II} and
\ion{Fe}{II} complexes are both correlated with $c$, and since
interstellar reddening cannot affect pEWs significantly, it seems that
it is the {\it intrinsic} colour of the SN which is correlated with
both of these pEWs. However, when using
either of these pEWs (as a proxy for $c$) combined with the pEW of the
\ion{Si}{II} $\lambda$4000 feature (as a proxy for $x_1$), the Hubble
diagram residuals are significantly larger than when simply using
$x_1$ and $c$.

The pEWs of both \ion{Si}{II} $\lambda$5972 and \ion{Si}{II}
$\lambda$6355 are well correlated with $x_1$ and correlated with $c$,
but the use of the \ion{Si}{II} $\lambda$5972 pEW does not improve
distance calculations. However, using the \ion{Si}{II} $\lambda$6355
pEW (along with $x_1$ and $c$) leads to an improvement in the WRMS
residuals at the 1.2$\sigma$ level. 
Finally, the \ion{Ca}{II} near-IR triplet is
correlated with $c$ and MLCS2k2 light-curve width parameter
$\Delta$. This feature and the \ion{O}{I} triplet have not
been investigated thoroughly in studies similar to this one since other
large SN~Ia spectral datasets often do not include these spectral
regions.

The \ion{Si}{II} ratio, used as a luminosity indicator previously
\citep[e.g.,][]{Nugent95,Benetti05,Hachinger06}, is found to be
well correlated with $\Delta m_{15}$. However, we caution that at a
given value of $\Delta m_{15}$, there can exist various
spectroscopically classified subtypes of SNe~Ia. This ratio is also
found to be correlated with observed \bvmax, which has been seen in other work
\citep{Altavilla09}. We also show that the \ion{Si}{II} ratio is
{\it not} an accurate proxy for $x_1$ when calculating distance
moduli. A model using $c$ and $\Re$(\ion{Si}{II}) performs
significantly worse than the usual $\left(x_1,c\right)$ model, 
contrary to the conclusion of \citet{Blondin11}.

On the other hand, the \ion{Ca}{II} ratio is found to be a good
indicator of light-curve width, as it is well correlated with the
MLCS2k2 $\Delta$ parameter and with $\Delta m_{15}$. The BSNIP
data also indicate that the SiS ratio is correlated with both $x_1$
and $c$, and distance models using $\Re$(SiS) with just $c$ or with
both $c$ and $x_1$ perform as well as the standard
$\left(x_1,c\right)$ model. Finally, we confirm the 
results of \citet{Hachinger06} that the SSi and SiFe ratios are both
accurate luminosity indicators, as they are both well correlated with
$\Delta m_{15}$.

Following \citet{Bailey09} and \citet{Blondin11}, we calculate Hubble
diagram residuals using models which include combinations of the
usual light-curve parameters (width and colour) and arbitrary
sets of flux ratios. 
A total of 17,822 different ratios of fluxes are used alone, with $x_1$,
with $c$, and with both $x_1$ and $c$ to investigate whether 
any of these models might improve the accuracy of SN~Ia distance 
measurements. No models utilising only a flux ratio or a flux ratio
and $x_1$ are found to decrease the Hubble residuals. A handful of
models using a flux ratio and $c$ are seen to perform as well as
the standard $\left(x_1,c\right)$ model. Interestingly, most of these 
best ratios are extremely close to the SiS ratio mentioned
above.

These results differ from those of the previous studies of \citet{Bailey09} 
and \citet{Blondin11}, both of which found that flux ratios alone or in
conjunction with light-curve information would usually perform better
than the $\left(x_1,c\right)$ model. This may be due to the fact that
our ``standard model'' (without a spectral indicator) already performs
significantly better than that of \citet{Bailey09} or
\citet{Blondin11}. The differences may also be caused by the larger
number of spectroscopically peculiar SNe~Ia in the BSNIP sample.

Finally, when combining a flux ratio with both $x_1$ and $c$, our 
top-performing ratio, $\mathcal{R}^c\left(3780/4580\right)$, decreases the
Hubble residuals by 10~per~cent, which is significant at the
2$\sigma$ level. The WRMS of the residuals using this model is
$0.130 \pm 0.017$~mag, as compared to $0.144 \pm 0.019$~mag when using 
the same sample with the standard $\left(x_1,c\right)$ model. This
Hubble diagram has one of the smallest scatters ever published and
at the highest significance ever seen in such a study. The wavelengths
involved in most of the best-performing ratios in the
$\left(x_1,c,\mathcal{R}^c\right)$ model approximately correspond to
the midpoint of the  \ion{Ca}{II}~H\&K feature and the border between
the \ion{Mg}{II} and \ion{Fe}{II} complexes. This supports previous
work which has shown that near-UV spectra of SNe~Ia contain
information related to SN~Ia luminosity which is not necessarily
captured in the photometry \citep[e.g.,][]{Foley08:uv}. 

\subsection{The Future}\label{ss:future}

New large-scale surveys are already obtaining SN~Ia data, and they
are observing to higher redshifts and gathering larger amounts of data
than what is in the 
BSNIP sample \citep[e.g.,][]{Rau09,Law09,Kaiser02}. Even larger
surveys at even higher redshifts are also planned (e.g., LSST, WFIRST). Many
more SNe~Ia will be discovered than can be rigorously observed; we are
quickly entering the age of SN research where we are limited by the 
follow-up observations. Thus, there must be significant effort put forth to 
determine the most efficient way to monitor and utilise such vast
quantities of objects. That is one of the major goals of BSNIP.

Soon, for the vast majority of objects, there will only be (at best) a
handful of photometric observations near maximum brightness. Those,
combined with a relatively low S/N spectrum near maximum, will likely
be all the follow-up observations we get. From the work presented here (and in
BSNIP~II) we have shown that there still is hope.

The pEW of the \ion{Si}{II} $\lambda$4000 feature is a good indicator
of light-curve width, and the pEWs of the \ion{Mg}{II} and \ion{Fe}{II} 
complexes are relatively good proxies for colour. Unfortunately, the
correlations between these spectral measurements and the corresponding
photometric 
properties is not perfect, and distance calculations that employ only
these spectroscopic measurements do not perform as well as the
standard model which uses light-curve width and colour. However, this
is still a promising avenue for further investigation using new
datasets that are even larger than BSNIP. Other correlations that
appear marginal in the BSNIP dataset, or models tested here that
performed only equally as well as the usual $\left(x_1,c\right)$
model, should also be reexamined in the future.

Occasionally, one will be fortunate enough to have sufficient photometric
observations to produce a light curve for which SALT2 (or another
light-curve fitter) is able to determine a width and colour. In
these cases it appears that the light-curve parameters can be combined
with a flux ratio from a spectrum near maximum brightness to 
improve the accuracy of SN~Ia distances. The best ratios for this, as
determined from the BSNIP data, are all near
$\mathcal{R}^c($\about3750/\about4550$)$.

This is all somewhat heartening for surveys that will discover and 
monitor SNe~Ia at higher redshifts. \ion{Si}{II} $\lambda$4000, the
\ion{Mg}{II} and \ion{Fe}{II} complexes, and
$\mathcal{R}^c($\about3750/\about4550$)$ all involve spectral features 
which are toward the blue end of the optical range. This is critical
for higher-$z$ surveys since, as pointed out in BSNIP~II, the red wing
of the typical, near-maximum \ion{Si}{II} $\lambda$6355 feature
becomes redshifted beyond \about1\,$\mu$m for $z \ga 0.6$. Furthermore,
as discussed multiple times in BSNIP~II, measuring fluxes and pEWs
directly from a spectrum is much easier and less reliant on smoothing
models or functional form assumptions than velocities, for example.

To quote the concluding paragraph of \citet{Blondin11}, ``Do spectra
improve distance measurements of SN~Ia? Yes, but not as much as we had
hoped.''  We have to agree with the
authors of this quote both on the objective part (i.e., spectra {\it
  do} improve distance measurements), as well as on the subjective
part (i.e., we hoped they would improve things {\it even more}).

\section*{Acknowledgments}

We thank S.~J.~Bailey, S.~Blondin, R.~J.~Foley, and K.~Mandel for
useful discussions and comments on earlier drafts of this work, as well
as the anonymous referee for comments and suggestions that improved
the manuscript.  
We especially thank J.~W.~Richards for teaching us statistics; he is a real \AA ngel.
We are grateful to the staff at the Lick and Keck Observatories for
their support. Some of the data utilised herein were obtained at the
W. M. Keck Observatory, which is operated as a scientific partnership
among the California Institute of Technology, the University of
California, and the National Aeronautics and Space Administration
(NASA); the observatory was made possible by the generous financial
support of the W. M. Keck Foundation. We wish to recognise
and acknowledge the very significant cultural role and reverence that
the summit of Mauna Kea has always had within the indigenous Hawaiian
community; we are most fortunate to have the opportunity to conduct
observations from this mountain.
A.V.F.'s group is supported by the NSF grant AST-0908886, DOE grants
DE-FC02-06ER41453 (SciDAC) and DE-FG02-08ER41563, and the TABASGO
Foundation. A.V.F. is grateful for the hospitality of the Aspen Center
for Physics, where this paper was finalised during the January 2012
program on ``The Physics of Astronomical Transients.''

\bibliographystyle{mn2e}

\bibliography{astro_refs}

\label{lastpage}

\end{document}